\def\switch#1#2{#2}  

\switch{
\documentclass[a4paper,10pt,onecolumn]{article} 
}{
\documentclass[a4paper,aps,prd,twocolumn,groupedaddress,nofootinbib,floatfix]{revtex4}
}

\usepackage{graphicx}

\switch{
\usepackage{natbib}
\bibpunct{(}{)}{;}{a}{}{,}
\addtolength{\oddsidemargin}{-1.5cm}
\setlength{\evensidemargin}{\oddsidemargin}
\addtolength{\textwidth}{3.0cm}
\addtolength{\topmargin}{-1.5cm}
\addtolength{\textheight}{3.0cm}
}{}

\newlength{\fgwidth}
\switch{
\setlength{\fgwidth}{3.0em} 
}{
\setlength{\fgwidth}{1cm} 
}

\newcommand{\G}{\mathcal{G}}
\newcommand{\eps}{\varepsilon}

\begin{document}
\switch{
\newcommand{\apj}{ApJ}
\newcommand{\nat}{Nature}
\newcommand{\prd}{Phys. Rev. D}
}{}
\newcommand{\physrep}{Phys. Rep.}
\newcommand{\apjl}{ApJL}
\newcommand{\mnras}{MNRAS}
\newcommand{\aap}{A\&A}
\newcommand{\apjs}{ApJS}
\newcommand{\qjras}{qras}
\newcommand{\araa}{ARAA}
\newcommand{\aj}{AJ}
\newcommand{\aapr}{AAPR}
\newcommand{\aaps}{AAPS}
\newcommand{\pasp}{PASP}

\title{Information field theory\\for cosmological perturbation reconstruction\\and non-linear signal analysis}

\switch{
\author{Torsten A. En{\ss}lin, Mona Frommert, Francisco S. Kitaura\\[1em]
\textit{Max-Planck-Institut f\"ur Astrophysik}\\
\textit{Karl-Schwarzschild-Str. 1}\\
\textit{Garching, Germany}}
}{
\author{Torsten A. En{\ss}lin}
\author{Mona Frommert}
\author{Francisco S. Kitaura}
\affiliation{Max-Planck-Institut f\"ur Astrophysik, Karl-Schwarzschild-Str. 1, 85741 Garching, Germany}
\date{\today}
}


\switch{
\maketitle
}{}

\begin{abstract} 
We develop \textit{information field theory} (IFT) as a means of Bayesian
inference on spatially distributed signals, the information fields.  A
didactical approach is attempted. Starting from general considerations
on the nature of 
measurements, signals, noise, and their relation to a physical reality, we
derive the information Hamiltonian, the source field, propagator, and
interaction terms. Free IFT reproduces the well known Wiener-filter
theory. Interacting IFT can be diagrammatically expanded, for which we provide
the Feynman rules in position-, Fourier-, and spherical harmonics space, and the 
Boltzmann-Shannon information measure.
The theory should be applicable in many fields. However, here, two cosmological
signal recovery problems are discussed in their IFT-formulation. 1)
Reconstruction of the cosmic large-scale structure matter distribution from
discrete galaxy counts in incomplete galaxy surveys within a simple model
of galaxy formation. We show
that a Gaussian signal, which should resemble the
initial density perturbations of the Universe, observed with a strongly
non-linear, incomplete and Poissonian-noise affected response, as the processes
of structure and galaxy formation and observations provide, can be
reconstructed thanks to the virtue of a response-renormalization flow
equation. 
2) We design a filter to detect local non-linearities in
the cosmic microwave background, which are predicted from some
Early-Universe inflationary scenarios, and expected due to measurement
imperfections. This filter is the optimal Bayes' 
estimator up to linear order in the non-linearity
parameter and can be used even to 
construct sky maps of non-linearities in the data. 
\end{abstract}

\switch{\newpage}{
\maketitle
}


\section{Introduction}

\subsection{Motivation}

The optimal extraction and restoration of information from data on spatially
distributed quantities like the cosmic \textit{large-scale structure} (LSS) or the \textit{cosmic
microwave background} (CMB) temperature fluctuations in cosmology, but also on
many other signals in physics and related fields, is essential for any
quantitative, data-driven scientific  inference. The problem of how to
design such methods possesses many technical and even conceptual
difficulties, which have led to a large number of recipes and methodologies.

Here, we address such problems from a strictly information
theoretical point of view. We show, as others have done before, that 
information theory for distributed quantities leads to a statistical
field theory, which we  
name \textit{information field theory} (IFT). In contrast to the
previous works, which mostly treat 
such problems on a classical field level, as will be detailed later,
here, we take full advantage  
of the existing field theoretical apparatus to treat interacting and
non-classical fields. Thus, we show how to use diagrammatic
perturbation theory and renormalization flows in order to construct
optimal signal recovering algorithms and to calculate moments of their
uncertainties. Non-classicality manifests itself as quantum and statistical
fluctuations in quantum and statistical field theory (QFT \& SFT), and
very similarly as  uncertainty in IFT.

The information theoretical perspective on signal inference problems
has technical advantages, since it permits to design information-yield
optimized algorithms and experimental setups. However, it also
provides deeper insight into the mechanisms of knowledge accumulation,
its underlying information flows, and its dependence on data models,
prior knowledge and assumptions than pure empirical evaluations of
ad-hoc algorithms alone could provide.

We therefore hope that our work is of interest for two types of
readers. The first are applied scientists, who are mainly interested
in the practical aspect of IFT since they are facing  a concrete
inverse problem for a spatially distributed quantity, especially but
not exclusively in cosmology. The second are more philosophical or
theoretically inclined scientists, for whom IFT may serve as a
framework to understand and classify many of the existing methods of
signal extraction and reception. Since we expect that many interested
readers are not very familiar with field theoretical formalisms, we
introduce some of its basic mathematical concepts. Due to this
anticipated non-uniform readership,  
not everything in this article might be of everyones interest, and
therefore we provide in the following a short overview on the
structure and content of the article.

\subsection{Overview of the work}

The remainder of this introduction section contains a detailed discussion 
of the previous work on signal inference theory as well as a very
brief introduction into the here relevant works on the cosmic LSS and the CMB.
The main part of this article falls into two categories: abstract IFT
and its application. 
The concepts of IFT are introduced in Sec. \ref{sec:concepts}, where
Bayesian methodology, the distinction of physical and information
fields, the definition of signal response and noise, as well the
design of signal spaces are discussed. The basic IFT formalism
including the free theory is introduced in Sec. \ref{sec:basics},
which, according to our judgement, summarizes and unifies the previous  
knowledge on IFT before this paper. An impatient reader, 
only interested in applying IFT and not worrying about concepts, 
may start reading in Sec. \ref{sec:basics}. 
From Sec. \ref{sec:interacting fields}
on the new results of this work are presented, starting with the discussion of 
interacting information fields, their Hamiltonians and Feynman
rules, and the Boltzmann-Shannon information measure. 
The normalisability of sensibly constructed IFTs is shown, as well the classical
information field equation is presented there. A step-by-step recipe
of how to derive and implement an IFT algorithms is also provided.

Details of the notation can be found, if not defined in the main text, in Appendix \ref{sec:notation}.

Applications of the theory are
provided in the following two sections, which can be skipped by a reader
interested only in the general theoretical framework. Although
specific inference problems are addressed, they should serve as a
blueprint for the tackling of similar problems. In
Sec. \ref{sec:poisson} 
the problem of the reconstruction of the cosmic matter distribution from galaxy
surveys is analyzed in terms of a Poissonain data model. In Sec. \ref{sec:fnl}
we derive an optimal estimator for non-Gaussianity in the CMB, and show how it
can be generalized to map potential non-Gaussianities in the CMB
sky. Our summary and outlook can be found in Sec. \ref{sec:summary}.

\subsection{Previous works}\label{sec:lit}

The work presented here tries to unify information theory and statistical field
theory in order to provide a conceptual framework in which optimal tools for
cosmological signal analysis can be derived, as well as for inference problems in
other disciplines. Below, we provide very brief introductions into each of the
required fields\footnote{This work has tremendously benefitted in a direct and indirect way from a large
number of previous publications in those fields. We, the authors, have to
apologize for being unable to give full credit to all relevant former works in
those fields for only concentrating on a brief summary of the papers more or
less directly influencing this work. This collection is obviously highly biased
towards the cosmological literature due to our main scientific interests and
expertise, and definitely incomplete.} (information theory, image reconstruction, statistical field
theory, cosmological large-scale structure, and cosmic microwave background),
for the orientation of non-expert readers. An expert in any of these
fields might decide to skip the corresponding sections.

\subsubsection{Information theory and Bayesian inference}

The fundament of information theory was laid by the work of \citet{Bayes} on
probability theory, in which the celebrated Bayes theorem was presented. The
theorem itself (see Eq. \ref{eq:Bayes}) is a simple rule for conditional
probabilities. It only unfolds its power for inference problems if used with
belief or knowledge states, described by conditional probabilities.

The advent of modern information theory is probably best dated by the work of Shannon
\citep{Shannon1948,1949mtc..book.....S} 
on the concept of information measure, being the negative Boltzmann-entropy,
and the work of Jaynes, combining the language of statistical mechanics and
Bayes probability theory and applying it to knowledge uncertainties
\citep{1957PhRv..106..620J, 1957PhRv..108..171J, jaynes1963,
1965AmJPh..33..391J, jaynes1968, 1982ieee...70..939J, 2004PhT....57j..76J}.
The required numerical evaluation of Bayesian probability integrals suffered
often from the curse of high dimensionality.  
The standard recipe against this, still in massive use today, is importance
sampling via Markov-Chain Monte-Carlo Methods (MCMC),  
following the ideas of  
\citet{metroplis}, 
\citet{hastings}, and 
\citet{gibbsamp}, 
where the latter authors already had image reconstruction applications in
mind. 
The Hamiltonian MCMC methods \citep{HMCMC}, in which the phase-space
sampling is partly following Hamiltonian dynamics, are also of
relevance here. There the Hamiltonian is introduced as the negative
logarithm of the probability, as we do in this work. 

With such tools, higher dimensional problems, as present in signal
restoration, could and can be tackled,  
however, for the price of getting stochastic uncertainty into the
computational results. For a recent review on image restoration
MCMC techniques, see \cite{2005cs........4037M}. 

The applications and extensions of these pioneering works are too numerous to
be listed here.  
Good monographs exist and the necessary references can be found there
\switch{%
:
\cite{toolsstatinf}, 
\cite{neal1993}, 
\cite{bayeschoice}, 
\cite{bayesdataanal}, and 
\cite{paramest}. 
}{%
\cite{toolsstatinf, 
neal1993, 
bayeschoice, 
bayesdataanal, 
paramest, 
2008arXiv0803.4089T}. 
}

\subsubsection{Image reconstruction in astronomy and elsewhere}\label{sec:lit:map}

The problem of image reconstruction from incomplete, noisy data is especially
important in astronomy, where the experimental conditions are largely set by
the nature of distant objects, weather conditions, etc., all mainly out of the
control of the observer, as well as in other disciplines like medicine and
geology, with similar limitations to arrange the object of observations for an
optimal measurement.  Some of the most prominent methods of image
reconstruction, which are based on a Bayesian implementation of an assumed data
model, are the Wiener-filter
\citep{1949wiener}, 
the Richardson-Lucy algorithm \citep{1972JOSA...62...55R, 1974AJ.....79..745L}, 
and the maximum-entropy image restoration \citep{1972JOSA...62..511F}\switch{%
\footnote{See also 
\cite{1972JOSA...62..511F}, 
\cite{1978Natur.272..686G}, 
\cite{1979MNRAS.187..145S}, 
\cite{1980MNRAS.191...69B}, 
\cite{1983CVGIP..23..113B}, 
\cite{1983iimp.conf..267G}, 
\cite{1984Natur.311..446S}, 
\cite{1984Natur.312..381T}, 
\cite{1984MNRAS.211..111S}, 
\cite{1986JMOp...33..287B}, 
\cite{gull1989}, 
\cite{gullskilling}, and 
\cite{1998mebm.conf....1S}.
}}{(see also 
\cite{
1978Natur.272..686G, 
1979MNRAS.187..145S, 
1980MNRAS.191...69B, 
1983CVGIP..23..113B, 
1983iimp.conf..267G, 
1984Natur.311..446S, 
1984Natur.312..381T, 
1984MNRAS.211..111S, 
1986JMOp...33..287B, 
gull1989, 
gullskilling,  
1998mebm.conf....1S}).
}

The Wiener filter can be regarded to be a full Bayesian image inference method
in case of Gaussian signal and noise statistics, as we will show in
Sect. \ref{sec:free}. It will be the working horse of the IFT formalism, since
the Wiener filter represents the algorithm to construct the exact field
theoretical expectation value given the data for an interaction-free
information Hamiltonian. The filter can be decomposed into two essential
information processing steps, first building the information source by
response-over-noise weighting the data, and then propagating this information
through the signal space, by applying the so called Wiener variance.

The Richardson-Lucy algorithm is a maximum-likelihood method to reconstruct
from Poissonian data and therefore is also of Bayesian origin. This method has
usually to be regularized by hand, by truncation of the iterative calculations,
against an over-fitting instability due to the missing (or implicitly flat)
signal prior. A Gaussian-prior based regularization was recently proposed by
\citet{2008MNRAS.389..497K}, and the implementation of a variant of this is
presented here in Sect. \ref{sec:classicalPoisson}.

Maximum entropy algorithms will not be the topic here, as well as not a number of other 
existing methods, which are partly within and partly outside the
Bayesian framework. They may be found in existing reviews on this
topic 
\citep[e.g.][]{1986ARA&A..24..127N, imagerest}. 

\subsubsection{Statistical and Bayesian field theory}

The relation of signal reconstruction problems and field theory was discovered
independently by several authors. In cosmology, a prominent work in this
directions is
\citet{1987ApJ...323L.103B}, 
in which the path integral approach was proposed to sample primordial density
perturbations with a Gaussian statistics under the constraint of existing
information on the large scale structure. The work presented here can be
regarded as a non-linear, non-Gaussian extension of this. 
Many methods from statistics and from statistical mechanics were of
course used even earlier, e.g. the usage of moment generating function
for cosmic density fields can already be found in \citet{1985ApJ...289...10F}. 

Simultaneously to  Bertschinger's work, 
\citet{1987PhRvL..58..741B, 1988PhRvL..61.1512B} 
argued that visual perception can be modeled as a field theory for the true
image, being distorted by noise and other data transformations, which are
summarized by a nuisance field. A probabilistic language was used, but no
direct reference to information theory was made, since not the optimal
information reconstruction was the aim, but a model for the human visual
reception system. However, this work actually triggered our research. 
 
\citet{1996PhRvL..77.4693B} 
applied a field theoretical approach to recover a probability distribution from data. Here, a Bayesian prior was used to regularize the 
solution, which was set up ad-hoc to enforce smoothness of the reconstruction,
obtained from the classical (or saddlepoint, or maximum a posteriori) solution
of the problem. However, an ``optimal'' value for the smoothness controlling
parameter was derived from the data itself, a topic also addressed by
\citet{2000AJ....120.2163S} and by a follow up publication to
ours \citep{PURE}. \citet{1996PhRvL..77.4693B} also recognized, as we do, that an IFT can
easily be non-local. 

Finally, the work of Lemm and coworkers\switch{%
\footnote{\citet{1998cond.mat..8039L, 1999physics..11077L,
2000FBS....29...25L, 2000quant.ph..2010L, 2000PhRvL..84.2068L,
2000PhRvL..84.4517L, 2000PhLA..276...19L, 2001AIPC..568..425L,
2001EPJB...20..349L, 2003quant.ph.12191L, 2005EPJB...46...41L}}
}{
\cite{1999physics..12005L, 2000FBS....29...25L, 2000PhRvL..84.2068L,
2000PhRvL..84.4517L, 2000PhLA..276...19L, 2001AIPC..568..425L,
2001EPJB...20..349L, 2005EPJB...46...41L}}
established a tight connection between statistical field theory and
Bayesian inference, and proposed the term
\textit{Bayesian field theory} (BFT) for this. 
However, we prefer the term {\it information field theory} since it
puts the emphasis on the relevant object, the information, 
whereas BFT refers to a method, Bayesian inference. 
The term
\textit{information field} is rather self-explaining, whereas the
meaning of a  \textit{Bayesian field} is not that obvious.
  
The applications considered by Lemm concentrate
on the reconstruction of probability fields over parameter spaces and
quantum mechanical potentials by means of the maximum a posteriori
equation. The extensive book summarizing the essential insights of
these papers, \cite{1999physics..12005L}, clearly states the
possibility of perturbative expansions of the field theory. However,
this is not followed up by these authors probably for reasons of the
computational complexity of the required algorithms. In contrast to
many of the previous works on IFT, which deal with ad-hoc priors, the
publication by 
\citet{1998cond.mat..8039L} 
is remarkable, since it provides explicit recipes of how to implement a priori
information in various circumstances more rigorously. 

The mathematical tools required to tackle IFT problems come
from SFT and QFT, which have a vast literature. We
have specially made use of the books of \citet{Binney1992},
\citet{PeskinSchroeder}, and \citet{2003qftn.book.....Z}. 

\subsubsection{Cosmological large-scale structure}\label{sec:lit:lss}

Our first IFT example in Sec. \ref{sec:poisson} is geared towards improving
galaxy-survey based cosmography, the reconstruction of the large-scale
structure matter distribution. We provide here a short overview on the relevant
background and works. 

The LSS of the matter distribution of the Universe is traced
by the spatial distribution of Galaxies, and therefore well observable. 
This structure is believed to have emerged from tiny, mostly Gaussian initial
density fluctuations of a relative strength of $10^{-5}$ via a
self-gravitational instability, partly counteracted by the expansion of the
Universe. The initial density fluctuations are believed to be produced during
an early inflationary epoch of the Universe, and to carry valuable information
about the inflaton, the field which drove inflation, in their
$N$-point correlation functions, to be extracted from the observational data. 

The onset of the structure formation process is well described by linear
perturbation theory and therefore to conserve Gaussianity, however, the later
evolution, the structures on smaller 
scales, and especially the galaxy formation require non-linear
descriptions. The observational situation is complicated by the fact that the
most important galaxy distance indicator, their redshift, is also sensitive to
the galaxy peculiar velocity, which causes the observational data on the
three-dimensional LSS to be partially degenerated. 
There are analytical methods to describe these effects%
\switch{%
\footnote{
Of special interest in this context may be the following works, and the papers they refer to:
\cite{1970A&A.....5...84Z}, 
\cite{1986ApJ...304...15B}, 
\cite{1986ApJ...310L..21M}, which already applies path-integrals,  
\cite{1980lssu.book.....P}, 
\cite{1987MNRAS.227....1K}, 
\cite{1990ApJ...362....1P}, 
\cite{1992ApJ...390L..61B}, 
\cite{1996ApJ...462...25Z}, 
\cite{hamilton-1998}, 
\cite{1999MNRAS.309..543B}, 
\cite{1999MNRAS.308....1B}, 
\cite{1999ApJ...520...24D}, 
\cite{2002astro.ph..6052Z}, 
\cite{2003MNRAS.341.1311S}, 
and
\cite{2004PhRvD..70h3007S}. 
}}{%
\footnote{%
Of special interest in this context may be \cite{1986ApJ...310L..21M},
which already applies path-integrals,   
\cite{1970A&A.....5...84Z, 
1986ApJ...304...15B, 
1980lssu.book.....P, 
1987MNRAS.227....1K, 
1990ApJ...362....1P, 
1992ApJ...390L..61B, 
1996ApJ...462...25Z, 
hamilton-1998, 
1999MNRAS.309..543B, 
1999MNRAS.308....1B, 
1999ApJ...520...24D, 
2002astro.ph..6052Z, 
2003MNRAS.341.1311S, 
2004PhRvD..70h3007S}, 
and the papers they refer to.}}, 
and also extensive work on $N$-body simulations of the
structure formation, the latter probably providing us with the most detailed
and accurate statistical data on the properties of the matter density
field \citep[e.g][]{2005Natur.435..629S}.

In recent years, it was recognized that the evolution of the cosmic density
field and its statistical properties can be addressed with 
field theoretical methods by virtue of renormalization flow
equations. Detailed semi-analytical calculations for the density field
time propagator, the two- and three- point correlation functions are
now possible due to this, which are expected to play an important role
in future approaches to reconstruct the initial fluctuations from the
observational data\switch{%
\footnote{Relevant references are 
\cite{2004A&A...421...23V, 2007A&A...476...31V, 2008A&A...484...79V}, 
\cite{2006PhRvD..73f3519C}, 
\cite{2006PhRvD..73f3520C}, 
\cite{2006PhRvD..74j3512M, 2006PhRvD..74l9901M, 2007PhRvD..75d3514M}, 
\cite{2006ApJ...651..619J, 2008arXiv0805.2632J}, 
\cite{2007JCAP...06..026M}, 
\cite{2007JPhA...40.6849G}, 
\cite{2008MPLA...23...25M}, 
\cite{2008PhRvD..77f3530M}, 
and
\cite{2008arXiv0806.0971P}. 
}}{
\cite{2004A&A...421...23V, 2007A&A...476...31V, 2008A&A...484...79V, 
2006PhRvD..73f3519C, 
2006PhRvD..73f3520C, 
2006PhRvD..74j3512M, 2006PhRvD..74l9901M, 2007PhRvD..75d3514M, 
2006ApJ...651..619J, 2008arXiv0805.2632J, 
2007JCAP...06..026M, 
2007JPhA...40.6849G, 
2008MPLA...23...25M, 
2008PhRvD..77f3530M, 
2008arXiv0806.0971P}} 
.

It was recognized early on, that the primordial density fluctuations can in
principle be reconstructed from galaxy observations
\citep{1987ApJ...323L.103B}. This has lead to a large development of various
numerical techniques for an optimal reconstruction\switch{%
\footnote{Some of the main references are:
\cite{1989ApJ...336L...5B, 1991ASPC...15...67B}, 
\cite{1989ApJ...344L..53P}, 
\cite{1990ApJ...364..349D}, 
\cite{1991ASPC...15..111K}, 
\cite{1991ApJ...380L...5H}, 
\cite{1992MNRAS.254..315W}, 
\cite{1992ApJ...391..443N}, 
\cite{1992ApJ...398..169R}, 
\cite{1993ApJ...405..449G}, 
\cite{1993ApJ...415L...5G}, 
\cite{1994A&A...291..697B}, 
\cite{1994ApJ...421L...1N}, 
\cite{1994ASPC...67..171L}, 
\cite{1994ApJ...423L..93L}, 
\cite{1995MNRAS.272..885F}, 
\cite{1995MNRAS.277..933S}, 
\cite{1995ApJ...449..446Z}, 
\cite{1995ApJ...453..533T}, 
\cite{1997MNRAS.285..793C}, 
\cite{1998ApJ...508..440N}, 
\cite{1998ApJ...504..601P}, 
\cite{1998ApJ...503..492S, 1998ApJ...506...64S}, 
\cite{1998ApJ...492..439B}, 
\cite{1999MNRAS.306..491T}, 
\cite{1999ApJ...515..471N}, 
\cite{1999ApJ...520..413Z}, 
\cite{2000ASPC..201..282G}, 
\cite{2000ApJ...544...21G}, 
\cite{2000MNRAS.316..464K}, 
\cite{2001ApJ...550..522B}, 
\cite{2001ApJ...552..413G}, 
\cite{2002Natur.417..260F}, 
\cite{2002MNRAS.331..901Z}, 
\cite{2003MNRAS.346..501B}, 
\cite{2003A&A...406..393M}, 
\cite{2004astro.ph.10063M}, 
\cite{2004MNRAS.349..425B}, 
\cite{2005ApJ...635L.113M}, 
and
\cite{2006MNRAS.365..939M}. 
}}{
\cite{1989ApJ...336L...5B, 1991ASPC...15...67B, 
1989ApJ...344L..53P, 
1990ApJ...364..349D, 
1991ASPC...15..111K, 
1991ApJ...380L...5H, 
1992MNRAS.254..315W, 
1992ApJ...391..443N, 
1992ApJ...398..169R, 
1993ApJ...405..449G, 
1993ApJ...415L...5G, 
1994A&A...291..697B, 
1994ApJ...421L...1N, 
1994ASPC...67..171L, 
1994ApJ...423L..93L, 
1995MNRAS.272..885F, 
1995MNRAS.277..933S, 
1995ApJ...449..446Z, 
1995ApJ...453..533T, 
1997MNRAS.285..793C, 
1998ApJ...508..440N, 
1998ApJ...504..601P, 
1998ApJ...503..492S, 1998ApJ...506...64S, 
1998ApJ...492..439B, 
1999MNRAS.306..491T, 
1999ApJ...515..471N, 
1999ApJ...520..413Z, 
2000ASPC..201..282G, 
2000ApJ...544...21G, 
2000MNRAS.316..464K, 
2001ApJ...550..522B, 
2001ApJ...552..413G, 
2002Natur.417..260F, 
2002MNRAS.331..901Z, 
2003MNRAS.346..501B, 
2003A&A...406..393M, 
2004astro.ph.10063M, 
2004MNRAS.349..425B, 
2005ApJ...635L.113M, 
2006MNRAS.365..939M}}
.
Many of them are based on a Bayesian approach, since they are implementations and extension of the Wiener filter. 
However, also other principles are used, like, e.g. the least action approach,
or Voronoi tessellation techniques\switch{%
\footnote{
E.g. 
\cite{1991QJRAS..32...85I}, 
\cite{1991MNRAS.250..519I}, 
\cite{1996MNRAS.279..693B}, 
\cite{2000A&A...363L..29S}, 
\cite{2001A&A...368..776R}, 
and
\cite{2006ChJAA...6..387Z}. 
}.}{
\cite[e.g.][]{1991QJRAS..32...85I, 
1991MNRAS.250..519I, 
1996MNRAS.279..693B, 
2000A&A...363L..29S, 
2001misk.conf..268V, 
2001A&A...368..776R, 
2006ChJAA...6..387Z}.} 
A discussion and classification of the various methods can be found in 
\cite{2008MNRAS.389..497K}.

Especially the Wiener filter methods were extensively applied to galaxy
survey data\switch{%
\footnote{Survey based reconstructions of the cosmic matter fields can be found in
\cite{1990ApJ...364..370B}, 
\cite{1991ApJ...372..380Y}, 
\cite{WienerFSL}, 
\cite{1995ApJ...454...15S}, 
\cite{1996ApJ...461L..17B}, 
\cite{1997MNRAS.287..425W}, 
\cite{1997ApJ...474..553Y}, 
\cite{1999AJ....118.1146S}, 
\cite{1999MNRAS.303..179N}, 
\cite{tegmark-1999-518}, 
\cite{2000ApJ...535L...5H}, 
\cite{2001ApJ...550...87G}, 
\cite{2002MNRAS.333..739M}, 
\cite{2004MNRAS.352..939E}, 
\cite{2004ogci.conf....5V}, 
\cite{2005ASPC..329..135H}, 
\cite{2005MNRAS.356.1168P}, 
and
\cite{2006MNRAS.373...45E}.} 
}{%
\footnote{Survey based reconstructions of the cosmic matter fields can be found in
\cite{1990ApJ...364..370B, 
1991ApJ...372..380Y, 
WienerFSL, 
1995ApJ...454...15S, 
1996ApJ...461L..17B, 
1997MNRAS.287..425W, 
1997ApJ...474..553Y, 
1999AJ....118.1146S, 
1999MNRAS.303..179N, 
tegmark-1999-518, 
2000ApJ...535L...5H, 
2001ApJ...550...87G, 
2002MNRAS.333..739M, 
2004MNRAS.352..939E, 
2004ogci.conf....5V, 
2005ASPC..329..135H, 
2005MNRAS.356.1168P, 
2006MNRAS.373...45E}.} 
}
and permitted partly to extrapolate the matter distribution into the \textit{zone
  of avoidance} behind the galactic disk\switch{\footnote{
\cite{1994ASPC...67..185H}, 
\cite{2000ASPC..218..173Z}, 
and
\cite{2000A&ARv..10..211K}.} 
 and to close the data-gap there, a topic
we also address in Sect. \ref{sec:poisson}.
}  
{ and to close the data-gap there, c.f.
\cite{1994ASPC...67..185H, 
2000ASPC..218..173Z, 
2000A&ARv..10..211K}
, a topic
we also address in Sect. \ref{sec:poisson}.
}

Another cosmological relevant information field to be extracted from galaxy
catalogues is the LSS power spectrum\switch{%
\footnote{\citep[E.g.][]{1994MNRAS.267.1020P, 
1996ApJ...465...34V, 1997ApJ...486...21Z, 1997PhRvL..79.3806T,
1999ApJ...511....5E}.}}{
\cite[e.g.][]{1994MNRAS.267.1020P, 
1996ApJ...465...34V, 1997ApJ...486...21Z, 1997PhRvL..79.3806T,
1999ApJ...511....5E}%
}.
This power is also measurable in the CMB, and for a long time the CMB provided
the best spectrum normalization
\citep{1992MNRAS.258P...1E, 1995ApJ...441L...9B}.

\subsubsection{Cosmic Microwave Background}\label{sec:lit:cmb}

Since our second example deals with the CMB, we give a brief overview on it and
on related inference methods.

The CMB reveals the statistical properties of the matter field at a time, when
the Universe was about 1100 times smaller in linear size than it is today. 
The photon-baryon fluid, which decouples at that epoch into neutral Hydrogen
and free streaming photons, has responded to the gravitational pull of the then
already forming dark matter structures. The photons from that epoch cooled due
to the cosmic expansion since then into the CMB radiation we observe today, and
carry information on the physical properties of the photon-baryon fluid of that
time like density, temperature and velocity. To very high accuracy, the
spectrum of the photons from any direction is that of a blackbody, with a mean
temperature of $2.7$ Kelvin and fluctuations of the order of $10^{-5}$ Kelvin,
imprinted by the primordial gravitational potentials at decoupling. 

Therefore, mapping these temperature fluctuations permits precisely to study
many cosmological parameters simultaneously, like the amount of dark matter
producing the gravitational potentials, the ratio of photons to baryons,
balancing the pressure and weight of the fluid, and geometrical and dynamical
parameters of space-time itself. The observations are technically challenging,
and therefore require sophisticated algorithms to extract the tiny signal of
temperature fluctuations against the instrument noise, but also to separate it
from other astrophysical foreground emission with the best possible accuracy.  

A number of such algorithms were developed\switch{%
\footnote{E.g.by
\cite{1992issa.proc..391J}, 
\cite{1994ApJ...432L..75B}, 
\cite{1997MNRAS.290..313M}, 
\cite{1997PhRvD..56.4514T}, 
\cite{1997ApJ...480L..87T}, 
\cite{1997ApJ...482..577D}, 
\cite{1998MNRAS.300....1H}, 
\cite{natoli}, 
\cite{2001A&A...374..358D}, 
\cite{maxima}, 
\cite{2004PhRvD..70h3511W}, 
\cite{2004ApJS..155..227E}, 
\cite{2004ApJ...609....1J}, 
\cite{mirage}, 
\cite{madam}, 
\cite{2006ApJS..162..401S}, 
\cite{2007ApJ...656..653L}, 
and
\cite{Hinshaw:2008kr}.} 
}{
\cite[e.g.][]{1992issa.proc..391J, 
1994ApJ...432L..75B, 
1997MNRAS.290..313M, 
1997PhRvD..56.4514T, 
1997ApJ...480L..87T, 
1997ApJ...482..577D, 
1998MNRAS.300....1H, 
natoli, 
2001A&A...374..358D, 
maxima, 
2004PhRvD..70h3511W, 
2004ApJS..155..227E, 
2004ApJ...609....1J, 
mirage, 
madam, 
2006ApJS..162..401S, 
2007ApJ...656..653L, 
Hinshaw:2008kr}}
,
which in many cases implement the Wiener filter. Thus, the required numerical
tools for an IFT treatment of CMB data are essentially available. 

The expected temperature fluctuations spectrum can be calculated from a linear
perturbative treatment of the Boltzmann equations of all dynamical active
particle species at this epoch, and fast computational implementations exists
permitting to predict it for a given set of cosmological parameters. Well known
codes for this task are publicly available\footnote{E.g. \texttt{cmbfast} 
(\switch{\texttt{http://cmbfast.org,
      http://ascl.net/cmbfast.html}}{\url{http://cmbfast.org},
    \url{http://ascl.net/cmbfast.html}}, \cite{1996ApJ...469..437S}),  
\texttt{camb} (\switch{\texttt{http://camb.info/}}{\url{http://camb.info/}},
\cite{Lewis:1999bs}), and \texttt{cmbeasy}
(\switch{\texttt{http://www.cmbeasy.org/}}{\url{http://www.cmbeasy.org/}},
\cite{2005JCAP...10..011D}).} and permit to extract information on cosmological
parameters from the measured CMB temperature fluctuation spectrum via
comparison to their predictions for a given parameter set. It was recognized
early on that this should happen in an information theoretically optimal way,
and Bayesian methods were therefore adapted in that area well before in other
astrophysical disciplines \citep[e.g.][]{1995ApJ...446...49B,
1997ApJ...480...22T, 1997PhRvD..55.5895T,Nolta:2008ih}.

The initial metric and density fluctuations, from which the CMB fluctuations
and the LSS emerged, are believed to be initially seeded by
quantum fluctuations of a hypothetical inflaton field, which should have driven
an inflationary expansion phase in the very early Universe 
\citep{1981PhRvD..23..347G, 1982PhLB..108..389L, 1982PhRvL..48.1220A, 
1982PhRvL..49.1110G, 1982PhLB..117..175S, 1983PhRvD..28..679B}.
The inflaton-induced fluctuations have a very Gaussian probability
distribution, however, some non-Gaussian features seem to be unavoidable in
most scenarios and can serve as a fingerprint to discriminate among them
\citep[e.g.][]{2001PhRvD..64h3005H, 2002PhRvD..66j3506B, 2004PhR...402..103B,
  2004JCAP...08..009B}. 
Observational tests on such non-Gaussianities based on the three-point
correlation function of the CMB data \citep[e.g.][]{2002ApJ...566...19K,
  2004PhRvD..70h3005B, 2005ApJ...634...14K, 2007ApJ...664..680Y,
  2008ApJ...678..578Y} 
were so far mostly negative, however not sensitive enough to seriously
constrain the possible theoretical parameter space of inflationary scenarios,
see 
e.g. \cite{2003ApJS..148..119K, 2008arXiv0804.0136C}. 
Recently, there has been the claim of a detection of such non-Gaussianities by
\citet{2008PhRvL.100r1301Y}
and a confirmation of this with better data and improved algorithms is
therefore highly desirable. In Sect. \ref{sec:fnl} we make a proposal for
improving the algorithmic side of this challenge. A recent review on the
current status of CMB-Gaussianity can be found in \cite{2008arXiv0805.4157M}.

\section{Concepts of information field theory}\label{sec:concepts}
\subsection{Information on physical fields}\label{sec:infophys}

In our attempts to infer the properties of our Universe from astronomical
observations we are faced with the problem of how to interpret incomplete,
imperfect and noisy data, draw our conclusions based on them and quantify the
uncertainties of our results. This is true for using galaxy surveys to map the
cosmic LSS, for the interpretation of the CMB, 
as well for many experiments in physical laboratories and
compilations of geological, economical, sociological, and biological data about
our planet. Information theory, which is based on probability theory and the
Bayesian interpretation of missing knowledge as probabilistic uncertainty,
offers an ideal framework to handle such problems. It permits to describe all
relevant processes involved in the measurement probabilistically, provided a
model for the Universe or the system under consideration is adopted. 

The states of such a model, denoted by the state variable $\psi$, are 
identified with the possible physical realities. They can have
probabilities $P(\psi)$ assigned to them, the so-called prior information. This
prior contains our knowledge about the Universe as we model it before any other
data is taken. For a given cosmological model, the prior may be the probability
distribution of the different initial conditions of the Universe, which
determine the subsequent evolution completely. Since our Universe is spatially
extended, the state variable will in general contain one or several fields,
which are functions over some coordinates $x$.

Also the measurement process is described by a data model which defines the
so-called likelihood, the probability $P(d|\psi)$ to obtain a specific dataset
$d$ given the physical condition $\psi$. In case the outcome $d$ of
the measurement is deterministic $P(d|\psi) = \delta(d - 
d[\psi])$, where $ d[\psi]$ is the functional dependence of the data on the
state. In any case, the probability distribution function of the data,
\begin{equation}
 P(d) = \int \mathcal{D}\psi\, P(d|\psi)\, P(\psi),
\end{equation}
is given in terms of a phase-space or path integral over all possible
realizations of $\psi$, 
to be defined more precisely later (Sect. \ref{sec:discret}).

A scientist is not actually interested in the total state of the Universe, but
only in some specific aspects of it, which we call the signal $s =
s[\psi]$. The signal is a very reduced description of the physical reality, and
can be any function of its state $\psi$, freely chosen according to the needs
and interests of the scientist or the ability and capacity of the measurement
and computational devices used. Since the signal does not contain the full
physical state, any physical degree of freedom which is not present in the
signal but influences the data will be received as probabilistic uncertainty,
or shortly noise. The probability distribution function of the signal, its
prior
\begin{equation}\label{eq:P(s)}
 P(s) = \int \mathcal{D}\psi\,\delta(s-s[\psi])\, P(\psi), 
\end{equation}
is related to that of the data via the joint probability
\begin{equation}\label{eq:P(d,s)}
 P(d,s) = \int \mathcal{D}\psi\,\delta(s-s[\psi]) \,P(d|\psi)\, P(\psi),
\end{equation}
from which the conditional signal likelihood
\begin{equation}\label{eq:likelihood}
P(d|s) =   P(d,s) / P(s)
\end{equation}
and signal posterior
\begin{equation}\label{eq:posterior}
P(s|d) =   P(d,s) / P(d)
\end{equation}
can be derived. 

Before the data is available, the phase-space of interest is spanned by the
direct product of all possible signals $s$ and data $d$, and all regions with non-zero
$P(d,s)$ are of potential relevance. Once the actual data $d_\mathrm{obs}$
have been taken, only a sub-manifold of this space, as fixed by the
data, is of further relevance. The probability function over this
sub-space is proportional to $P(d=d_\mathrm{obs},s)$, and needs just
to be renormalized by dividing by 
\begin{eqnarray}
 \int \mathcal{D}s \, P(d_\mathrm{obs},s) \!\!&=&\!\!\int\! \mathcal{D}s \int
 \! \mathcal{D}\psi \,\delta(s-s[\psi]) \,P(d_\mathrm{obs}|\psi)\, P(\psi)
 \nonumber\\ \!\!&=&\!\!  \int\! \mathcal{D}\psi \,P(d_\mathrm{obs}|\psi)\,
 P(\psi) = P(d_\mathrm{obs}),
\end{eqnarray}
which is the unconditioned probability (or evidence) of that
data. Thus, we find the resulting information of the data to be the
posterior distribution $P(s|d_\mathrm{obs}) =
P(d_\mathrm{obs},s)/P(d_\mathrm{obs})$. 
This posterior is the fundamental mathematical object from which all our
deductions have to be made. It is related via Bayes's theorem \citep{Bayes} to the
usually better accessible signal likelihood, 
\begin{equation}\label{eq:Bayes}
P(s|d) =   P(d|s)\,P(s) / P(d),
\end{equation}
which follows from Eqs. \ref{eq:likelihood} and \ref{eq:posterior}.

The normalization term in Bayes's theorem, the evidence $ P(d)$, is now also fully
expressed in terms of the joint probability of data and signal,  
\begin{equation}
 P(d) = \int \mathcal{D}s\, P(d,s),
\end{equation}
and the underlying physical field $\psi$ basically becomes invisible at this
stage in the formalism. The evidence plays a central role in Bayes inference,
since it is the likelihood of all the assumed model parameters. Combining this
parameter-likelihood with parameter-priors one can start Bayesian inference on
the model classes.

\subsection{Signal response and noise}\label{sec:response&noise}

If signal and data depend on the same underlying physical properties,
there may be correlations between the two, which can be expressed in
terms of signal response $R$ and noise $n$ of the data as 
\begin{equation}
 d = R[s] + n_s.
\end{equation}
We have chosen two different ways of denoting the dependence of response and
noise on the signal $s$, in order to highlight that the response
should embrace most of the reaction of the data to the signal, whereas
the noise should be as independent as possible. We ensure this by
putting the linear correlation of the data with the signal fully into
the response. The response is therefore the part of the data which
correlates with the signal 
\begin{equation}
\label{eq:signal response}
 R[s] \equiv \langle d \rangle_{(d|s)} \equiv  \int \mathcal{D}d\, d \, P(d|s),
\end{equation}
and the noise is just defined as the remaining part which does not:
\begin{equation}
 n_s \equiv d - R[s] = d - \langle d \rangle_{(d|s)} .
\end{equation}
Although the noise might depend on the signal, as it is well known for
example for Poissonian processes, it is -- per definition -- linearly
uncorrelated to it, 
\begin{equation}
 \langle n_s \, s^\dagger \rangle_{(d|s)} =(\langle d \rangle_{(d|s)} - R[s])\,
 s^\dagger =0\, s^\dagger =0, 
\end{equation}
whereas higher order correlation might well exist and may be further
exploited for their information content. The dagger denotes complex
conjugation and transposing of a vector or matrix. 

These definitions were chosen to be close to the usual language in signal
processing and data analysis. They permit to define signal response and noise
for an arbitrary choice of the signal $s[\psi]$. No direct causal connection
between signal and data is needed in order to have a non-trivial response,
since both variables just need to exhibit some couplings to a common sub-aspect
of $\psi$. The above definition of response and noise is however not unique,
even for a fixed signal definition, since any data transformation $d' = T[d]$
can lead to different definitions, as seen from
\begin{equation}
 R'[s] \equiv \langle d' \rangle_{(d|s)} = \langle T[d] \rangle_{(d|s)} \neq
 T[\langle d \rangle_{(d|s)}] = T[R[s]].
\end{equation}
Exceptions are some unique relations
between signal and state, $P(\psi|s)= \delta(\psi - \psi[s])$, and maybe a
few other very special cases. Thus, the concepts of signal response
and therewith defined noise depend on the adopted coordinate system in
the data space. This coordinate system can be changed via a data
transformation $T$, and the transformed data may exhibit better or
worse response to the signal. Information
theory aids in designing a suitable data transformation, so that the
signal response is maximal, and the signal noise is minimal,
permitting the signal to be best recovered. Thus, we may aim for an
optimal $T$, which yields 
\begin{equation}\label{eq:trafo}
T[d] =  \langle s \rangle_{(s|d)} .
\end{equation}
We define the posterior average of the signal, $m_d =  \langle
 s \rangle_{(s|d)}$, to be the map of the
signal given the data $d$ and call $T$ a \textit{map-making-algorithm}
if it fulfills Eq. \ref{eq:trafo} at least approximately. As a
criterion for this one may require that the signal response of a
map-making-algorithm, 
\begin{equation}
 R_T[s] \equiv \langle T[d]\rangle_{(d|s)},
\end{equation}
is positive definite with respect to signal variations as stated by
\begin{equation}
 \frac{\delta R_T[s]}{\delta s} \ge 0.
\end{equation}
This ensures that a map-making algorithm will respond with a
non-negative correlation of the map to any signal feature, with
respect to the noise ensemble. 
In general, $T$ will be a non-linear operation on the data, to be
constructed from information theory if it should be optimal in the
sense of Eq. \ref{eq:trafo}. In any case, the fidelity of a signal
reconstruction can be characterized by the quadratic signal uncertainty,
\begin{equation}\label{eq:Tfidelity}
 \sigma_{T,d}^2 = \langle (s-T[d])\, (s-T[d])^\dagger\rangle_{(s|d)},
\end{equation}
averaged over typical realizations of signal and noise. Of special interest is
the trace of this 
\begin{equation}\label{eq:TraceTfidelity}
 \mathrm{Tr}(\sigma_{T,d}^2) 
= \int \!\! dx\, \langle | s_x-T_x[d]|^2\rangle_{(s|d)},
\end{equation}
since it is the expectation value of the squared Lebesgue-$L^2$-space distance
between a signal reconstruction and the underlying signal. 
Requesting a map making algorithm to be optimal with respect to
Eq. \ref{eq:TraceTfidelity}, implies $T[d] = \langle s \rangle_{(s|d)}$ and
therefore it to be optimal in an information theoretical sense according to
Eq. \ref{eq:trafo}. 

The uncertainty $\sigma_{T,d}^2 $ depends on $d$, since in Bayesian
 inference one averages over the 
 posterior, which is conditional to the data. The frequentist
 uncertainty estimate, which is the expected uncertainty 
 of any estimator before the data is obtained, is given by an average
 over the joint probability function: 
\begin{equation}\label{eq:Tfidelityfreq}
 \sigma_{T}^2 = \langle (s-T[d])\, (s-T[d])^\dagger\rangle_{(d,s)}.
\end{equation}
The latter is a good quantity to characterize the overall performance
of an estimator, whereas  $\mathrm{Tr}(\sigma_{T,d}^2)$ is a more
precise indicator of the actual estimator performance  
for a given dataset. As we will see in our IFT applications, data
dependence of the uncertainty is a common feature of non-linear
inference problems. 

An illustrative example should be in order. Suppose our data is an exact copy
of a physical field, $d = \psi$, our signal the square of the latter, $s =
\psi^2$, and the physical field obeys an even statistics, $P(\psi) = P(-\psi)
$. Then, the signal response is exactly zero, $R[s]=0$, and the data contains
only noise with respect to the chosen signal, $d =n_s$. Thus, we have chosen a
bad representation of our data to reveal the signal. If we, however, introduce
the transformation $d'= T[d] = d^2$, we find a perfect response, $R'[s]= s$,
and zero noise, $n'_s=0 $. 

In this case, finding the optimal map-making
algorithm was trivial, but in more complicated situations, it can not be
guessed that easily. Since the response and noise definitions depend on the
signal definition, some thoughts should be given to how to choose the
signal in a way that it can be well reconstructed.

\subsection{Signal design}\label{sec:design}

For practical reasons one will usually choose $s$ according to a few
guidelines, which should simplify the information induction process: 
\begin{enumerate}
\item The functional form of $s[\psi]$ should best be simple, steady,
analytic, and if possible linear in $\psi$, permitting to use the signal $s$ to
reason about the state of reality $\psi$. 
\item  The degrees of freedom of $s$ should be related to the ones of
the data $d$ in the sense that cross correlations exist which permit
to deduce properties of $s$ from $d$. Signal degrees of freedoms, which are
insensitive to the data, will only be constrained by the prior and
therefore just contain a large amount of uncertainty. This adds to
the error budget, and should be avoided as far as possible. 
\item The choice of $s[\psi]$ should also be lead by mathematical
convenience and practicality. In the examples presented in this work,
simple signals are chosen which permit to guess good approximations
for signal likelihood $P(d|s)$ and prior $P(s)$ without the need to
develop the full physical theory starting with $P(\psi)$. 
\end{enumerate}

To give a more specific example, we assume a cosmological model in
which the reality is thought to be solely characterized by the 
primordial dark matter density 
distribution $\psi(x)$, from which all observable cosmological
phenomena like galaxies derive in a deterministic way. The coordinate
$x$ may refer to the comoving coordinates at some early epoch of the
Universe. Although the LSS of the matter
distribution at a later time may predominantly depend on the initial
large-scale modes, and is reflected in the galaxy distribution, the
actual positions of the individual galaxies also depend in a
non-trivial way on the small-scale modes. Due to the discreteness of
our observable, the galaxy positions, it may be impossible to
reconstruct these small scale modes. Therefore it could be sensible to
define a signal $s[\psi]= F\,\psi$, with $F$ being a linear low-pass
filter, which suppresses all small-scale structures. This signal may be
reconstructible with high precision, whereas any attempt to
reconstruct $\psi$ directly would be plagued by a larger error budget,
since all the data-unconstrained small-scale modes represent
uncertainties to a reconstruction of $\psi$, but not to one of $s$
being defined as a low pass filtered version of $\psi$.  

\subsection{Signal moment calculation}\label{sec:moments}

The information of some data $d$ on a signal $s$ defined over some set
$\Omega$, which in most applications will be a manifold  
like a sub-volume of the $\mathrm{R}^n$, or the sphere in case of a CMB
signal, is completely contained in the posterior $P(s|d)$ of the
signal given the data.\footnote{We are mostly dealing with scalar
fields, however, multi-component, vector or tensor fields can be
treated analogously, and many of the equations just have to
be re-interpreted for such fields and stay valid.}  
The expectation value of $s$ at some location $x\in\Omega$, and higher
correlation functions of $s$ can all be obtained from the posterior by
taking the appropriate average: 
\begin{eqnarray}
 \langle s(x_1) \cdots s(x_n) \rangle_d & \equiv &  \langle s(x_1) \cdots
 s(x_n) \rangle_{(s|d)}\nonumber\\
 &\equiv& \int\!\!\mathcal{D}s\, s(x_1) \cdots s(x_n)\, P(s|d).
\end{eqnarray}

The problem is that often neither the expectation values nor even the posterior
are easily calculated analytically, even for fairly simple data
models. Fortunately, there is at least one class of data models for which the
posterior and all its moments can be calculated exactly, namely in case the
posterior turns out to be a multivariate Gaussian in $s$. In this case
analytical formulae for all moments of the signal are known and are in
principle computable. Technically, one is still often facing a huge, but linear
inverse problem. However, in the last decades a couple of computational
high-performance map-making techniques were developed to tackle such problems
either on the sphere, for CMB research, or in flat spaces with one, two or
three dimensions, for example for the reconstruction of the cosmic LSS
(detailed references are given in Sect. \ref{sec:lit}).  The purpose
of this work is to show how to expand other posterior distributions around the
Gaussian ones in a perturbative manner, which then permits to use the existing
map-making codes for the computation of the resulting diagrammatic perturbation
series. Since the diagrammatic perturbation series in Feynman-diagrams are well
known and understood in QFT and SFT, the
most economical way is to reformulate the information theoretical problem in a
language which is as close as possible to the former two theories. Thereby,
many of the results and concepts become directly available for signal inference
problems. Moreover, it seems that expressing the optimal signal estimator in
terms of Feynman diagrams immediately provides computationally efficient
algorithms, since the diagrams encode the skeleton of the minimal necessary
computational information flow.

\subsection{Signal and data spaces}\label{sec:space}
\subsubsection{Discretisation and continuous limit}\label{sec:discret}

Both, the signal and the data space may be continuous, however, in
practice will most often be discrete since digital data processing
only permits to chose a discretized representation of the distributed
information. The space in which the data and signal discretisation
happens can be chosen freely, and of course can be as well a Fourier,
wavelet or spherical harmonics space. Even if we would like to analyze
a continuous signal, the computationally required discretisation will
force an implicit redefinition of our actual signal to be the
discretely sampled version of that continuous signal, and this
discretisation step should also be part of the data model, if it has
the potential to significantly affect the analysis 
\citep[e.g. see][]{2009arXiv0901.3043J}. 

Although discretisation implies some information loss it also has an
advantage. We can just assume discretisation and therefore read all
scalar and tensor products as being the usual, component-wise ones, 
now just in high-, but finite-dimensional vector spaces. 

To be concrete, let $\{x_i\} \subset \Omega$ be a discrete set of
$N_{\mathrm{pix}}$ pixel positions, each of which has a volume-size
$V_i$ attributed to it, then the scalar product of two discretized
function-vectors $f= (f_i)$, and $g= (g_i)$ sampled at these points
via $f_i = f(x_i)$, and $g_i = g(x_i)$ could be defined by  
\begin{equation}\label{eq:scalar product}
 g^\dagger f \equiv \sum_{i=1}^{N_{\mathrm{pix}}} V_i\,{g_i}^*\, f_i.
\end{equation}
The asterix denotes complex conjugation. This scalar product has the continuous limit
\begin{equation}\label{eq:scalar product cont}
 g^\dagger f \longrightarrow \int \! dx \, {g}^*(x)\, f(x).
\end{equation}

In many cases the actual volume normalization in Eq. \ref{eq:scalar product} does not matter for
final results, since it usually cancels out, and therefore $V_i$ is
often dropped completely for equidistant sampling of signal and data
spaces. The volume terms also disappear for a scalar product involving a function 
which is discretized via volume integration, $f_i = \int_{V_i}dx\,f(x)$, e.g. 
the number of counts within the cell $i$.
Anyhow, higher order tensor products are defined analogously.

The path integral of a functional $F[f] \equiv F(f_1, \ldots,
f_{N_{\mathrm{pix}}})$ over all realizations of such a discretized
field $f$ is then just a high-dimensional volume integral, with as
many dimensions as pixels: 
\begin{equation}\label{eq: path integral definition}
 \int \mathcal{D}f \, F[f] \equiv  \left( \prod_ {i=1}^{N_{\mathrm{pix}}} \int
 df_i \right) \,F(f_1, \ldots, f_{N_{\mathrm{pix}}}). 
\end{equation}
This definition of a finite-dimensional path integral is well normalized, since
in case that we want to integrate over a probability distribution over $f$,
which is separable for all pixels, $ P(f) = \prod_{i=1}^{N_{\mathrm{pix}}}
P_i(f_i)$, as e.g. for white and Poissonian noise, we find
\begin{equation}\label{eq:path integral norm definition}
\langle 1 \rangle_{(f)} = \int \mathcal{D}f \, P(f) =
\prod_{i=1}^{N_{\mathrm{pix}}} \underbrace{\int df\,P_i(f)}_{= 1} =1.
\end{equation}

Although, in real data-analysis applications, it is practically never
required to perform the continuous limit 
$N_{\mathrm{pix}}\rightarrow \infty$ with $V_i \rightarrow 0$ for all
$i$, we stress that this limit can formally be taken and is well
defined even for the path integral, as we argue in more detail in
Sec. \ref{sec:normalisability}. The basic argument is that suitable
signals could and should be defined in such a way that path-integral
divergences, which plague sometimes QFT, can easily be avoided by
sensible signal design. Practically, the existence of a well-defined
continuous limit of a well-posed IFT implies that two numerical
implementations of a signal reconstruction problem, which differ in
their space discretisation on scales smaller than the structures of the
signal, can be expected to provide identical results up to a small
discretisation difference, which vanishes with higher
discretisation-resolution.

\subsubsection{Parameter spaces}

 In many applications, the signal space is identified with the
 physical space or with the sphere of the sky. However, IFT can also
 be done over parameter spaces. In Sec. \ref{sec:fnl}, a field theory
 over the sphere will implicitly define the knowledge state for an
 unknown parameter of that theory, which can be regarded again to
 define an information theory for that parameter. The latter is an IFT
 in case that the parameter has spatial variations. 

However, there are also functions defined over a parameter space,
$\Omega_{\mathrm{parameter}}= \{ p\}$ for some parameter $p$, which
one might want to obtain knowledge on from incomplete data. A very
import one is the probability distribution of the parameter given the
observational data, $P(p|d)$, which defines our parameter-knowledge
state. This function may only be incompletely known and therefore
require an IFT approach for its reconstruction and interpolation. Such
incomplete knowledge on the function could be due to incomplete
numerical sampling of its function values because of large
computational costs and the huge volumes of multi-dimensional
parameter spaces. Or, there might be another unknown nuisance parameter
$q$ in the problem, which induces an uncertainty in $P_{(p|d)} = P(p|d)$
and therefore an IFT over all possible realizations of this  knowledge
state field function via 
\begin{equation}
 P[P_{(p|d)}] =   \int \mathcal{D} P_{(p|d)}\, \delta\left[P_{(p|d)} - \!\int\!
 dq \, P(p,q|d)\right]. 
\end{equation}
In case that $q$ is a field, the marginalisation integral in the delta
functional also becomes a path-integral.  Probabilistic decision
theory, based on knowledge state as expressed by probability functions
on parameters, has to deal with such complications. 
For inference directly on $p$, and not on the knowledge state
$P_{(p|d)} $, the marginalized probability 
\begin{equation}
 P{(p|d)} = \!\int\! dq \, P(p,q|d)
\end{equation}
contains all relevant information, and that will be sufficient for
most inference applications, and especially for the ones in this work.

\section{Basic formalism}\label{sec:basics}

\subsection{Information Hamiltonian}

We argued that the posterior $P(s|d)$ contains all available
information on the signal. 
Although the posterior might not be easily accessible mathematically,
we assume in the following that the prior $P(s)$ of the signal before
the data is taken as well as the likelihood of the data given a signal
$P(d|s)$ are known or at least can be Taylor-Fr\'echet-expanded around
some reference field configuration $t$. Then Bayes's theorem permits
to express the posterior as 
\begin{equation}
 P(s|d) = \frac{P(d,s)}{P(d)} = \frac{P(d|s)\,P(s)}{P(d)} \equiv \frac{1}{Z}\, e^{-H[s]}\,.
\end{equation}
Here, the Hamiltonian
\begin{equation}
 H[s] \equiv H_d[s] \equiv -\log \left[ P(d,s) \right] =-\log \left[ P(d|s)\,P(s) \right] ,
\end{equation}
the evidence of the data
\begin{equation}
 P(d) \equiv \int \!\!\mathcal{D}s\; P(d|s)\,P(s) = \int \!\!\mathcal{D}s\;
  e^{-H[s]} \equiv Z,
\end{equation}
and the partition function $Z\equiv Z_d $ were introduced. It is
extremely convenient to include a moment generating function into the
definition of the partition function
\begin{equation}\label{eq:Z_d[J]}
 Z_d[J] \equiv \int \!\!\mathcal{D}s\;  e^{-H[s]+J^\dagger s}.
\end{equation}
This means $P(d) = Z=Z[0]$, but also permits to calculate any moment
of the signal field via Fr\'echet-differentiation of Eq. \ref{eq:Z_d[J]},
\begin{equation}
 \langle s(x_1) \cdots s(x_n) \rangle_d = \left. \frac{1}{Z} \, 
\frac{\delta^n\, Z_d[J]}{\delta J(x_1) \cdots \delta J(x_n)} \right|_{J=0}\,.
\end{equation}
Of special importance are the so-called connected correlation
functions or cumulants 
\begin{equation}\label{eq:Es}
 \langle s(x_1) \cdots s(x_n) \rangle_d^\mathrm{c} \equiv \left. 
\frac{\delta^n\, \log Z_d[J]}{\delta J(x_1) \cdots \delta J(x_n)} \right|_{J=0}\,,
\end{equation}
which are corrected for the contribution of lower moments to a
correlator of order $n$. For example, the connected mean and
dispersion are expressed in terms of their unconnected counterparts
as: 
\begin{eqnarray}
  \langle s(x)  \rangle_d^\mathrm{c} &=& \langle s(x)  \rangle_d,\nonumber\\
 \langle s(x) \,s(y) \rangle_d^\mathrm{c} &=& 
\langle s(x) \, s(y) \rangle_d - \langle s(x)  \rangle_d\, \langle s(y) \rangle_d,
\end{eqnarray}
where the last term represents such a correction.
For Gaussian random fields all higher order connected correlators vanish:
\begin{equation}
 \langle s(x_1) \cdots s(x_n) \rangle_d^\mathrm{c} = 0
\end{equation}
for $n>2$.
For non-Gaussian random fields, they are in general non-zero, and for
later usage we provide the connected three- and four-point functions,
\begin{eqnarray}\label{eq:34point}
 \!\!\!\!\langle s_x s_y s_z \rangle_d^\mathrm{c} \!\!&=& \!\!
\langle (s_x\! - \bar{s}_x)(s_y\! - \bar{s}_y) (s_z\! - \bar{s}_z) \rangle_d,\nonumber\\
\!\!\!\!\!\!\!\!\!\!\!\!\langle s_x s_y s_z s_u\rangle_d^\mathrm{c} \!\!&=& \!\!
\langle (s_x\! - \bar{s}_x) (s_y\! - \bar{s}_y) (s_z\! - \bar{s}_z)  (s_u \!- \bar{s}_u)\rangle_d\!\!\!\! \!\!\!\!\nonumber\\
\!\!&-&\!\! \langle s_x s_y \rangle_d^\mathrm{c}\langle s_z s_u \rangle_d^\mathrm{c}
- \langle s_x \,s_z \rangle_d^\mathrm{c}\langle s_y\,s_u \rangle_d^\mathrm{c}\nonumber\\
\!\!&-&\!\! \langle s_x \,s_u \rangle_d^\mathrm{c}\langle s_y\,s_z \rangle_d^\mathrm{c},
\end{eqnarray}
where we used $s_x = s(x)$ and defined $\bar{s}_x = \langle s(x)  \rangle_d$.

The assumption that the Hamiltonian can be Taylor-Fr\'echet expanded in
the signal field permits to write 
\begin{equation}\label{eq:generic Hamiltonian0}
H[s] =  \frac{1}{2}\,s^\dagger D^{-1}\, s - j^\dagger s + H_0 +
\sum_{n=3}^{\infty} \frac{1}{n!} \,\Lambda^{(n)}_{x_1\ldots x_n}\,
s_{x_1}\cdots  s_{x_n}. 
\end{equation}
Repeated coordinates are thought to be integrated over.
The first three Taylor coefficients have special roles. The
constant $H_0$ is fixed by the normalization condition of the joint
probability density of signal and data. If $H_d'[s]$ denotes some
unnormalised Hamiltonian, its normalization constant is given by 
\begin{equation}\label{eq:H0norm}
 H_0 = \log\, \int \!\mathcal{D}s \int \!\mathcal{D}d \; e^{-H_d'[s]}.
\end{equation}
Often $H_0$ is irrelevant unless different models or hyperparameters
are to be compared.  

We call the linear coefficient $j$ information source. This term is usually
directly and linearly related to the data. The quadratic coefficient,
$D^{-1}$, defines the information propagator $D(x,y)$, which
propagates information on the signal at $y$ to location $x$, and
thereby permits, e.g., to partially reconstruct the signal at
locations where no data was taken. Finally, the anharmonic tensors
$\Lambda^{(n)}$ create interactions between the modes of the free,
harmonic theory. Since this free theory will be the basis for the full
interaction theory, we first investigate the case $\Lambda^{(n)} =0$.

\subsection{Free theory}\label{sec:free}
\subsubsection{Gaussian data model}
For our simplest data model we assume a Gaussian signal with prior
\begin{equation}\label{eq:GaussPrior}
P(s) = \G(s,S) \equiv \frac{1}{|2\pi\, S|^\frac{1}{2}}\, \exp\left( -\frac{1}{2}
s^\dagger S^{-1} s \right) , 
\end{equation}
where $S=\langle s\,s^\dagger \rangle$ is the signal covariance. The
signal is assumed here to be processed by nature and our measurement
device according to a linear data model 
\begin{equation}
d = R\, s + n.
\end{equation}
Here, the response $R[s] = R \,s$ is linear in and the noise $n_s = n$ is 
independent of the signal $s$. The linear response matrix $R$ of
our instrument can contain window and selection functions,
blurring effects, and even a Fourier-transformation of the signal
space, if our instrument is an interferometer. Typically, the
data-space is discrete, whereas the signal space may be continuous. In
that case the $i$-th data point is given by 
\begin{equation}
d_i = \int \! dx\, R_i(x)\, s(x) + n_i.
\end{equation}

We assume, for the moment, but not in general, the noise to be
signal-independent and Gaussian, and therefore distributed as 
\begin{equation}
P(n|s) = \G(n,N),
\end{equation}
where $N= \langle n\, n^{\dagger} \rangle$ is the noise covariance
matrix. Since the noise is just the difference of the data to the
signal-response, $n=d-R\,s$, the likelihood of the data is given by 
\begin{equation}
\label{eq:PsPosterior}
P(d|s) = P(n= d-R\,s|s) = \G(d-R\,s,N),
\end{equation}
and thus the Hamiltonian of the Gaussian theory is 
\begin{eqnarray}
H_\mathrm{\G}[s] &=& - \log\left[ P(d|s) \,P(s)\right]\nonumber\\
& =& - \log\left[ \G(d-R\,s,N) \, \G(s,S)\right]\nonumber\\
&=&\frac{1}{2} s^{\dagger} D^{-1} s - j^{\dagger} s + H_0^\mathrm{\G}\,.
\label{eq:freeHamiltonian}
\end{eqnarray}
Here
\begin{equation}\label{eq:D}
D= \left[ S^{-1} + R^{\dagger} N^{-1} R\right]^{-1} 
\end{equation}
is the propagator of the free theory. The information source,
\begin{equation}
j = R^{\dagger} N^{-1} d,
\end{equation}
depends linearly on the data in a response-over-noise weighted fashion
and reads 
\begin{equation}
j(x)  = \sum_{ij} {R^*_i(x)} N^{-1}_{ij}\, d_j
\end{equation}
in case of discrete data but continuous signal spaces. Finally, 
\begin{equation}
 H_0^\mathrm{\G} = \frac{1}{2}\, d^\dagger\,N^{-1}\,d + \frac{1}{2}\,\log\left(
 |2\,\pi\,S|\,|2\,\pi\,N| \right) 
\end{equation}
has absorbed all $s$-independent normalization constants.

The partition function of the free field theory,
\begin{eqnarray}
Z_\mathrm{\G}[J] \!&=& \!
\int \!\mathcal{D}s \, e^{-H_\mathrm{\G}[s] + J^{\dagger}s} \\
\!&=& \!
\int \!\mathcal{D}s \, \exp\left\lbrace -\frac{1}{2} s^{\dagger} D^{-1} s +
(J+j)^{\dagger} s\, - H_0^\mathrm{\G} \right\rbrace\! ,\nonumber
\label{ZJ4} 
\end{eqnarray}
is a Gaussian path integral, which can be calculated exactly, yielding
\begin{equation}
\label{eq:ZdfreeTheory}
Z_\mathrm{\G}[J] = \sqrt{|2\pi\, D|}\,
\exp\left\lbrace +\frac{1}{2} (J+j)^{\dagger} D (J+j) -H_0^\mathrm{\G} \right\rbrace\!.
\end{equation}
The explicit partition function permits to calculate via Eq. \ref{eq:Es}
the expectation of the signal given the data, in the following called
the map $m_d$ generated by the data $d$: 
\begin{eqnarray}
\label{eq:WFmap} 
m_d &=& \langle s \rangle_{d}  = \left. \frac{\delta \log Z_G}{\delta J}\right|_{J=0} =D\,j\\
&=& \underbrace{\left[ S^{-1} + R^{\dagger} N^{-1} R\right]^{-1} R^{\dagger} N^{-1}}_{F_{\mathrm{WF}}} d. \nonumber
\end{eqnarray}
The last expression shows that the map is given by the data after
applying a generalized Wiener filter, $m_d = F_{\mathrm{WF}} \,d$. The
propagator $D(x,y)$ describes how the information on the density field
contained in the data at location $x$ propagates to position $y$:
$m(y) = \int dx \,D(y,x)\, j(x)$.  

The connected autocorrelation of the signal given the data,
\begin{equation}
\langle s s^{\dagger}\rangle_{d}^{c} = D = \left[ S^{-1} + R^{\dagger} N^{-1} R\right]^{-1},
\end{equation}
is the propagator itself. All higher connected correlation functions
are zero. Therefore, the signal given the data is a Gaussian random
field around the mean $m_d$ and with a variance of the residual error 
\begin{equation}
r = s-m_d
\end{equation}
provided by the propagator itself, as a straightforward calculation shows:
\begin{equation}\label{eq:WFrr} 
\langle r r^{\dagger} \rangle_{d} = \langle s s^{\dagger}\rangle_{d} - \langle
s \rangle_{d} \langle s^{\dagger} \rangle_{d} = \langle s
s^{\dagger}\rangle_{d}^{c} =  D. 
\end{equation}
Thus, the posterior should be simply a Gaussian given by
\begin{equation}
P(s|d) = \G(s-m_d,D).
\end{equation}
As a test for the latter equation, we calculate the evidence of the
free theory via 
\begin{eqnarray}
P(d) &=& \frac{P(d|s)\,P(s)}{P(s|d)} = \frac{\G(d-R\,s,N)\, \G(s,S)}{\G(s-D\,j,D)}
\nonumber\\ 
&=& \left( \frac{|D|/|S|}{|2\pi\,N|}\right)^{\frac{1}{2}} \exp\left\lbrace
\frac{1}{2} (j^\dagger D\, j - d^{\dagger} N^{-1} d ) \right\rbrace, 
\end{eqnarray}
which is indeed independent of $s$ and also identical to
$Z_\mathrm{\G}[0]$, as it should be.

\subsubsection{Free classical theory}

The Hamiltonian permits to ask for \textit{classical} equations
derived from an extremal principle. This is justified, on the one
hand, as being just the result of a the saddlepoint approximation of
the exponential in the partition function. On the other hand, the
extrema principle is equivalent to the maximum a posteriori (MAP)
estimator, which is quite commonly used for the construction of
signal-filters. An exhaustive introduction into and discussion of the
MAP approximation to Gaussian and non-Gaussian signal fields is
provided by \citet{1999physics..12005L}. 

The classical theory is expected to capture essential features of the
field theory. However, if the field fluctuations are able to probe
phase space regions away from the maximum in which the Hamiltonian (or
posterior) has a more complex structure, deviations between classical
and field theory should become apparent. 

Extremizing the Hamiltonian of the free theory (Eq. \ref{eq:freeHamiltonian})
\begin{equation}
\left. \frac{\delta H_G}{\delta s}\right|_{s=m} =  D^{-1} m - j \equiv 0
\end{equation}
we get the classical mapping equation $m=Dj$, which is identical to
the field theoretical result (Eq. \ref{eq:WFmap}).  

It is also possible to measure the sharpness of the maximum of the
posterior by calculating the Hessian curvature matrix 
\begin{equation}
\mathcal{H}_\mathrm{\G}[m] = \left. \frac{\delta^2\, H[s]}{\delta
  s^2}\right|_{s=m} = D^{-1}. 
\end{equation}
In the Gaussian approximation of the maximum of the posterior, the
inverse of the Hessian is identical to the covariance of the residual 
\begin{equation}
\langle r\,r^{\dagger} \rangle = \mathcal{H}^{-1}[m] = D,
\end{equation}
which for the pure Gaussian model is of course identical to the exact
result, as given by the field theory (Eq. \ref{eq:WFrr}).

\section{Interacting Information Fields}\label{sec:interacting fields}
\subsection{Interaction Hamiltonian }
\subsubsection{General Form}

All results of the free theory presented so far are well-known within the
field of signal reconstruction. IFT reproduces them elegantly, and is
therefore of pedagogical value. However, the new results presented in
the rest of this paper arise as soon as one leaves the free theory.  
Non-Gaussian signal or noise, a non-linear response, or a signal
dependent noise create anharmonic terms in the Hamiltonian. These
describe interactions between the eigenmodes of the free Hamiltonian. 

We assume the Hamiltonian can be Taylor expanded in the
signal fields, which permits to write 
\begin{equation}\label{eq:generic Hamiltonian}
H[s] =  \underbrace{\frac{1}{2}\,s^\dagger D^{-1}\, s - j^\dagger s +
  H_0^\mathrm{\G}}_{H_\mathrm{\G}[s]} +  \underbrace{\sum_{n=0}^{\infty}
  \frac{1}{n!} \,\Lambda^{(n)}_{x_1\ldots x_n}\, s_{x_1}\cdots
  s_{x_n}}_{H_\mathrm{int}[s]}.
\end{equation}
Repeated coordinates are thought to be integrated over. In contrast to 
Eq. \ref{eq:generic Hamiltonian0} we have now included perturbations
which are constant, linear and quadratic in the signal field, 
because we are summing from $n=0$. This
permits to treat certain non-ideal effects perturbatively. For example
if a mostly position-independent propagator gets a small position
dependent contamination, it might be more convenient to treat the
latter perturbatively and not to include it into the propagator used
in the calculation. Note further, that all coefficients can be assumed
to be symmetric with respect to their
coordinate-indices.\footnote{This means $D_{x\, y} 
=D_{y\, x}$ and $ \Lambda^{(n)}_{x_{\pi(1)}\ldots x_{\pi(n)}}
=\Lambda^{(n)}_{x_1\ldots x_n}$ with $\pi$ any permutation of
$\{1,\ldots, n\}$, since even non-symmetric coefficients would
automatically be symmetrized by the integration over all repeated
coordinates. Therefore, we assume in the following that such a
symmetrization operation has been already done, or we impose it by
hand  before we continue with any perturbative calculation by applying 
\begin{displaymath}
 \Lambda^{(n)}_{x_1\ldots x_n} \longmapsto \frac{1}{n!}\,
 \sum_{\pi\in\mathcal{P}_n} \Lambda^{(n)}_{x_{\pi(1)}\ldots x_{\pi(n)}}\,. 
\end{displaymath}
This clearly leaves any symmetric tensor invariant if $\mathcal{P}_n$
is the space of all permutations of $\{1,\ldots, n\}$. }

Often, it is more convenient to work with a shifted field $\phi =
s-t$, where some (e.g. background) field $t$ is removed from s. The
Hamiltonian of $\phi$ reads  
\begin{eqnarray}\label{eq:shifted Hamiltonian}
H'[\phi] &=&  \underbrace{\frac{1}{2}\,\phi^\dagger D^{-1}\, \phi - j'^\dagger
  \phi + H'_0}_{H'_G[\phi]} \nonumber\\ 
&&
+  \underbrace{\sum_{n=0}^{\infty} \frac{1}{n!} \,{\Lambda'}^{(n)}_{x_1\ldots
  x_n}\, \phi_{x_1}\cdots  \phi_{x_n}}_{H'_\mathrm{int}[\phi]}, 
\;\mbox{with}\nonumber\\
H'_0 &=& H_0^\mathrm{\G} - j^\dagger t +\frac{1}{2} \, t^\dagger D^{-1} t,\\
j' &=& j - D^{-1}\, t,\;\mbox{and}\nonumber\\
 {\Lambda'}^{(m)}_{x_1\ldots x_m} &=& \sum_{n=0}^{\infty}
 \frac{1}{n!}\Lambda^{(m+n)}_{x_1\ldots x_{m+n}} \, t_{x_1} \cdots
 t_{x_n}.\nonumber 
\end{eqnarray}

\subsubsection{Feynman rules}

Since all the information on any correlation functions of the fields is
contained in the partition sum and can be extracted from it, only the
latter needs to be calculated: 
\begin{eqnarray}
 Z[J] \!&=&\! \!\int\! \mathcal{D}s\, e^{-H[s]+J^\dagger s} \nonumber\\
\!&=&\! \int\! \!\mathcal{D}s\, \exp\left[-\!\sum_{n=0}^{\infty} \frac{1}{n!}
  \,\Lambda^{(n)}_{x_1\ldots x_n}\, s_{x_1}\cdots  s_{x_n}\right] \,
e^{-H_G[s]+J^\dagger s} \nonumber\\ 
\!&=&\! \exp\left[-\!\sum_{n=0}^{\infty} \frac{1}{n!}
 \,\Lambda^{(n)}_{x_1\ldots x_n}\, \frac{\delta}{\delta J_{x_1}}\cdots
 \frac{\delta}{\delta J_{x_n}}\right] \nonumber\\ 
 \! &\times &\! \!\int\! \mathcal{D}s\, e^{-H_G[s]+J^\dagger s} \nonumber\\ 
\!&=&\! \exp\left[-H_\mathrm{int}[ \frac{\delta}{\delta J}] \right] \,Z_G[J]. 
\end{eqnarray}
There exist well known diagrammatic expansion techniques for such
expressions \cite[e.g.][]{Binney1992}. The expansion terms of the logarithm of the partition
sum, from which any connected moments can be calculated,  are
represented by all possible connected diagrams build out of lines 
(\includegraphics[width=0.9\fgwidth,bb=0 0 128 6]{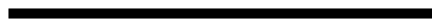}),
vertices (with a number of legs connecting to lines, like
\includegraphics[width=0.4\fgwidth,bb=0 0 40 17]{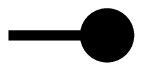},
\includegraphics[width=0.65\fgwidth,bb=0 0 63 17]{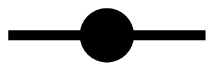},
\includegraphics[width=0.55\fgwidth,bb=0 0 57 40]{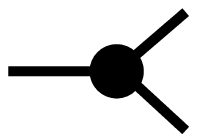},
\includegraphics[width=0.4\fgwidth,bb=0 0 42 42]{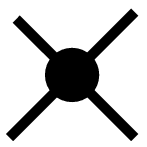}, 
...)
and without any external line-ends (any line ends in a vertex). 
These diagrams are interpreted according to the
following Feynman rules: 
\begin{enumerate}
 \item Open ends of lines in diagrams correspond to external
 coordinates and are labeled by such. Since the partition sum in
 particular does not depend on any external coordinate, it is
 calculated only from summing up closed diagrams. However, the field
 expectation value $m(x) = \langle s(x) \rangle_{(s|d)} = d\log
 Z[J]/dJ(x)|_{J=0}$ and higher order correlation functions depend on
 coordinates and therefore are calculated from diagrams with one or
 more open ends, respectively.  
\item A line with coordinates $x'$ and $y'$ at its end represents the
propagator $D_{x'\,y'}$ connecting these locations. 
\item Vertices with one leg get an individual internal, integrated
coordinate $x'$ and represent the term
$j_{x'}+J_{x'}-\Lambda^{(1)}_{x'}$. 
\item Vertices with $n$ legs represent the term 
  $-\Lambda^{(n)}_{x_1'\ldots x_n'} $, where each individual leg is
  labeled by one of the internal coordinates $x_1',\ldots,\, x_n'$. This more
  complex vertex-structure, as compared to QFT, is a
consequence of non-locality in IFT.  
\item All internal (and therefore repeatedly occurring) coordinates
are integrated over, whereas external coordinates are not. 
\item Every diagram is divided by its symmetry factor, the number 
of permutations of vertex legs leaving the topology invariant, as described in
any book on field theory \cite[e.g.][]{Binney1992}.
\end{enumerate}
The $n$-th moment of $s$ is generated by taking the $n$-th derivative
of $\log Z[J]$ with respect to $J$, and then setting $J=0$. This
correspond to removing $n$ end-vertices from all diagrams. For
example, the first four diagrams contributing to a map ($m=\langle s
\rangle_{(s|d)}$) are  
\begin{eqnarray}\label{eq:complicatedDiagram}
 \includegraphics[width=\fgwidth]{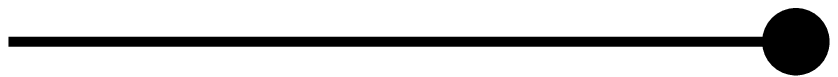} \!&=&\! D\,j = D_{xy}\, j_y
 \nonumber\\ 
 \!&\equiv & \! \int \! dy\, D(x,y)\, j(y),\nonumber\\
\includegraphics[width=\fgwidth]{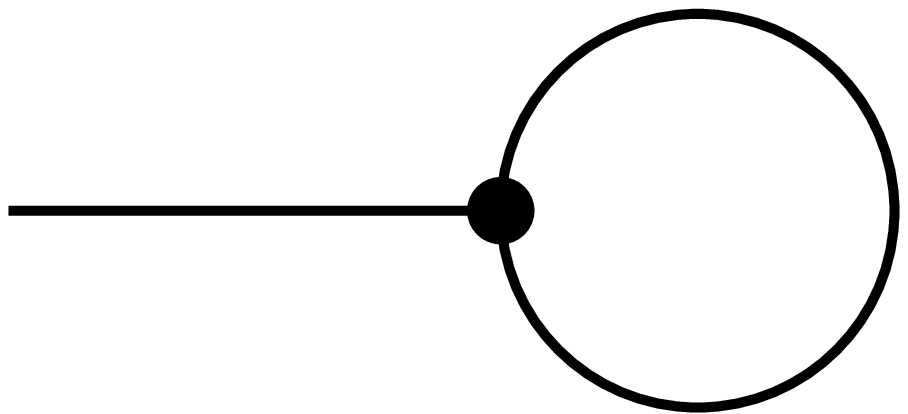} \!&=&\! - \frac{1}{2}\,D\,
\Lambda^{(3)}[\cdot, D] = -\frac{1}{2} D_{xy}\, \Lambda^{(3)}_{yzu}\, D_{zu}
\nonumber\\ 
\!& \equiv &\! -\frac{1}{2}\, \! \int \! dy\, D_{xy} \, \! \int \! dz\, \! \int
\! du\, \Lambda^{(3)}_{xyu}\,D_{zu},\nonumber\\ 
\includegraphics[width=\fgwidth]{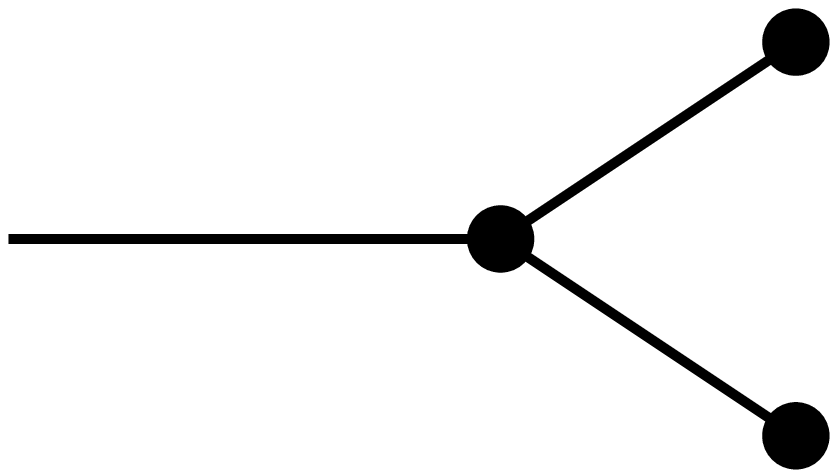} 
\!&=&\! -\frac{1}{2}\,D\, \Lambda^{(3)}\,[\cdot,D\,j, D\,j]\nonumber\\
\!&=&\! -\frac{1}{2} D_{xy}\,  \Lambda^{(3)}_{yuz} \, D_{zz'}\,j_{z'}\,
D_{uu'}\,j_{u'}\\ 
\!&\equiv&\! -\frac{1}{2}  \! \int \! dy\, D_{xy} \! \int \! dz \! \int \! du\,
\Lambda^{(3)}_{yzu} \nonumber\\ 
\!&\times&\! \int \! dz'\,  D_{zz'}\,j_{z'} \! \int \! du'\, D_{uu'}\,j_{u'},\;\mbox{and}\nonumber\\
\includegraphics[width=\fgwidth]{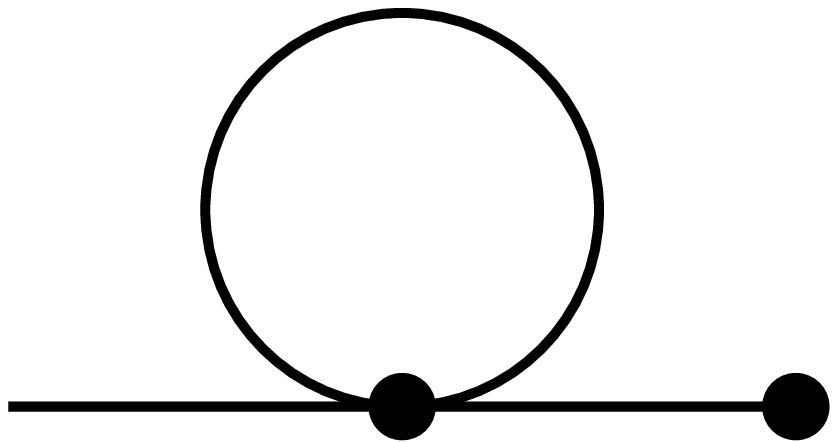} \!&=&\! -\frac{1}{2}\,
D\,\Lambda^{(4)}[\cdot, D,\,D\,j] \nonumber\\ 
\!&=&\!
-\frac{1}{2} D_{xy}\, \Lambda^{(4)}_{yzuv}\,D_{zu}\, D_{vv'}\,j_{v'}
\nonumber\\ 
\!&\equiv&\! - \frac{1}{2} \! \int \! dy\, D_{xy}\, \! \int \! dz\,\! \int \!
du\,\! \int \! dv\, \Lambda^{(4)}_{yzuv}\,D_{zu}\, \nonumber\\ 
\!&\times&\!
\! \int \! dv'\, D_{vv'}\,j_{v'}.\nonumber
\end{eqnarray}
Here we have assumed that any first and second order perturbation was
absorbed into the data source and the propagator, thus $\Lambda^{(1)}
= \Lambda^{(2)} = 0$. Repeated indices are assumed to be integrated
(or summed) over. 

\subsubsection{Local interactions and Fourier space rules}
In case of purely local interactions
\begin{equation}
 \Lambda^{(n)}_{x_1\ldots x_n} = \lambda_n(x_1)\, \delta(x_1-x_2)\cdots\delta(x_1-x_n)
\end{equation}
the interaction Hamiltonian reads
\begin{equation}\label{eq:genericHamiltonian}
 H_{\mathrm{int}} = \sum_{m=0}^{\infty} 
 \frac{1}{m!}\,\lambda_m^\dagger s^m 
\end{equation}
and the expressions of the Feynman diagrams simplify considerably. The
fourth Feynman rule can be replaced by 
\begin{enumerate}
 \item[4.]  Vertices with $n$ lines connected to it  are associated
 with a single internal coordinate $x'$ and represent the term
 $-\lambda_n(x')$. 
\end{enumerate}
For example, the last loop diagram in Eq \ref{eq:complicatedDiagram} becomes
\begin{equation}
\includegraphics[width=\fgwidth]{fg/fg_a4.eps} = -\frac{1}{2} \! \int \!
dy\, D_{xy}\,  \lambda_4(y)\,D_{yy}\,\! \int \! dz\, D_{yz}\,j_{z}. 
\label{eq:complicatedDiagramsimpler} 
\end{equation}

In case of local interactions, it can be helpful to do the calculations in
Fourier space, for which the Feynman rules can be obtained by inserting a
real-space identity operator $1 = F^\dagger F$ in between any scalar product
and assigning the inverse Fourier transformation $F^\dagger$ to the left 
and the forward transform $F$ to the right term, e.g.
\begin{displaymath}
 D\, j = F^\dagger \underbrace{F\,  D\, F^\dagger}_{D'} \,\underbrace{F\,
 j}_{j'} =   F^\dagger \, D'\,j'. 
\end{displaymath}
This yields:
\begin{enumerate}
\item An open end of a line has an external momentum coordinate $k$, and gets
an $\int dk\,e^{-\,i\,k\,x}/(2\pi)^n$ applied to it, if real space functions
are to be evaluated.
\item A line connecting momentum $k$ with momentum $k'$ corresponds to a
directed propagator between these momenta: $D_{kk'}=D(k,k')$.
\item A data source vertex is  $(j+J-\lambda_1)(k'')$, where $k''$ is the momentum at
the data-end of the line.
\item A vertex with $m>1$ lines with momentum labels
$k_1, \ldots, k_m$ is $ -\lambda_m(k_0)  
     (2\pi)^n \, \delta(\sum_{i=0}^{m} k_i)$.
\item An internal end of a line has an internal (integrated) momentum
coordinate $k'$. Integration means a term $\int \!dk'/(2\pi)^n$ in front
of the expression. 
\item The expression gets divided by the symmetry factor of its diagram.
\end{enumerate}

Here, $j(k) = (F\,j)(k) = \int dx \, j(x)\, e^{i\,k\,x}$, $D(k,k') =
$ $(F\,D\,F^{\dagger})(k,k') =$ $ \int dx \int dx' D(x,x')\,e^{i\,(k\,x -
k'\,x')}$, etc.  are the Fourier-transformed information source, propagator,
etc., respectively

Note, that momentum directions have to be taken into account.  The momenta that
go into a vertex, data source or open end get a positive sign in the
delta-function of momentum conservation, the ones that go out of a vertex get a
minus sign.

\subsubsection{Simplistic interaction Hamiltonians}\label{sec:simplisticHamiltoian}

In order to have a toy case, which permits analytic calculations, we
introduce a simplistic Hamiltonian by requiring the data model to be
translational invariant and all interaction terms to be local.  This
is the case whenever the signal and noise covariances are fully
characterized by power 
spectra over the same spatial space, 
\begin{eqnarray}
S(k,q) &=& (2\pi)^n\, \delta(k-q) \,P_{S}(k),\\
N(k,q) &=& (2\pi)^n\, \delta(k-q) \,P_{N}(k),
\end{eqnarray}
with
$P_s(k) = \langle |s(k)|^2 \rangle /V$, and $P_n(k) = \langle |n(k)|^2
\rangle /V$, where $V$ is the volume of the system. 
We assume further that the signal processing can be completely
described by a convolution with an instrumental beam,
\begin{equation}
 d(x) = \int
dy\, R(x-y) \, s(y) + n(x),
\end{equation}
 where the response-convolution kernel has a
Fourier power spectrum  $P_R(k) = |R(k)|^2$ (no factor $1/V$). 
In this case $D$ can be fully described by a power spectrum:
\begin{eqnarray}
D(k,q) &=& (2\pi)^n\, \delta(k-q) \,P_{D}(k),
\end{eqnarray}
with
$P_D(k) = (P_S^{-1}(k) + P_R(k)\, P_N^{-1}(k))^{-1}$.

The locality of the interaction terms requires  
$\lambda_{m} = \mathrm{const}$ beside translational invariance and therefore
the interaction Hamiltonian reads  
\begin{eqnarray}\label{eq:H_int_const_lambda}
H_{\rm int}[s] \!&=&\! \sum_{m=1}^{\infty} \frac{ \lambda_m }{m!}\,\int \! dx \, s^m(x)\\
\!&=&\! \sum_{m=1}^{\infty} \frac{\lambda_m}{m!}\, \left( \prod_{i=1}^m \int \! \frac{dk_i}{(2\pi)^n}\,
s_{k_i} \right) (2\pi)^n \delta(\sum_{j=1}^m k_j)\nonumber
\end{eqnarray}

In that case, the Feynman rules simplify considerably. For the
interaction Hamiltonian of Eq. \ref{eq:H_int_const_lambda}, the Feynman
rules are now:
\begin{enumerate}
\item{unintegrated $x$-coordinate: $\exp(-i\,k\,x)$ (if real space
functions are to be evaluated)},  
\item{propagator: $P_D(k)$,}
\item{data source vertex: $(j + J - \lambda_{1})(k)$,}
\item{vertex with $m>1$ lines: $-\lambda_m$,}
\item{imply momentum conservation at each vertex: $(2\pi)^n
\delta(\sum_{i=1}^m k_i))$, and
integrate over every internal momentum: $\int \frac{dk}{(2\pi)^n}$,}
\item{and divide by the symmetry factor.}
\end{enumerate}

\subsubsection{Feynman rules on the sphere}

For CMB reconstruction and analysis, but presumably also for terrestrial applications,
 the Feynman rules on the sphere
$\Omega = S^2$ are needed and therefore provided in Appendix \ref{sec:sphere}.

\subsection{Normalisability of the theory}\label{sec:normalisability}

In contrast to QFT, IFT should be properly normalized and not
necessarily require any renormalization procedure. The reason is that
IFT is not a low-energy limit of some unknown high-energy theory, but
can be set up as the full (high-energy) theory. The  Hamiltonian is
just the logarithm of the joint probability function of data and
signal, $H_d[s] \equiv -\log \left[ P(d,s) \right]$, and therefore
well defined and properly normalized if the latter is. Only if ad-hoc
Hamiltonians are set up, or if approximations lead to ill-normalized
theories, normalization should be an issue. 

However, since we are trying a perturbative expansion of the
theory, there is no guarantee that all individual terms are providing
finite results. For example in QFT, simple loop diagrams are known to
be divergent and require renormalization. In the following we
investigate a simplistic, but representative case of IFT, which shows
that such problems are generally not to be expected. 

Let us adopt the simplistic situation described in
\ref{sec:simplisticHamiltoian} and estimate a simple loop diagram for
which we assume for notational convenience $\lambda_3 =-2\,(2\pi)^n\,\lambda'$
(with $\lambda'>0$): 
\begin{eqnarray}
\includegraphics[width=\fgwidth]{fg/fg_a3.eps} \!&=&\! - \frac{1}{2}\,D\,
\lambda_3\, \widehat{D}  \\ 
\!&= &\!   \lambda' \int\! dk \int\! dk'\, \delta(k+k'-k')\, P_D(k)\,P_D(k')\
e^{ikx} \nonumber\\ 
\!&\le &\! \lambda'\, P_D(0) \int\! dk'\,P_S(k') = \lambda'\, V\,
P_D(0)\,\langle s^2(x) \rangle, \nonumber 
\end{eqnarray}
where $V$ is the volume of the system. Here and in the following,
$\widehat{C}$ denotes the diagonal of the matrix $C$.

Thus, as long the signal field is of bounded variance, the loop
diagram is convergent due to $P_D\le P_S$ for all $k$.
Even a signal of
unbounded variance would not lead to a divergent loop diagram if
$\int\!dk\, (P_N/P_R)(k)$ is finite, since we also have   $P_D\le
P_N/P_R$. 
A bounded variance signal is very natural, especially in a cosmological
setting.\footnote{The cosmological signal of primary interest, 
the initial density fluctuations as revealed by the
large-scale-structure and the CMB, is
expected to exhibit a suppression of small-scale power due to the 
free-streaming of dark matter particles before they became
non-relativistic. Also the CMB temperature fluctuations are damped on small
scales, due to free streaming of photons around the time of recombination.}

Finally, since a signal as an information field can be chosen freely,
we can define it to be the filtered version of the physical field
(e.g. dark matter distribution or CMB fluctuations), so that only
modes of sufficiently bound variance are present in it.  Since we have
the freedom to chose information fields, which are mathematically well
behaved, we can therefore ensure convergence of expressions. 

Although this is not a general proof of normalisability of the theory,
which is beyond the scope of this paper, it should provide confidence
in the well-behavedness of the formalism in sensible applications. The price to
be payed for 
this well-behavedness is the more complex structure of the propagator,
which, in comparison to QFT, even in simplistic cases can be
non-analytical and require numerical evaluation.

\subsection{Expansion around the classical solution}
\subsubsection{General case}
The classical solution of the Hamiltonian in Eq. \ref{eq:generic
Hamiltonian} is provided by its minimum,  
\begin{equation}
 \frac{\delta H}{\delta s_x} = D^{-1}_{x\,y} s_y -j_x  +  \sum_{m=1}^{\infty}
 \frac{1}{m!}\,\Lambda^{(m+1)}_{x\, x_1\ldots x_m}\, s_{x_1}\ldots s_{x_m}  = 0.
\end{equation}
This leads to the equation for the classical field
\begin{equation}
  s^{\mathrm cl}_y = D_{y\,x} \, \left( j_x  -  \sum_{m=1}^{\infty}
 \frac{1}{m!}\,\Lambda^{(m+1)}_{x\, x_1\ldots x_m}\, s^{\mathrm cl}_{x_1}\ldots
 s^{\mathrm cl}_{x_m} \right), 
\end{equation}
which one can try to solve iteratively. 

\subsubsection{Local interactions}
For simplicity, we concentrate for a moment on the case of purely
local interactions, for which the equation for the classical field
$s_\mathrm{cl}$ is 
\begin{equation}\label{eq:classicalEq}
 s_\mathrm{cl} = D \left( j - \sum_{m=1}^{\infty}
 \frac{\lambda^\dagger_{m+1}}{m!}\, s_\mathrm{cl}^{m} \right).
\end{equation}
Iterating this equation and rewriting the resulting terms as Feynman
diagrams shows that the classical solution contains the
tree-diagrams. The loop diagrams can be added by investigation of the
non-classical uncertainty field $\phi = s-s_\mathrm{cl}$.  

A non-classical expansion of the information field around the
classical field is possible by inserting 
$ s = s_\mathrm{cl} + \phi$
into the Hamiltonian (Eq. \ref{eq:genericHamiltonian}). Reordering terms
according to the powers of the field $\phi$ leads to its Hamiltonian 
\begin{eqnarray}\label{eq:Hamiltonian'}
 H'[\phi] &\equiv& H[s_\mathrm{cl} + \phi]\nonumber\\
&=& H_0' + \frac{1}{2} \,\phi^\dagger {D'}^{-1} \,\phi 
- j'^{\dagger} \phi + \sum_{m=3}^{\infty}
 \frac{1}{m!}\,{\lambda'_m}^\dagger \phi^m, \nonumber\\
\mbox{with}&&\nonumber\\
\label{eq:lambda'local}
 \lambda'_n &\equiv& \sum_{m = 0}^\infty \frac{\lambda_{n+m}}{m!}\, s_\mathrm{cl}^{m},\\
H'_0 &\equiv& H[s_\mathrm{cl}] = H_0  + \frac{1}{2}\, {s_\mathrm{cl}}^\dagger
D^{-1} s_\mathrm{cl} + \lambda'_0 ,\nonumber\\ 
j' &\equiv& j - \lambda'_1 - D^{-1} \,s_\mathrm{cl} ,\; \mbox{ and}\;
%
D' \equiv (D^{-1} + \widehat{\lambda'_2})^{-1}\,.\nonumber
\label{eq:D'}
\end{eqnarray}
In case $s_\mathrm{cl}$ is exactly the classical solution,
Eqs. \ref{eq:classicalEq} and \ref{eq:lambda'local} imply that $j' =
0$. Thus, there are no one-line internal vertices in any
Feynman-graphs of the $\phi$-theory, and only loop-diagrams contribute
uncertainty-corrections\footnote{We propose the term {\it
uncertainty-corrections} in order to describe the influence of the
spread of the probability distribution function around its
maximum. The uncertainty-corrections are the information field
theoretical equivalent to quantum-corrections in quantum field
theories.} to any information theoretical estimator. For example, the
uncertainty-corrections to the classical map estimator are given by 
\begin{eqnarray}\label{eq:uncertainty-loops-corr}
 \delta m &=& m_d - s_\mathrm{cl} = \langle \phi \rangle_d\\
&=& 
\includegraphics[width=\fgwidth]{fg/fg_a3.eps} + 
\includegraphics[width=\fgwidth]{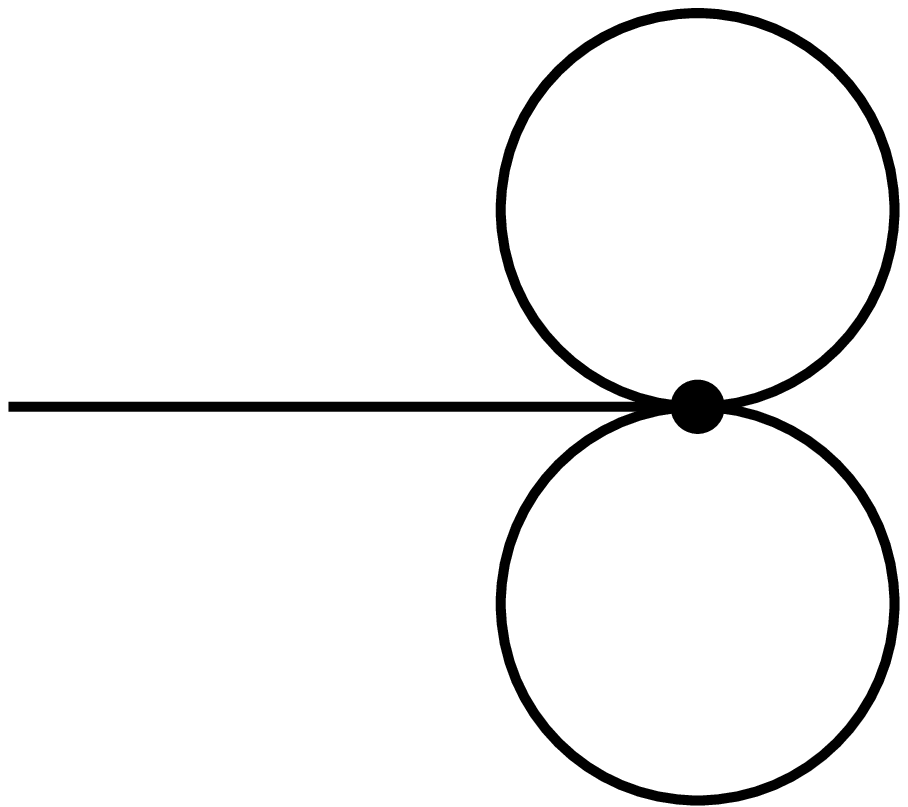} + 
\includegraphics[width=\fgwidth]{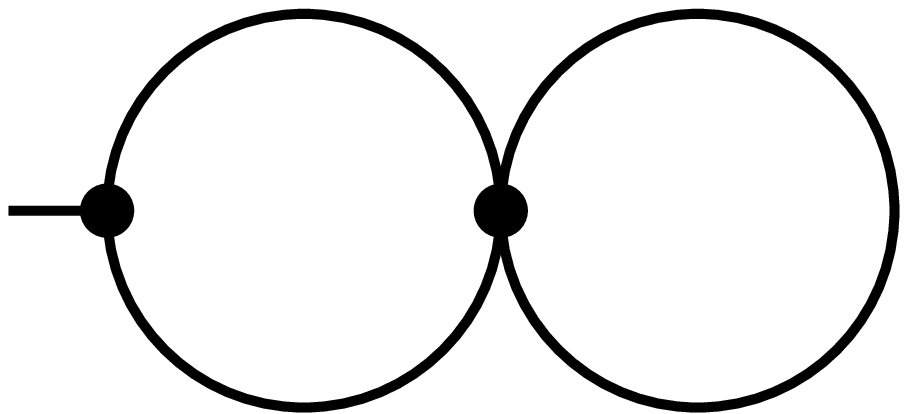} + 
\includegraphics[width=\fgwidth]{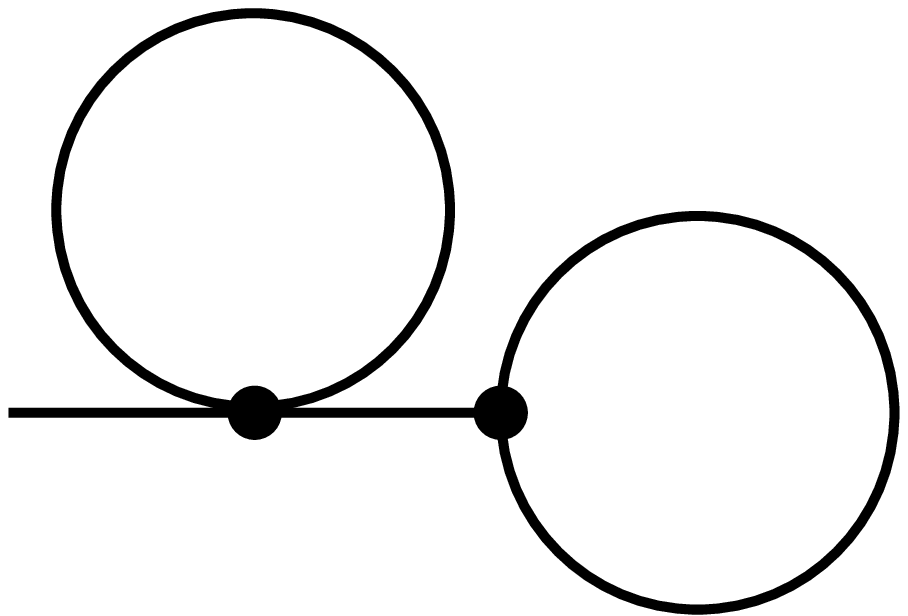} + 
\includegraphics[width=\fgwidth]{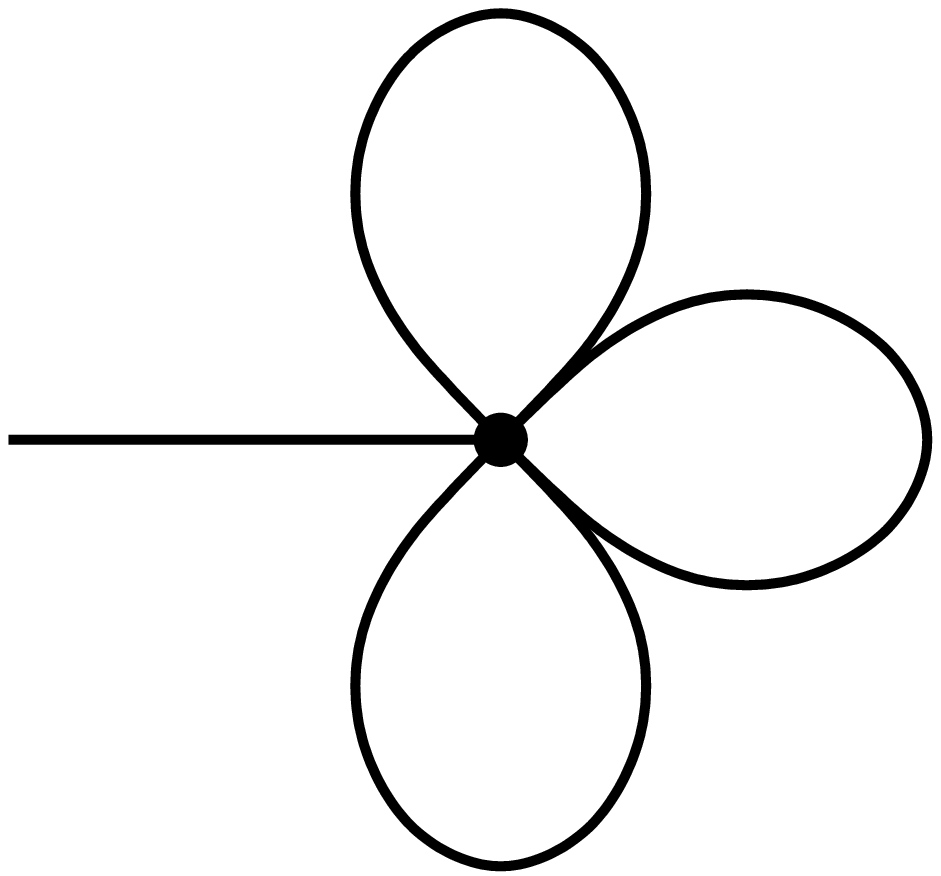} + 
\ldots \nonumber
\end{eqnarray}
However, in case $s_\mathrm{cl}$ is not (exactly) the classical
solution, may this due to a truncation error of an iteration scheme to
solve for the  classical field, or may $s_\mathrm{cl}$ be chosen for a
completely different purpose, Eq. \ref{eq:Hamiltonian'}
provides the correct field theory for $\phi = s - s_\mathrm{cl}$
independent of the nature of $s_\mathrm{cl}$. In case of a truncation
error, incorporating diagrams with data-source terms $j'$ into any
computation will permit to correct the inaccuracy of $s_\mathrm{cl}$
in a systematic way. 

\subsection{Boltzmann-Shannon Information}\label{sec:it}
\subsubsection{Helmholtz free energy}

Information fields carry information on distributed physical quantities. The
amount of signal-information should be measurable in information units like bits and bytes.
This is possible by adopting the
Boltzmann-Shannon information measure of negative
entropy. The entropy of a signal probability function measures the phase-space volume 
available for signal uncertainties, and
therefore the constraintness of the remaining uncertainties. 
Thus we define
\begin{eqnarray}
I_d &\equiv& \int \! \mathcal{D} s \, P(s|d)\, \log P(s|d) \nonumber\\
&=& - \int\! \mathcal{D} s \, \frac{1}{Z} \, e^{-H[s]} \,
(H[s] + \log Z)\nonumber\\
&=& - \langle H[s] \rangle_{d} - \log Z.
\end{eqnarray}
as the information measure.
Introducing 
\begin{eqnarray}
Z_\beta[d,J] &=& \int\! \mathcal{D}s\, \exp\left\lbrace -\beta\,(H[s] -
J^{\dagger} s)\right\rbrace\!, \mathrm{\;and}\nonumber \\ 
F_\beta[d,J] &=& -\frac{1}{\beta}\, \log Z_\beta[d,J], 
\end{eqnarray}
of which the latter is the Helmholtz free energy as a function of the
inverse temperature $\beta$, we can write 
\begin{equation}\label{eq:Information} 
I_d =  - \log Z_1[d,0] - \langle H[s] \rangle_{d} =
- \left. \frac{\partial F_\beta[d,J] }{\partial \beta}\right|_{\beta=1,\,J=0}, 
\end{equation}
as can be verified by a direct calculation. The first expression for
$I_d$ in Eq. \ref{eq:Information} is equivalent to the well known thermodynamic 
relation $F = E - T\, S_\mathrm{B}$ with the internal energy
$E=\langle H[s] \rangle_{d}$, the Boltzmann entropy $ S_\mathrm{B} = -
I_d$ and the temperature, which is set here to $T=1$. 
The second expression actually holds even if the
Hamiltonian is improperly normalized, e.g. $H_0$ can be chosen
arbitrarily if $Z_\beta[d,J]$ is calculated consistently with this
choice.

The Helmholtz free energy $F_\beta[J]$ 
is also the generator of all connected
correlation functions of the signal $\langle s_{x_1} \cdots
s_{x_n}\rangle_{(s|d)}^\mathrm{c} =$$  - \delta^n F_\beta[d,J] /\delta
J_{x_1} \cdots \delta J_{x_n}|_{\beta=1,\,J=0}$. It
can be calculated as follows: 
\begin{eqnarray}
F_\beta &=& -\frac{1}{\beta}\, \log \left(
\frac{Z_{\beta}^\mathrm{\G}[J]}{Z_{\beta}^\mathrm{\G}[J]} \int
\mathcal{D}s\,e^{-\beta \, H_\mathrm{int}[s] } \, e^{-\beta \, (H_\mathrm{\G}[s]
  - J^\dagger s) } \right)\nonumber \\ 
&=&  -  \frac{1}{\beta}\, \log  Z_{\beta}^\mathrm{\G}[J]  - \frac{1}{\beta}\,
\log \left\langle    e^{-\beta \, H_\mathrm{int}[s] }
\right\rangle_{(s|J+j,\mathrm{\G})} \!\!\!\!\!\!\!\!\!\!\!\!\!\!\!\!\!\!\!\!\!\!\! , 
\end{eqnarray}
where the average in the last term is over the Gaussian probability
function $P_{J, \beta}^\mathrm{\G}[s] =  \exp(-\beta\,(
H_\mathrm{\G}[s] - J^\dagger s))/ Z_{\beta}^\mathrm{\G}[J]$. This term can be calculated by using
the well-known fact that the logarithm of the sum of all possible
connected and unconnected diagrams with only internal coordinates (or
without free ends), as generated by the exponential function of the
interaction terms, is given by the sum of all connected diagrams
\citep{Binney1992}. For example, a free theory, perturbed by small,
up-to-fourth-order interaction terms (all being proportional to some
small parameter $\gamma$), has 
\begin{eqnarray}\label{eq:F_beta_explicit}
F_\beta[J] &=& \underbrace{H_0^\mathrm{\G}+\Lambda^{(0)}}_{H_0} -  
\beta^{-1}\, \left[ 
\includegraphics[width=0.5\fgwidth]{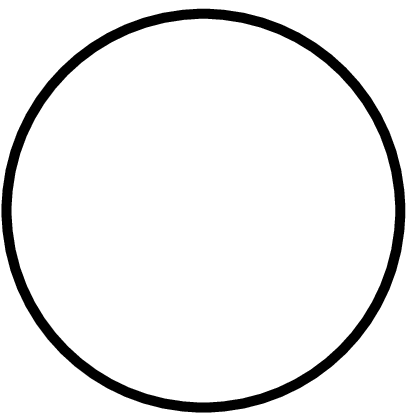} +
\includegraphics[width=\fgwidth]{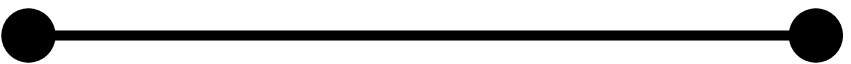} +
\includegraphics[width=0.5\fgwidth]{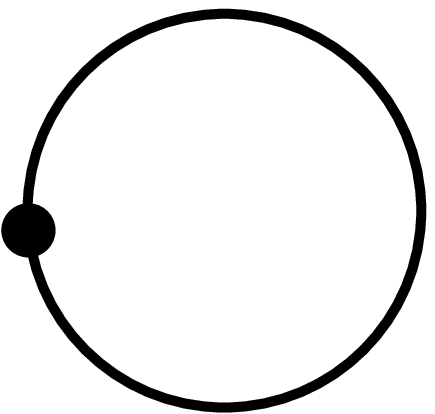} +
\includegraphics[width=\fgwidth]{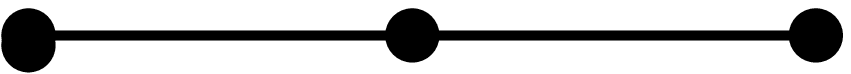} 
\right.
\nonumber \\ &&
\!\!\!\!\!\!\!\!\!\!\!\!\!\!\!\!\!\!\!\!\!\!\!\!\!\!\!\!\!\!\!\!
\left.
+ \includegraphics[width=\fgwidth]{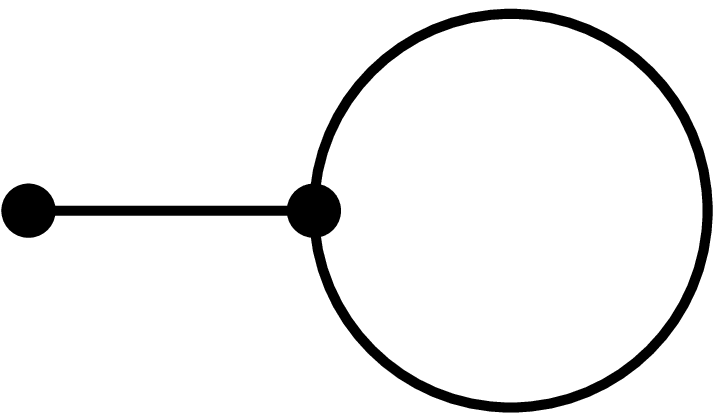} +
\includegraphics[width=0.8\fgwidth]{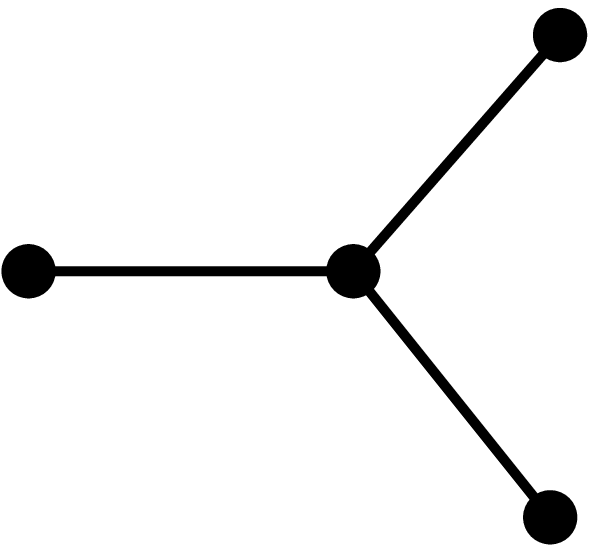} +
\includegraphics[width=\fgwidth]{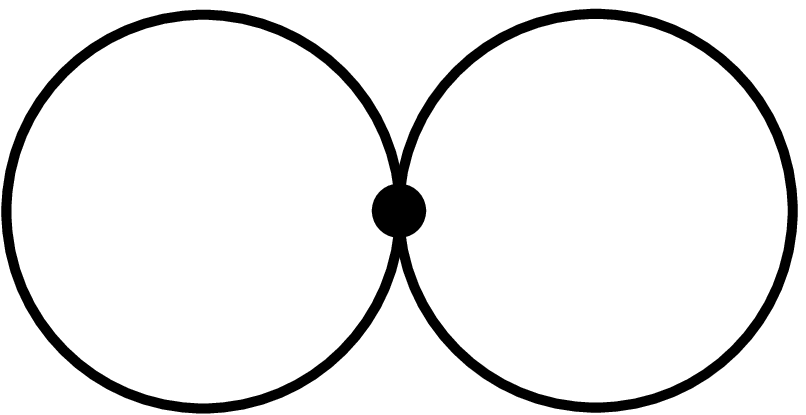} +
\includegraphics[width=0.8\fgwidth]{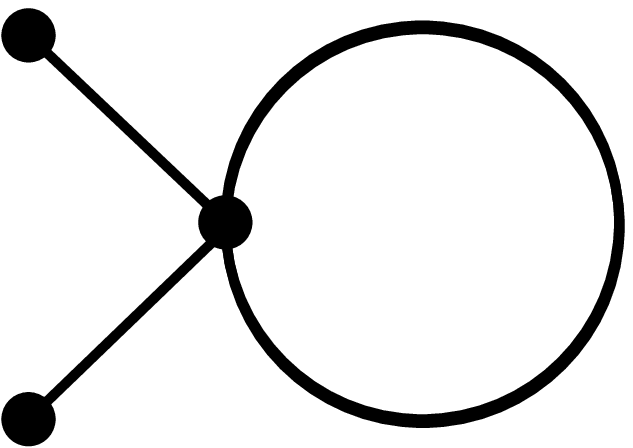} +
\includegraphics[width=0.8\fgwidth]{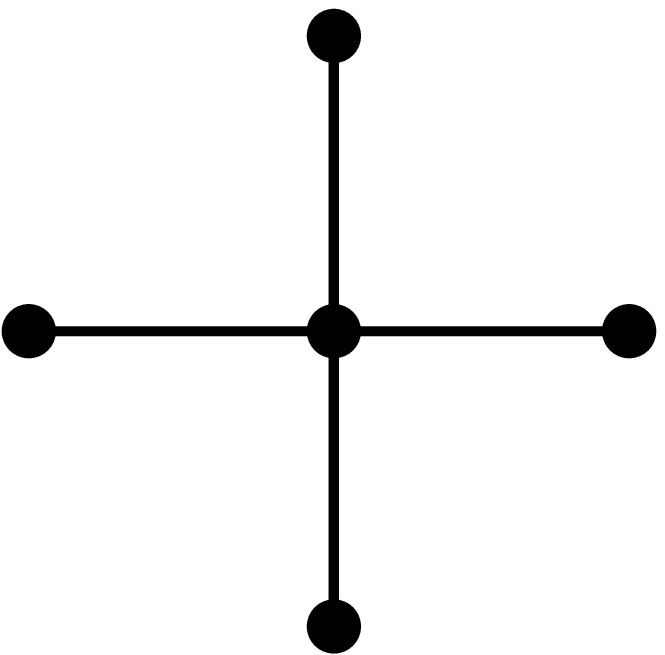}\right] + \mathcal{O}(\gamma^2),
\end{eqnarray}
where an information source vertex reads $\beta\,(J+j-\Lambda^{(1)})$,
an internal vertex with $n$ lines $\beta\, \Lambda^{(n)} $, and the
propagator $\beta^{-1}\, D$. Finally, we have defined
\begin{displaymath}
  \includegraphics[width=0.5\fgwidth]{fg/fg_d1.eps}  = \frac{1}{2}\, \log
  |2\pi\, D\,\beta^{-1}| =\frac{1}{2}\, \mathrm{Tr} (\log (2\pi\,
  D\,\beta^{-1})). 
\end{displaymath}
Thus, we have
\begin{eqnarray}\label{eq:F_beta_explicit2}
F_\beta[J] &=& H_0 -
\frac{1}{2\,\beta}\, \mathrm{Tr} (\log (2\pi\, D\,\beta^{-1}))
+ \frac{1}{2\,\beta}\, \Lambda^{(2)}[D]\nonumber\\
&&
+ \frac{1}{2}\,(J+j-\Lambda^{(1)})^\dagger
(D+\Lambda^{(2)})\,(J+j-\Lambda^{(1)}) \nonumber\\ 
&&
+ \frac{1}{2\,\beta} \Lambda^{(3)}[D,m_J] + \frac{1}{3!}\Lambda^{(3)}[m_J,
  m_J,m_J] \nonumber\\ 
&&
+ \frac{1}{8\,\beta^2}\Lambda^{(4)}[D,D]
+\frac{1}{4\,\beta}\Lambda^{(4)}[D,m_J, m_J] \nonumber\\ 
&&
 + \frac{1}{4!}\Lambda^{(4)}[m_J,m_J,m_J, m_J] 
+ \mathcal{O}(\gamma^2),
\end{eqnarray}
where we introduced the zero-order map $m_J = D\, (J+j)$ for
notational convenience. 
The power of $\beta$ associated with the different diagrams in
Eq. \ref{eq:F_beta_explicit} is given by the number of vertices minus
the number of propagators minus one. Thus, all tree-diagrams are of
order $\beta^0$, the one-loop diagrams are of order $\beta^{-1}$ and
the two loop diagram of order  $\beta^{-2}$, and only the latter two
affect the information content: 
\begin{eqnarray}\label{eq:I_explicit}
I_d\!\! &=&\!\!  -\left[ \frac{\varrho }{2}+ 
\includegraphics[width=0.5\fgwidth]{fg/fg_d1.eps} +
\includegraphics[width=0.5\fgwidth]{fg/fg_d2.eps} +
\includegraphics[width=\fgwidth]{fg/fg_d9.eps} +
\includegraphics[width=\fgwidth]{fg/fg_d8.eps} +
\includegraphics[width=0.8\fgwidth]{fg/fg_d10.eps}\right] +
\mathcal{O}(\gamma^2)\!\!\!\!\!\!\!\!
\nonumber \\ 
\!\!&=&  \!\!
\frac{1}{2}\,\left[ -\mathrm{Tr} (1+\log (2\pi\, D))
+  \Lambda^{(2)}[D]
%
+  \Lambda^{(3)}[D,m_0] \right.\nonumber\\
\!\!&&\!\!
\left.
+ \frac{1}{2}\Lambda^{(4)}[D\otimes (D + m_0\,m_0^\dagger)]\right]  
+ \mathcal{O}(\gamma^2),
\end{eqnarray} 
where $\varrho = \mathrm{Tr}(1)$, $\beta=1$, $J=0$, and thus $m_0 =
D\, j$.\footnote{Here, we introduced the symmetrized tensor product 
$A \,\otimes \, B$ of an $n$-rank tensor $A$ and an $m$-rank tensor $B$,
which has the property
\begin{displaymath}
 (A \,\otimes \, B  )_{x_1\ldots 
x_{n+m}}\!=\! 
\frac{1}{m!}\!\! \! \! 
\sum_{\pi \in \mathcal{P}_{n+m}}\!\!\! A_{x_{\pi(1)}\ldots x_{\pi(n)}} \,
B_{x_{\pi(n+1)}\ldots x_{\pi(n+m)}}, 
\end{displaymath}
with $ \mathcal{P}_{l}$ being the set of permutations of $\{1, \ldots,
l \}$.
}

\subsubsection{Free theory}\label{sec:freeinfo}

To obtain the information content of the free theory, we can set
$\gamma =0$ in Eqs. \ref{eq:F_beta_explicit2} and \ref{eq:I_explicit}
or use Eq. \ref{eq:ZdfreeTheory} with the replacements $J\rightarrow
\beta \,J$, $j\rightarrow \beta \,j$, $D\rightarrow \beta^{-1}D$, and
$H_0 \rightarrow \beta \,H_0$. In both cases we find identically 
\begin{eqnarray}
F_\beta[ J] \!\!\!&=&\!\! \!H_0^\mathrm{\G}-\frac{1}{2} (J+j)^\dagger D \, (J+j)
- \frac{1}{2\beta} 
\mathrm{Tr} \log\left( \frac{2\,\pi}{\beta} D \right)\!, \mathrm{and}\nonumber\\
I_d \!\!\!&=&\!\!\! - \frac{1}{2} \, \mathrm{Tr} \left( 1+ \log\left( 2\,\pi\,
D \right) \right). 
\end{eqnarray}
Very similarly, one can calculate the information prior to the data, which
turns out to be 
\begin{equation}
 I_0 = - \frac{1}{2} \, \mathrm{Tr} \left( 1+ \log\left( 2\,\pi\, S \right) \right).
\end{equation}
Thus, the data-induced information gain is
\begin{eqnarray}\label{eq:data induced information gain}
\Delta I_d  &=& I_d - I_0 = \frac{1}{2} \, \mathrm{Tr} \left( \log\left( S\,
D^{-1} \right) \right)\nonumber\\ 
&=& \frac{1}{2} \, \mathrm{Tr} \left( \log\left( 1 + S\,R^\dagger N^{-1}\,R \right) \right).
\end{eqnarray}
The information gain depends on the signal-response-to-noise ratio $Q
\equiv R\,S\,R^\dagger N^{-1}$, also shortly denoted by the
measurement fidelity or quality. The information increases linearly
with $Q$ as long as $Q\ll 1$,  
but levels off to a logarithmic increase for $Q\gg 1$. 

We note, that for the free theory only the information gain does not 
depend on the actual data realization. 

\subsection{IFT Recipe}

A typical IFT application will aim at calculating a model evidence $P(d)$, the
expectation value of a signal given the data, the map $m(x)=\langle s(x)
\rangle_{(s|d)}$ of the signal, or its variance $\sigma_s^2(x,y) = \langle
(s(x)-m(x))\,(s(y)-m(y)) \rangle_{(s|d)}$ as a measure of the signal
uncertainty.  The general recipe for such applications can be summarized as
following:
\begin{itemize}
\item Specify the signal $s$ and its prior probability distribution $P(s)$. If
 the signal is derived from a physical field $\psi$, of which a prior statistic
 is known, the distribution of $s=s[\psi]$ is induced according to Eq. \ref{eq:P(s)}.
\item Specify the data model in terms of a likelihood $P(d|s)$ conditioned on
  $s$. Again, if the data are related to an underlying physical field $\psi$,
  the likelihood is given by Eq. \ref{eq:likelihood}.
\item Calculate the Hamiltonian $H_d[s] = -\log(P(d,s))$, where $P(d,s)=
  P(d|s)\,P(s)$ is the joint probability, and expand it in a Taylor-Fr\'echet
  series for all degrees of freedom of $s$. Identify the coefficients of the
  constant, linear, quadratic, and $n^\mathrm{th}$-order terms with the
  normalization $H_0$, information source $j$, inverse propagator $D^{-1}$, and
  $n^\mathrm{th}$-order interaction term $\Lambda^{(n)}$, respectively, as
  shown in Eq. \ref{eq:generic Hamiltonian0} or \ref{eq:generic Hamiltonian}.
\item Draw all diagrams, which contribute to the quantity of interest,
  consisting of vertices, lines, and open-ends up to some order in complexity
  or some small ordering parameter. The log-evidence is given by the sum of all
  connected diagrams without open ends, the expectation value of the signal by
  all connected diagrams with one open end, and the signal-variance around this
  mean by all connected diagrams with two open ends.
\item Read the diagrams as computational algorithms specified by the Feynman
  rules in Sect. \ref{sec:interacting fields}, and implement them by using
  linear algebra packages or existing map-making codes for the information
  propagator and vertices. The required discretisation is outlined in
  Sect. \ref{sec:discret}. Information on how to implement the required matrix
  inversions efficiently can be found in the literature given in
  Secs. \ref{sec:lit:map}, \ref{sec:lit:lss}, and \ref{sec:lit:cmb} and
  especially in \citep{2008MNRAS.389..497K}.
\item If the resulting non-linear data transformation (or filter) has the
  required accuracy, e.g. to be verified via Monte-Carlo simulations using
  signal and data realizations drawn from the prior and likelihood,
  respectively, an IFT algorithm is established.
\item In case that too large interaction terms in the Hamiltonian prevent a
  finite number of diagrams to form a well performing algorithm, a re-summation
  of high order terms is due. This can be achieved by the saddle point
  approximation (classical solution, maximum a posteriori estimator), or even
  better by a detailed renormalization-flow analysis along the lines outlined
  in Sect. \ref{sec:responserenorm}.
\end{itemize}

\section{Cosmic large-scale structure via galaxy surveys}\label{sec:poisson}

\subsection{Poissonian data model and Hamiltonian}

Many datasets suffer from Poisson noise, which is
non-Gaussian and signal dependent, and therefore
well suited to test  IFT in the non-linear regime. For example, the
cosmological LSS is traced by galaxies, which may be assumed
to be generated by a Poisson process. On large-scales, the expectation value
of the galaxy density follows that of the underlying (dark) matter
distribution. The aim of cosmography is to recover the initial density
field from the shot-noise contaminated galaxy data. Currently, large galaxy
surveys are conducted in order to chart the cosmic matter distribution in three
dimensions. Improving the galaxy based LSS reconstruction
techniques and understanding their uncertainties better is therefore
an imminent and important goal. Optimal techniques to
reconstruct Poissonian-noise affected signals are also crucial for other
problems, since e.g. imaging with photon detectors plays an important
role in astronomy and other fields. Here, we outline how such problems
can be treated, by discussing a specific data model motivated by the
problem of  large-scale-structure reconstruction from galaxies. 
For this problem we work out the 
optimal estimator and show its superiority numerically.
A more general discussions of models of galaxy and structure formation and 
references to relevant works was given in Sect. \ref{sec:lit:lss}.

In order to treat the Poissonian case in a convenient fashion, we
subdivide the physical space into small cells with volumes $\Delta V$, and
assume that a cell located at $x_{i}$ has an expected number of
observed galaxies  
\begin{equation}\label{eq:muii}
\mu_{i} \approx \kappa\, (1+b\, s(x_{i}))
\end{equation}
with $\kappa = \bar{n}_{\mathrm{g}}\,\Delta V$ being the cosmic average number
of galaxies per cell and $b$ being the bias of the galaxy over-density
with respect to the dark matter overdensity $s$, still assumed to be a
Gaussian random field (Eq. \ref{eq:GaussPrior}). However, this data
model has two shortcomings. First, too negative fluctuations of the
Gaussian random field with  $s<-1$ lead to negative expectation
values, for which the Poissonian statistics is not defined. Second, the
mean density of observable galaxies $\kappa$ and their bias parameter
$b$ are constant everywhere, whereas in reality both exhibit spatial
variations.\footnote{Such variations are due to the geometry of the
observational survey sky coverage, due to a galaxy selection function which 
decreases with distance from the observer, and due to a changing
composition of the galaxy population. The latter distance-effects are
caused by the cosmic evolution of galaxies and by the changing
observational detectability of the different types with distance. We
note, that an observed sample of galaxies, which was selected 
deterministically or stochastically from a
complete sample e.g. by their luminosity due to instrumental
sensitivity, still possesses a
Poissonian statistics, if the original distribution does.} Although
being now spatially inhomogeneous, we assume $\kappa$ and $b$ to be known
for the moment and to incorporate all above observational effects. 

To cure the above mentioned shortcomings we replace Eq. \ref{eq:muii}
by a non-linear and non-translational invariant model: 
\begin{equation}\label{eq:muiifinal}
\mu_{i} = \kappa(x_i)\, \exp(b(x_i)\, s(x_{i})),
\end{equation}
where $\kappa$ and $b$ may depend on position in a known way, and the
unknown Gaussian field $s$, the log-matter density, may exhibit
unrestricted negative 
fluctuations. Note that $\mu$ is the signal response, by our
definition in Eq. \ref{eq:signal response}, since $\mu[s] = \langle d
\rangle_{(d|s)}$. We call $\kappa$ the \textit{zero-response}, since
$\mu[0] = \kappa$. It should be stressed that the data model in
Eq. \ref{eq:muiifinal} is just a convenient choice for illustration
and proof-of-concept purposes, and is easily exchangeable with more
realistic, and even non-local data models. However, this log-normal 
data model was originally proposed by
\citet{1991MNRAS.248....1C}, 
investigated for constrained realizations by
\citet{1995MNRAS.277..933S} 
and \citet{2001PASP..113.1009V} 
and seems to reproduce the statistics of LSS simulations much better
than the often used normal distribution of the
overdensity \citep{2009arXiv0903.4693N}. 

Having chosen a Poissonian process to populate the Universe and our
observational data with galaxies according to the underlying log-density
field $s$, the likelihood is 
\begin{eqnarray}
 P(d|s) &=& \prod_{i}\frac{\mu_i^{d_i}}{d_i!} e^{-\mu_i} \\
&=& \exp\left\lbrace \sum_i \left[ d_i\,\log \mu_i - \mu_i - \log(d_i!)\right] \right\rbrace , \nonumber
\end{eqnarray}
where $d_i$ is the actual number of galaxies observed in cell
$i$. Since $P(s) = \G(s,S)$, the Hamiltonian is given by 
\begin{eqnarray}
\label{eq:PoissonHamiltonian} 
H_d[s] &=& -\log P(d,s) =  -\log P(d|s)-\log P(s) \nonumber\\
&=&    - d^\dagger b\,s  + \kappa^\dagger \exp(b\,s) + H'_0 + \frac{1}{2}\,
s^\dagger S^{-1} s \nonumber\\ 
&=&   \frac{1}{2}\, s^\dagger D^{-1} s - j^\dagger s + H_0 + \sum_{n=3}^\infty
\frac{1}{n!}\, \lambda_n^\dagger \, s^n,\; \mbox{with}\nonumber\\ 
D^{-1}&=& S^{-1} +\widehat{\kappa\, b^2}, \\
j &=& b\,(d-\kappa),\nonumber\\
H_0 &=& \frac{1}{2}\,\log(|2\pi\,S|)+(\kappa+\log(d!))^\dagger 1 - d^\dagger
\log \kappa,\nonumber\\ 
\mbox{and}&&\nonumber\\
\lambda_n &=& \kappa\,b^n.\nonumber
\end{eqnarray}
The hat on a scalar field denotes that it should be read as a matrix,
which is diagonal in position space (see
Appendix \ref{sec:notation}). 
A few remarks should be in order. Comparing the propagator to the one of our
Gaussian theory one can read off an inverse noise term $M= R^\dagger
N^{-1} R = \widehat{\kappa\, b^2}$. Thus the
effective (inversely response weighted) noise decreases with
increasing mean galaxy number and bias, and seems to be infinite in regions
without data ($\kappa =0$) without causing any problem for the formalism.

The information source $j$ increases with increasing response (bias) of the
data (galaxies) to the signal (density fluctuations). However, it certainly
vanishes for zero response ($b=0$) or in case that the observed galaxy counts
match the expected mean at a given location exactly. 
Finally,
the interaction terms $\lambda_n$ are local in position space, and
vanish with decreasing $b$ and $\kappa$. The latter parameter is under
the control of the data analyst, since it is proportional to the
volume of the individual pixel sizes, and therefore can be made
arbitrarily small by choosing a more fine grained resolution in signal
space. However, this would not change the convergence properties of
the series since any interaction vertex has then to be summed over a
correspondingly larger number of pixels within a coherence patch of
the signal, which exactly compensates for the smaller coefficient.%
\footnote{$\kappa$ seems to control 
the stiffness of the later introduced response renormalization flow 
equation and its values is therefore numerically relevant. A lower $\kappa$, 
due to  a finer space pixelisation, results in a less stiff and better
behaved equation.}
The bias, in contrast,  is set by nature and can be regarded as a power
counting parameter, which provides naturally a numerical hierarchy
among the higher order vertices and diagrams for $b^2 S <1$.
Note that $j = \mathcal{O}(b)$.  

\subsection{Galaxy types and bias variations}

Real galaxies can be cast into different classes, which all differ in terms of
their luminosities, bias factors, and the frequencies with which they are
found in the Universe. Although we are not going to investigate this
complication in the following, it should be explained here how all the
formulae in this section can easily be reinterpreted, in order to incorporate
also the different classes of galaxies. 

The galaxies can be characterized by a type-variable $L\in
\Omega_\mathrm{type}$, which may be the intrinsic luminosity, the
morphological galaxy type, or a multi-dimensional combination of all
properties which determine the galaxy type's spatial distributions via
a $L$-dependent bias $b_L$, and their detectability as encoded in
$\kappa_L$. The data space is now spanned by $\Omega_\mathrm{data} =
\Omega_\mathrm{space} \times \Omega_\mathrm{type}$, and also $\mu$,
$\kappa$ and $b$ can be regarded as functions over this space. 

Performing the same algebra as in the previous section, just taking
the larger data-space into account, we get to exactly the same
Hamiltonian, as in Eq. \ref{eq:PoissonHamiltonian}, if we interpret
any term containing $d$, $\kappa$ and $b$ to be summed or integrated
over the type variable $L$. Thus, we read 
\begin{eqnarray}
\label{eq:Poisson Bias Type}
 j(x) \!\!\!&=&\!\!\! \left( b\,(d - \kappa)\right)(x) \equiv \!\!  \int \! \!
 d L\, b_L(x)\, (d_L(x) - \kappa_L(x)),\nonumber \\ 
D^{-1}_{xy} \!\!\!&=&\!\!\! \left( S^{-1} + \widehat{\kappa\, b^2} \right)
 _{xy} \equiv S^{-1}_{xy} +  1_{xy} \! \! \int \! \! dL\, \kappa_L(x)
 b^2_L(x),\nonumber \\ 
\lambda_n(x)  \!\!\!&=& \!\!\! \left(\kappa\,b^n\right)(x) \equiv \!\!  \int \!
 \! dL\,\kappa_L(x)\, b^n_L(x),\;\mbox{and}\\ 
\mu [s](x) \!\!\!&=&\!\!\!  \left(\kappa\, e^{b\,s}\right) (x) \equiv   \!\!
 \int \! \! dL\, \kappa_L(x) \, e^{b_L(x)\, s(x)} =  \! \! \int \!\!  dL\,
 \mu_L[s](x),\nonumber 
\end{eqnarray}
which all live in $\Omega_\mathrm{space}$ solely, so that the
computational complexity of the matter distribution reconstruction
problem is not affected at all, and only a bit more book-keeping is
required in its setup. 

A few observations should be in order. In case of all galaxies having
the same bias factor, Eq.  \ref{eq:Poisson Bias Type} is simply a
marginalization of the type variable $L$, and any differentiation of
the various galaxy types is not necessary. Since all known galaxy types
seem to have $b\sim \mathcal{O}(1)$, such a marginalization seems to
be justified, and explains why LSS reconstructions,
which applied this simplification, are relatively successful, although
the different galaxy masses, luminosities, and frequencies vary
by orders of magnitude. 
As our numerical experiments below reveal, the data, and therefore
the reconstructability of the density field, exhibit a sensitive
dependence on the bias for $s$-fluctuations with unity
variance.\footnote{This is found for our specific data model $\mu
\propto \exp (b\, s)$, however, should also apply for other
models, which somehow have to keep $\mu \ge 0$ even for $b\,s < -1 $} 
Such a variance is indeed observed on scales below $\sim 10$ Mpc in
the galaxy distribution, and therefore the galaxy type-dependent bias variation does
indeed matter. Larger galaxies, which have
larger biases, therefore provide per galaxy a slightly larger
information source ($j\propto b$), less shot noise ($R^\dagger N^{-1}R
\propto b^2$), and increasingly larger higher-order interaction terms
($\lambda_n \propto b^n$) in comparison to smaller galaxies. However,
smaller galaxies are much more numerous by orders of magnitude, and
therefore provide the largest total contribution to the information source,
noise reduction and most low-order interaction terms. Thus, the latter
will dominate and therefore permit a reasonable accurate matter
reconstruction from an inhomogeneous galaxy survey using a single bias
value. Nevertheless, improvements of the bias treatment
are possible by applying the recipes described here.

\subsection{Non-linear map making}

The map, the expectation of our information field $s$ given the data,
is to the lowest order in interaction   
\begin{eqnarray}\label{eq:mPoisson}
m_1 &=&  \includegraphics[width=\fgwidth]{fg/fg_a1.eps} 
+ \includegraphics[width=\fgwidth]{fg/fg_a3.eps} 
+ \includegraphics[width=\fgwidth]{fg/fg_a2.eps}
 + \includegraphics[width=\fgwidth]{fg/fg_a4.eps} 
+ \mathcal{O}(b^6) \nonumber\\
&=&
D_{xy}\, j_y - \frac{1}{2} D_{xy}\, b^3_y\,\kappa_y\,D_{yy} 
- \frac{1}{2} D_{xy}\,  b^3_y\,\kappa_y\,(D_{yz}\,j_z)^2
\nonumber\\
&& -  \frac{1}{2} D_{xy}\,  b^4_y\,\kappa_y\,D_{yy}\,D_{yz}\,j_z+ \mathcal{O}(b^6) 
\end{eqnarray}
or in compact notation
\begin{equation}
 m_1 = m_0 - \frac{1}{2} \, D\, \widehat{b^3\,\kappa} \left(\widehat{D} +
 m_0^2 + \widehat{\widehat{D}b} \, m_0\right)  +
 \mathcal{O}(b^6). 
\end{equation}
It is apparent, that the non-linear map making formula contains
corrections to the linear map $m_0= D\, j$. The first two correction
terms are always negative, reflecting the fact that our non-linear
data model has non-symmetric fluctuations in the data with respect to
the mean.
The last correction term is oppositely
directed to the linear map, thereby correcting for the
curvature in the signal response.

\begin{figure*}[tbh]
\centering
\includegraphics[width=0.80\textwidth,bb= 135 200 480 560]{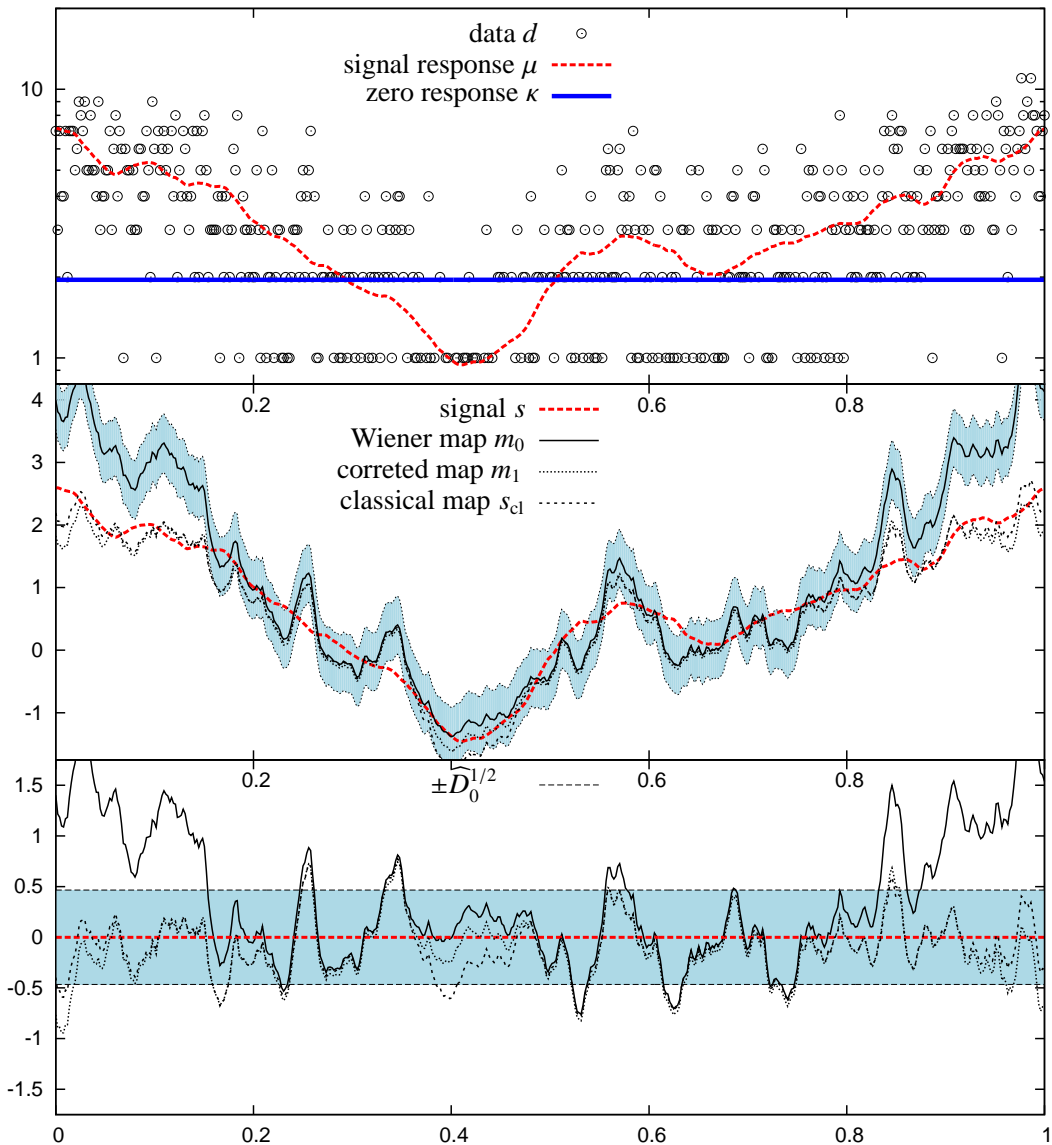}
\caption{Poissonian-reconstruction of a signal with unit variance and
 correlation length $q^{-1}=0.04$, observed with slightly non-linear
 response ($b=0.5$, resolution: 513 pixels
 per unit length, zero-signal galaxy density: 1000 galaxies per unit
 length). \textbf{Top:} data $d$, signal response $\mu$,
 and zero-response $\kappa$. \textbf{Middle:} signal $s$, linear
 Wiener-filter reconstruction $m_0 = D\,j$, its one-sigma error interval
$m_0 \pm \widehat{D}^{1/2}$, next order reconstruction $m_1$
 according to Eq. \ref{eq:mPoisson}, and classical solution
 $s_\mathrm{cl}$ according to
 Eq. \ref{eq:classicalEqPoisson}. Although the linear Wiener is reconstructing 
well at most locations, the nonlinear response requires the
 perturbative corrections  
present in $m_1$ or the classical solution in regions of high signal strength.
\textbf{Bottom:} The residuals, the deviations of
 $m_0$, $m_1$, $s_\mathrm{cl}$ from the signal, and the Wiener-variance $\pm \widehat{D}^{1/2}$.} 
 \label{fig:PoissonSmallBias}
\end{figure*}

A one-dimensional, numerical example is displayed in
Fig. \ref{fig:PoissonSmallBias}. There, the signal was realized to
have a power spectrum $P_s(k) \propto (k^2 + q^{2})^{-1}$, with a
correlation length $q^{-1}=0.04$. The normalization was chosen such
that the auto-correlation function is $\langle s(x) \, s(x+r)
\rangle_{(s)} = \exp(-|q\, r|)$ and therefore the signal dispersion
is unity, $\langle s^2 \rangle_{(s)} =1$. The data are generated by a
Poissonian process from $\kappa_s = \kappa\, \exp(b\,s)$ with $b =
0.5$. All three displayed reconstructions exhibit less power than the
original signal, as it is expected since the reconstruction is
conservative, and therefore biased towards zero. 

The non-linear correction to the naive map $m_0$ should not be too
large, otherwise higher order diagrams have to be included. In the
case displayed in Fig. \ref{fig:PoissonSmallBias}, $b=0.5$ ensured
that the linear corrections were mostly going into the right
direction. However, in case $b\approx 1$ there is no obvious ordering
of the importance of the different interaction vertices, and numerical
experiments reveal that the first order corrections strongly
overcorrect the linear map $m_0=D\,j$. In such a case
interaction re-summation techniques should be used to incorporate as
many higher order interaction terms as possible. One very powerful
re-summation is provided by the classical solution, as developed
below, which contains all tree-diagrams simultaneously. This solution,
also show in Fig. \ref{fig:PoissonSmallBias}, is very close to $m_1$
in this case. 

\begin{figure*}[tb]
\centering
\includegraphics[width=0.8\textwidth,bb= 135 200 480 560]{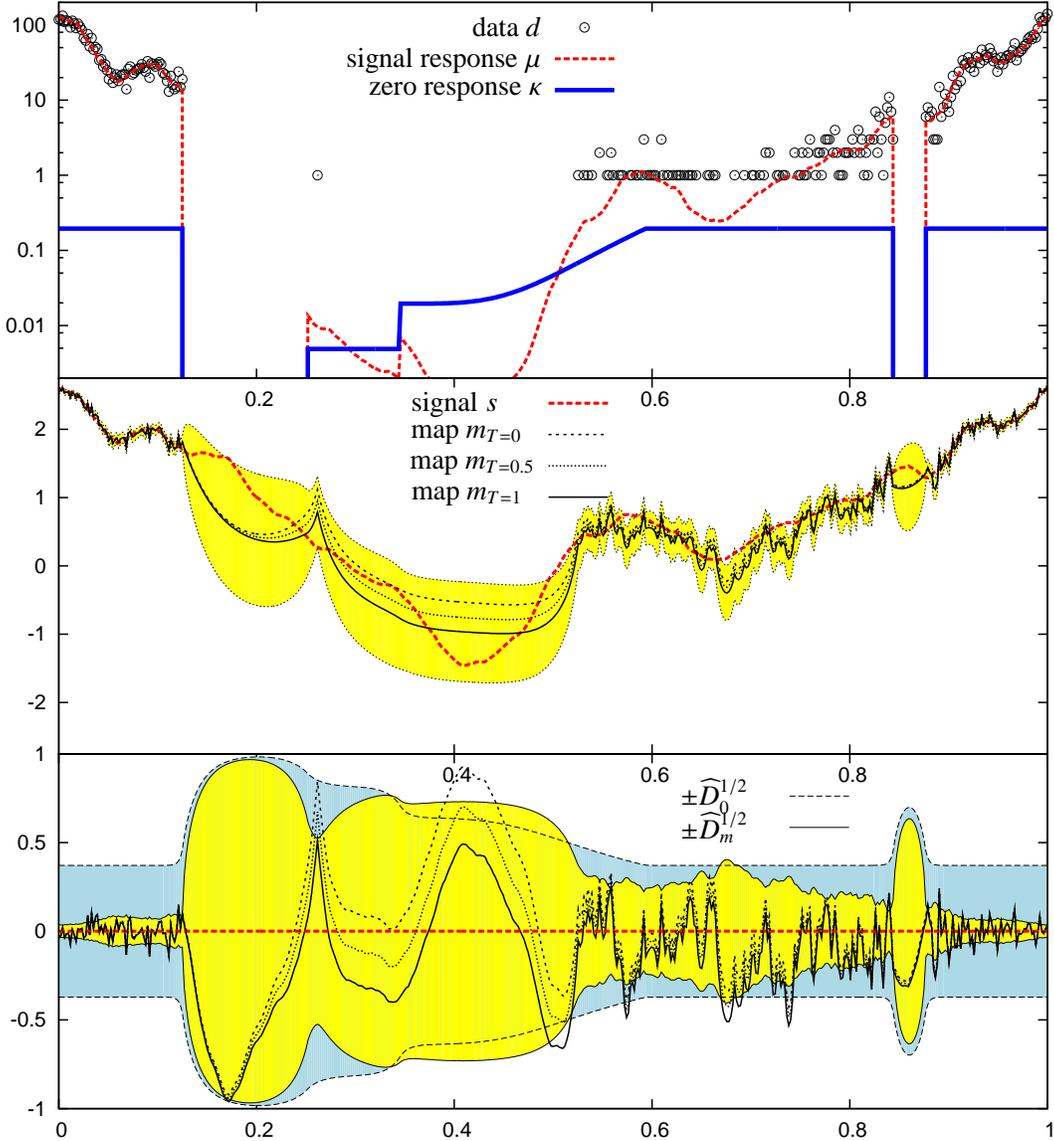}
\caption{Poissonian-reconstruction of the same signal realization as
 in Fig. \ref{fig:PoissonSmallBias} (unit variance and correlation
 length $q^{-1}=0.04$), observed now with a strongly non-linear  response 
($b=2.5$, resolution: 512 pixels per unit length, zero-signal galaxy density:
100 galaxies per unit length where mask is one) through a complicated
 mask. \textbf{Top:} data $d$, signal response $\mu$, and
 zero-response $\kappa$. \textbf{Middle:} signal $s$, classical solution
$s_\mathrm{cl} = m_{T=0}$, intermediate solution $m_{T=0.5}$ and  
renormalization-based reconstruction $m_{T=1}$ with uncertainty
 interval $m_{T=1}\pm \widehat{D}_{T=1}^{1/2}$, and mask
 $\kappa/(n_\mathrm{g}\,\Delta V)$. The linear 
Wiener-filter reconstruction $m_0$ as well as its next order corrected version
$m_1$ are not displayed, since they are partly far outside the displayed
area. \textbf{Bottom:} Deviations of the three reconstructions from the signal,
and the original and the renormalized uncertainty estimates
 $\pm \widehat{D}_0^{1/2}$ and $\pm \widehat{D}_{T=1}^{1/2}$,
 respectively. Note, that in the regions with many observed galaxies, 
the high signal to noise ratio can be seen in the narrowness of
$\widehat{D}_{T=1}^{1/2}$, which is significantly smaller than the
 data-unaffected $\widehat{D}_0^{1/2}$ at these locations. }
 \label{fig:PoissonLargeBias}
\end{figure*}

\subsection{Classical solution}\label{sec:classicalPoisson}

The classical signal field or MAP solution is given by
Eq. \ref{eq:classicalEq}, which reads in this case 
\begin{eqnarray}\label{eq:classicalEqPoisson}
 s_\mathrm{cl} &=& D \left( j - \sum_{m=2}^{\infty}
 \frac{b^{m+1}}{m!}\, \kappa \, s_\mathrm{cl}^{m} \right)\nonumber\\
&=& D \, b \left( d - \kappa\, \left(e^{b\,s_\mathrm{cl}} 
-b\,s_\mathrm{cl}\right)
 \right)\\
&=& S\, b\, (d
 -\underbrace{\kappa\,e^{b\,s_\mathrm{cl}}}_{\kappa_{s_\mathrm{cl}}}). \nonumber 
\end{eqnarray}
The last expression motivates to introduce the expected number of
galaxies given the signal $s$: 
\begin{equation}
 \kappa_s = \kappa\, e^{b\,s}.
\end{equation}
Also alternative forms of the MAP equation can be derived, for example
one, which is especially suitable for large $j$: 
\begin{equation}
 s_\mathrm{cl} = \frac{1}{b} \, \log\left[ \frac{j- S^{-1}s_\mathrm{cl}
 }{\kappa\,b}\right] = \frac{1}{b} \, \log\left[ \frac{d}{\kappa} -1 -
 \frac{S^{-1}s_\mathrm{cl} }{\kappa\,b}\right]. 
\end{equation}
This may be solved iteratively, while ensuring that
$s_\mathrm{cl}^{(i)} \le S\, j$ at all iterations $i$ with equality
only where $\kappa =0$.  
This form of the classical field equation has some similarities to the
naive inversion of the response formula, $\langle d \rangle_{(d|s)}=
\kappa\,\exp (b\,s)$, which yields 
\begin{equation}\label{eq:snaive}
 s_\mathrm{naive} =  \frac{1}{b} \, \log\left[ \frac{d}{\kappa}\right],
\end{equation}
a formula one can only dare to use in regimes of large $d$. 
Since  $s_\mathrm{naive}$ contains the full noise of the data, a suitable 
naive map may be given by $m_\mathrm{naive} = S\,s_\mathrm{naive}$, after 
some fix for the locations without galaxy counts.
The classical solution, however,  is more conservative than this naive data
inversion, in that there is a damping term, $ {S^{-1}s_\mathrm{cl}
}/({\kappa\,b})$, compensating a bit the influence of too large data
points.  

Those equations permit to calculate the classical solution if suitable
numerical regularization schemes are applied, since naive iterations
can easily lead to numerical divergences in the non-linear case. 

One way of doing this is by turning the classical equation 
(Eq. \ref{eq:classicalEqPoisson}) into a dynamical system. 
Its initial conditions are given by a well solvable linear or even trivial
problem to which non-linear complications are added successively during 
an interval of some pseudo-time. The endpoint of this dynamics is then the 
required solution. The meaning of the pseudo-time depends on the way it was 
set up. In any case, it can just be regarded as a mathematical trick
to generate a differential equation, which might be easier to solve numerically 
than the original problem.

For example, a pseudo-time $\tau$ can be introduced by setting $j(\tau) =
\tau\, j$. Thus,  the information source is successively injected into
an initially trivial field state, $s_\mathrm{cl}(0)=0$. This allows to
set up a differential equation for $s_\mathrm{cl}(\tau)$ by taking the
time derivative of Eq. \ref{eq:classicalEqPoisson}, 
\begin{equation}
\label{eq:scltau}
 \dot{s}_\mathrm{cl} = D_{s_\mathrm{cl}}\, j\;\;\mbox{with}\;\;
 D_{s_\mathrm{cl}}= \left(S^{-1} + \kappa_{s_\mathrm{cl}}\,b^2\right)^{-1}, 
\end{equation}
which has to be solved for $s_\mathrm{cl}(1)$ starting from
$s_\mathrm{cl}(0)=0$. This equation is very appealing, since it
looks like Wiener-filtering an incoming information stream $j$ and
accumulating the filtered data, while simultaneously tuning the filter
$D_{s_\mathrm{cl}(\tau)}$
to the accumulated knowledge on the signal $s_\mathrm{cl}(\tau)$ and thereby implied
Poissonian-noise structure. Thus, it is a nice example system for
continuous Bayesian learning and also illustrates how different
datasets can successively be fused into a single knowledge basis. 

Map-making algorithms with a higher fidelityare
possible by not only investigating the maximum of the posterior, but
by averaging the signal $s$ over the full support of $P(s|d)$. 
Anyhow, we can assume that a good approximation $t\approx
s_\mathrm{cl}$ to the classical solution can be
achieved. Figs. \ref{fig:PoissonSmallBias} and
\ref{fig:PoissonLargeBias} display classical solutions for slightly
and strongly non-linear Poissonian inference problems. Especially the
second example shows that the classical solution 
can be improved in regions of large uncertainty (see region between
$x=0.2$ and $0.5$ in Fig. \ref{fig:PoissonLargeBias}, where apparently
better estimators exist) for 
missing uncertainty loop diagrams, which contain information about the
non-Gaussian structure of the posterior $P(s|d)$ away from
$s_\mathrm{cl}$.

\subsection{Uncertainty-loop corrections}

Now, we see how the missing uncertainty loop corrections can be added
to the classical solution.  
These corrections can be derived from the Hamiltonian of the
uncertainty-field $\phi = s-t$,  
\begin{eqnarray}\label{eq:PoissonuncertaintyHamil}
 H_t[\phi] &=& \frac{1}{2}\, \phi^\dagger D_t^{-1} \phi - j_t^\dagger\phi  +
 \kappa_t^\dagger g(b\, \phi) + H_{0,t},\;\;\mbox{where} \nonumber\\ 
D_t^{-1} &=& S^{-1} + b^2\,\widehat{\kappa}_t, \nonumber\\
j_t &=& b\,(d- \kappa_t) - S^{-1} t,\\
g(x) &=& e^x -1 -x - \frac{1}{2}\,x^2 = \sum_{m=3}^\infty \frac{x^m}{m!}, \nonumber
\end{eqnarray}
and $H_{0,t}$ is a momentarily irrelevant normalization
constant. Again, we have permitted for a non-zero $j_t$, since $t$
might not be exactly the classical solution.  

It is interesting to note that the interaction coefficients in this
Hamiltonian, $\lambda_t^{(m)} = \kappa_t\, b^m$, all reflect the
expected number of galaxies given the reference field $t$. Thus, the
replacement $\kappa_0 \rightarrow \kappa_t$ would provide us with the
shifted field Hamiltonian, as defined in Eq. \ref{eq:shifted Hamiltonian},
except for the term $-S^{-1}t$ in $j_t$. It
turns out, that this term is some sort of counter-term, which
accumulates the effect of the non-linear interactions.

We see that effective interaction terms arise when relevant
parts of the solution are absorbed in the background field $t$. A
similar approach is desirable for the loop diagrams. Instead of
drawing and calculating all possible loop diagrams, we want to absorb
several of them simultaneously into effective coefficients. For each
vertex of the Poissonian Hamiltonian with $m$ legs, there exist diagrams
in any Feynman-expansion, in which a number of $n$ simple loops are
added to this vertex. Such an  $n$-loop enhanced $m-$vertex is given by 
\begin{equation}
\includegraphics[width=0.75\fgwidth]{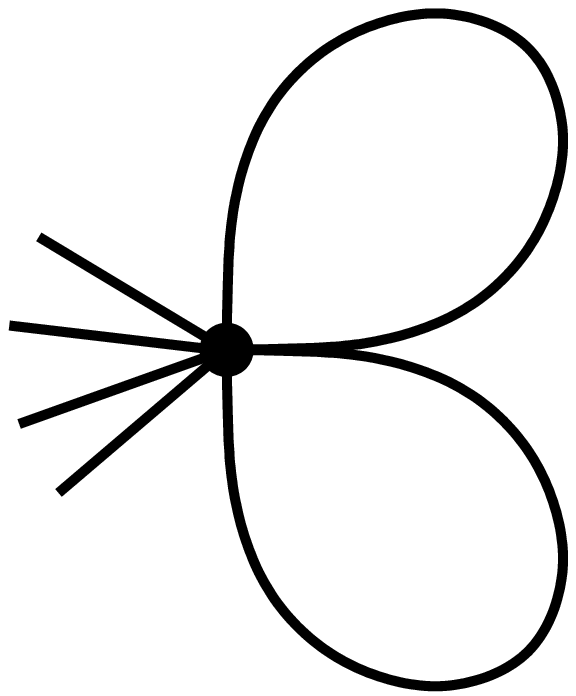} = 
\frac{-1}{2^n\,n!}\,\lambda_t^{(m+2n)} \,\widehat{D_t}^n = \frac{-1}{2^n\,n!}\,
\kappa_t\, b^{m+2n}\,\widehat{D_t}^n. 
\end{equation}
All these diagrams can be re-summed into an effective interaction vertex, via
\begin{eqnarray}
 \lambda_t^{(m)}  &\rightarrow &  {\lambda'}_t^{(m)} =
%
\kappa_t\, b^m \sum_{n=0}^\infty  \frac{1}{2^n\,n!}\, b^{2n}\,\widehat{D}^n
\nonumber \\
&=& \kappa_t \, \exp\left(\frac{b^2}{2}\, \widehat{D}\right) \, b^m\\
&=&\kappa_{t + \frac{b^2}{2} \widehat{D}} \, \,b^m = \lambda^{(m)}_{t +
  \frac{b^2}{2} \widehat{D}}.\nonumber  
\end{eqnarray}
Thus, this re-summation is effectively equivalent to the replacement
\begin{equation}\label{eq:replaceKappa}
\kappa_{t}\rightarrow \kappa_{t+b\, \widehat{D}/2},
\end{equation}
which reflects the larger expected response to a reference field $t$
due to the uncertainty fluctuations around it. Those fluctuations pick
up the asymmetric shape of  the exponential term in the Hamiltonian,
where the larger response to positive fluctuations is not fully
compensated by the lower response to negative fluctuations. One might
wonder, if the simple replacement rule in Eq. \ref{eq:replaceKappa}
could supplement the classical solution with the missing uncertainty
loop corrections. Thus we ask, if the modified classical equation 
\begin{equation}
\label{eq:fixpm}
 m = b\, S\,(d - \kappa_{m+b\, \widehat{D}/2})
\end{equation}
together with a self-constitently determined propagator
\begin{equation}
\label{eq:fixpM}
 D^{-1} = S^{-1}+ b^2\,  \widehat{\kappa}_{m+b\, \widehat{D}/2}
\end{equation}
could provide the mean field given the data. A more rigorous
renormalization calculation will show that this is indeed the case,
within some approximation.


The loop-corrected density and propagator permit to construct estimators 
for the dark matter density itself,
\begin{equation}
 \varrho = \varrho_0 \, e^{c\, s},
\end{equation}
instead of its logarithm, $s$. Here
$c$ fixes the relation between $s$ and $\varrho$, and  $\varrho_0$ being the cosmic median dark 
matter density. Translating our log density map into the density results in the naive density estimator
\begin{equation}\label{eq:rhonaive}
 m_\varrho^\mathrm{naive} = \varrho_0 \, e^{c\, m},
\end{equation}
which is not optimal in the sense of minimal rms deviations. The proper estimator would  be
\begin{equation}\label{eq:rhomap}
 m_\varrho = \langle \varrho_0 \, e^{c\, s} \rangle_{(s|d)} = 
\varrho_0 \, e^{c\, m + {c^2}\,\widehat{D}/2},
\end{equation}
which contains uncertainty loop corrections accounting for the shift of the mean under the non-linear 
transformation between log-density and density.

\begin{figure*}[tbh]
 \centering
\includegraphics[width=0.49\textwidth]{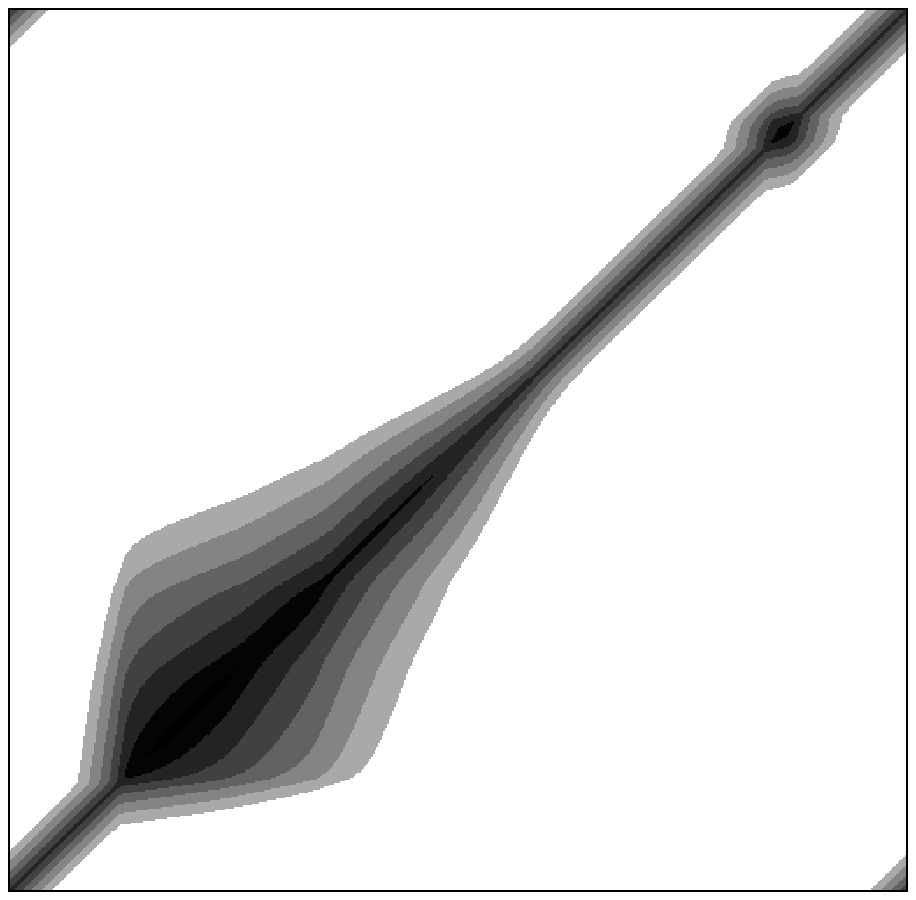}
\includegraphics[width=0.49\textwidth]{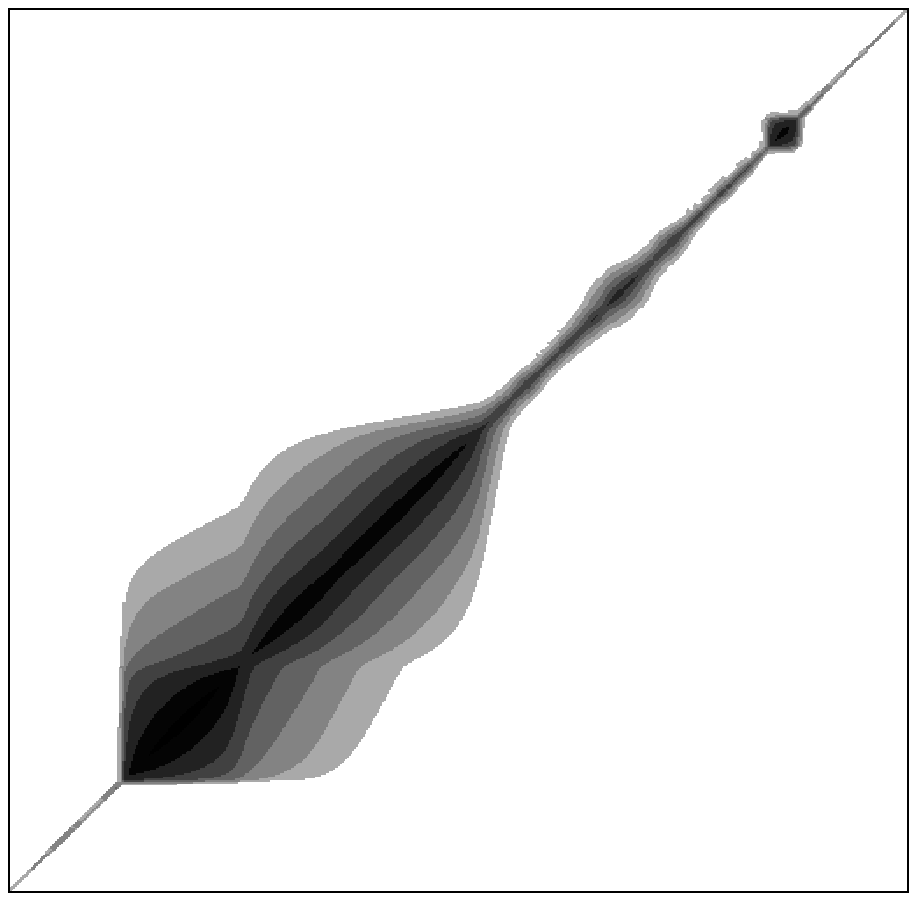}
 \caption{The original propagator $D_0 =
 (S^{-1}+\widehat{\kappa_0\,b^2})^{-1}$ (left) and the final of the
 renormalization flow $D$ (Eq.  
\ref{eq:renormalizationwithT}, right) in logarithmic grey
 scaling for the data displayed in
 Fig. \ref{fig:PoissonLargeBias}.
The values of the diagonals show the
 local uncertainty variance (in Gaussian approximation) before
 ($\widehat{D_0}$) and after ($\widehat{D}$) the data is analyzed,
 respectively. 
The bottom left and top right corners exhibit
 non-vanishing propagator values due to the assumed periodic spatial
 coordinate, which puts these corners close to the two others on the
 matrix diagonal.} 
 \label{fig:Dprop} 
\end{figure*}

\subsection{Response renormalization}\label{sec:responserenorm}

Since we are dealing with a $\phi^\infty$-field theory, the zoo of
loop diagrams is quite complex, and forms something like a
\textit{Feynman foam}. In order not to get stuck in the multitude of
this foam, we urgently require a trick to keep either the maximal
order of the diagrams low, or to limit the number of vertices per
diagram, or both. We have basically two handles on any
interaction term $\lambda_n = \kappa \, b^n$, the bias $b$ and the
zero-response $\kappa$. We concentrate on the response, since it enters the Hamiltonian in a linear way and also the data can be regarded to be proportional to $\kappa$. Thus, the full Hamiltonian
\begin{equation}
 H[s] = \frac{1}{2}\, s^\dagger S^{-1} s - b\,d^\dagger s + \kappa_0 \, e^{b\,s}
\end{equation}
can be regarded to be proportional to the response, except for the prior term and also constant terms we immediately drop here and in the following.

Let us assume that prior to any data analysis we have an initial guess $m_0$ for the signal with some Gaussian uncertainty characterized by the covariance $D_0$. This can be expressed via a Hamiltonian of the form
\begin{equation}
 H_0[s] = \frac{1}{2}\, (s-m_0)^\dagger D_0^{-1} (s - m_0),
\end{equation}
which defines a probability density via $P_0(s) \propto \exp(-H_0[s])$. In case the prior should be our initial guess, we have $m_0 = 0$ and $D_0 = S$, but we need not restrict ourself to this case. Now, we want to anticipate step by step the information of the full problem, and forget our initial guess with the same rate. This can be modeled by adopting an affine parameter $\tau$, which measures how much we exposed ourself to the full problem. For each $\tau$, which we regard as a pseudo-time, our knowledge state is described by an Hamiltonian $H_\tau$. Increasing $\tau$ by some small amount $\eps$ should therefore lead to the next knowledge state characterized by
\begin{equation}
 H_{\tau+\eps} = H_\tau[s] + \eps \,\left( H[s] - H_\tau[s]\right).
\end{equation}
This equation just models an asymptotical approach to the correct Hamiltonian.
If the initial guess was the prior, one sees that for infinitesimal steps $\eps$ the knowledge flow corresponds to tuning up all terms proportional to $\kappa$, 
\begin{displaymath}
 H_\tau [s] = \frac{1}{2}\, s^\dagger S^{-1} s + \left( 1 - e^{-\tau}\right) \, \left(- b\,d^\dagger s + \kappa_0 \, e^{b\,s}\right) \rightarrow
H[s] .
\end{displaymath}
This motivates the term {\it response renormalization} for this kind of continuous learning system, into which the information source as well the interactions are fed with the same rate.

The trick for the renormalization procedure is to approximate the knowledge state at each moment $\tau$ to be of Gaussian shape and therefore the Hamiltonian to be free (quadratic in the signal). Thus we set
\begin{equation}
 H_\tau[s] = \frac{1}{2}\, (s-m_\tau)^\dagger D_\tau^{-1} (s - m_\tau),
\end{equation}
where $m_\tau$ and $D_\tau = (S^{-1} + M_\tau)^{-1}$ are the mean and dispersion of the field given the acquired knowledge at time $\tau$, respectively.

These have to be updated when the next learning step is to be performed. The next Hamiltonian, before it being again replaced by a free one, is 
\begin{eqnarray}
 H_{\tau+\eps}[\phi] &=& \frac{1}{2}\, \phi^\dagger D_\tau^{-1} \phi \nonumber
\\
&+&  \eps \left( (S^{-1} m_\tau - b \, d)^\dagger \phi - \frac{1}{2}\, \phi^\dagger M_\tau \phi + \kappa_{m_\tau} \, e^{b\,\phi} \right)\nonumber\\
&=&  \frac{1}{2}\, \phi^\dagger D_\tau^{-1} \phi + \eps \sum_{n=1}^{\infty} \frac{1}{n!} \,\lambda_n\, \phi^n,
\end{eqnarray}
if expressed for the momentarily uncertainty field $\phi = s - m_\tau$.
Here, the perturbative expansion coefficients are given by
\begin{eqnarray}
 \lambda_1 &=& \kappa_{m_\tau} \, b + S^{-1} m_\tau - b \,d, \nonumber\\
 \lambda_2 &=& \kappa_{m_\tau} \, b^2 - \widehat{M}_\tau, \;\mbox{and}\nonumber\\
 \lambda_n &=&  \kappa_{m_\tau} \, b^n\;\mbox{for}\; n>2, \nonumber
\end{eqnarray}
assuming for simplicity that $M_\tau$ is diagonal. This is a save restriction, since we will see that for $\tau \rightarrow \infty$ this is the case asymptotically, even for a non-diagonal  initial $M_0$. Thus we can require that our initial guess was also of this form.

In order to approximate this Hamiltonian by a free one, we have to calculate the shifted mean field and its connected two-point correlation function, the full propagator. To first order in $\eps$ only leaf diagrams with a
single perturbative interaction vertex contribute to the perturbed
expectation value of $\phi$: 
\begin{eqnarray}\label{eq:leafe diagrams}
 \langle \phi \rangle_{(s|d)}^{(\tau+\eps)}\! \!\!&=& \!\!\!
\includegraphics[width=\fgwidth]{fg/fg_a1.eps} +
\includegraphics[width=\fgwidth]{fg/fg_a3.eps} + 
\includegraphics[width=\fgwidth]{fg/fg_b2.eps} + 
\includegraphics[width=\fgwidth]{fg/fg_b5.eps} + 
\; \ldots \nonumber\\
 \!\!\!&=& \!\! \!\varepsilon\,D_\tau\,\left[ b\,d - S^{-1} m_\tau - b\, \kappa_{m_\tau}\,e^{b^2\,\widehat{D}_\tau/2}\right]. 
\end{eqnarray}
Note, that only odd interaction terms shift the
expectation value $m_{\tau + \eps} =  m_\tau + \langle \phi \rangle_{(s|d)}^{(\tau+\eps)}$.
The even ones do not exert any net forces in
the vicinity of $\phi_\tau =0$ since they represent a
potential which is mirror symmetric about this point.

The renormalized propagator $D_ {\tau + \eps}$ is given by the connected two-point correlation function $ \langle \phi \phi^\dagger \rangle_{(s|d)}^{(\tau+\eps)}$, and this is up to linear order in $\eps$ 
\begin{eqnarray}
 \langle \phi \phi^\dagger \rangle_{(s|d)}^{(\tau+\eps)}\!\!&=& \includegraphics[width=\fgwidth]{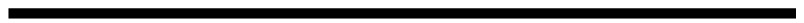}+\includegraphics[width=\fgwidth]{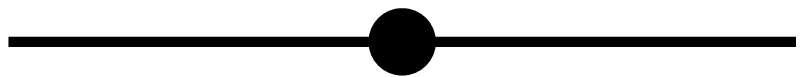}+\includegraphics[width=\fgwidth]{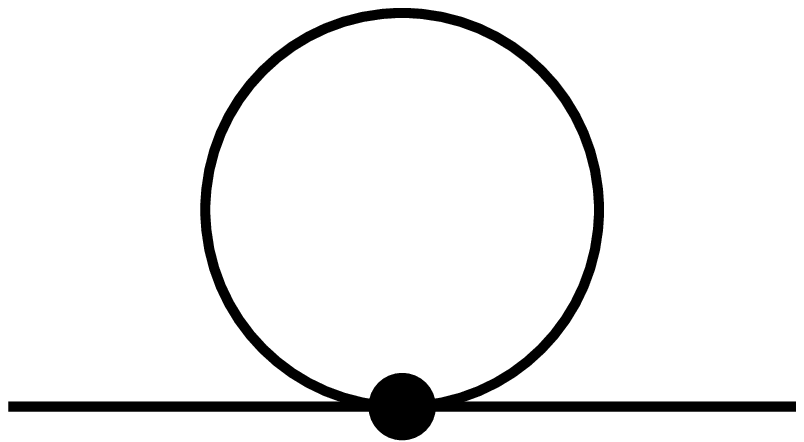}+ \includegraphics[width=\fgwidth]{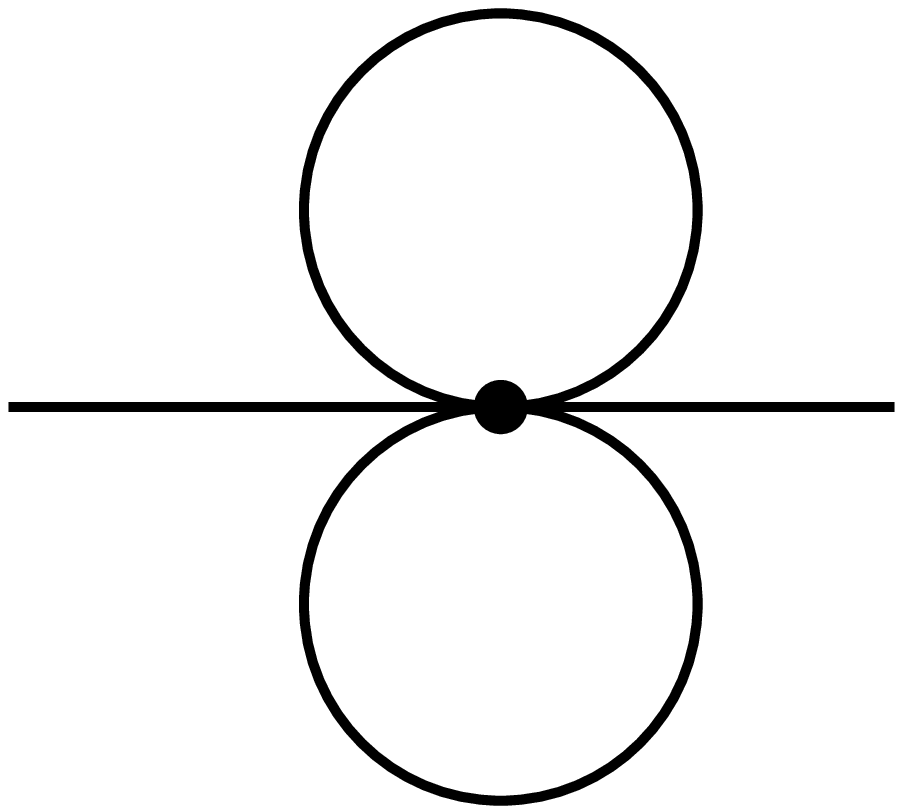}+\ldots  \nonumber\\
&=& \! D_\tau + \eps \, D_\tau \left ( M - b^2\, \kappa_{m_\tau}\,e^{b^2\,\widehat{D}_\tau/2}\right) \!D_\tau 
\end{eqnarray}

Rewriting this for an update of $M_\tau$ we find up to linear order in $\eps$
\begin{equation}
 M_{\tau+\eps} =  (1-\eps)\,M_\tau  - \eps\, b^2\, \kappa_{m_\tau}\,e^{b^2\,\widehat{D}_\tau/2}.
\end{equation}
Taking the limit $\eps \rightarrow 0$ yields the integro-differential system
\begin{eqnarray}
 \dot{m} &=& D\, \left(b\,d - b\,\kappa_0\,e^{b\, m + b^2\,\widehat{D}/2} - S^{-1}m \right)\nonumber\\
\dot{M} &=& b^2 \kappa_0\,e^{b\, m + b^2\,\widehat{D}/2} - M,\;\mbox{and}\\
D &=& \left(S^{-1} + \widehat{M}\right)^{-1}.\nonumber
\end{eqnarray}
This converges at a fix point, which we previously guessed in Eqs. \ref{eq:fixpm} and \ref{eq:fixpM} for our uncertainty-loop enhanced classical equation.

The classical and the renormalization flow fix point equations can be unified:
\begin{eqnarray}\label{eq:renormalizationwithT}
 m &=& b\, S\, \left(d - \kappa_{b\, m + T\, b\,\widehat{D}/2}\right),\nonumber\\
D &=& \left(S^{-1} + \widehat{\kappa}_{b\, m + T\, b\,\widehat{D}/2}\right)^{-1},
\end{eqnarray}
with $T=0$  and $T=1$ for the classical and renormalization result, respectively.

The parameter $T$ is more than a pure convenience. If we would have introduced a temperature $T$ at the beginning, via $P(d,s|T) = \exp (-H_d[s]/T)$,  Eq. \ref{eq:renormalizationwithT} would have been the result of the renormalization flow calculation. And the classical limit naturally corresponds to the zero temperature regime, in which the field expectation value is not affected by any uncertainty fluctuations since the system is at its absolute energy minimum.

\begin{figure*}[tbh]
 \centering
\includegraphics[width=0.8\textwidth,bb = 130 260 480 500]{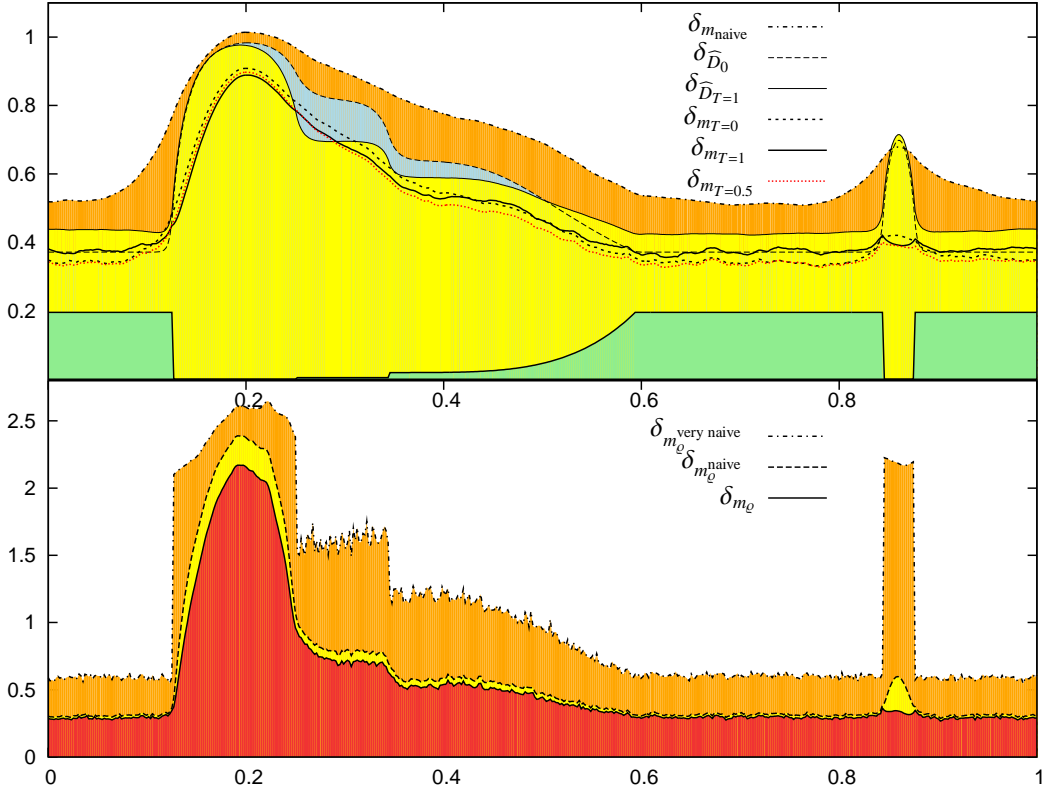}
 \caption{{\bf Top:}
Statistical reconstruction error from $1000$ signal and data realizations Curves are, roughly in order from top (bad performance) to bottom (good performance): error $\delta_{m_\mathrm{naive}} = \langle (s-m_\mathrm{naive})^2 \rangle_{(d,s)}^{1/2}$ of the signal-covariance-convolved naive map $m_{\mathrm{naive}} = S\, s_{\mathrm{naive}}$ (see Eq. \ref{eq:snaive}), expected Wiener-uncertainty $\delta_{\widehat{D}_0} = \widehat{D}_0^{1/2}$, averaged renormalized uncertainty $\delta_{\widehat{D}_{T=1}} = \langle \widehat{D}_{T=1} \rangle_{(d,s)}^{1/2}$, error of the classical map $\delta_{m_{T=0}} = \langle (s-m_{T=0})^2 \rangle_{(d,s)}^{1/2}$, error of the renormalized map $\delta_{m_{T=1}} = \langle (s-m_{T=1})^2 \rangle_{(d,s)}^{1/2}$, and error of the intermediate map $\delta_{m_{T=0.5}} = \langle (s-m_{T=0.5})^2 \rangle_{(d,s)}^{1/2}$. The lowest curve without label is $\kappa$. {\bf Bottoms:}
Error variance of estimators for the density, $\varrho = e^s$, namely $\delta_{m_\varrho^\mathrm{very\;naive}} = \langle ( \varrho - e^{m_\mathrm{naive}})^2\rangle_{(d,s)}^{1/2}$, $\delta_{m_\varrho^\mathrm{naive}} = \langle ( \varrho - m_\varrho^\mathrm{naive})^2\rangle_{(d,s)}^{1/2}$and $\delta_{m_\varrho} = \langle ( \varrho - m_\varrho)^2\rangle_{(d,s)}^{1/2}$ (see Eqs. \ref{eq:rhonaive} and \ref{eq:rhomap}).
} 
 \label{fig:reconstructionstatistics} 
\end{figure*}

An example of such reconstructions can be seen in Fig. \ref{fig:PoissonLargeBias},
and its uncertainty structures in Fig. \ref{fig:Dprop}. Here, the renormalization equation  indeed seems to provide a better result compared to the classical one. However, a statistical comparison of the two reconstructions using 1000 realization of the signal and data in Fig. \ref{fig:reconstructionstatistics} shows that there is at most a marginal difference. This may be surprising, since the classical and renormalization solution are quite distinct, and the latter is always lower than the former. One might therefore ask, if the two are bracketing the correct solution. And indeed, intermediate solutions constructed using $T=1/2$ perform better than the ones for $T=0$ and $T=1$, as can be seen in Fig. \ref{fig:reconstructionstatistics}.

If neither $T=0$ nor $T=1$ provide the optimal reconstruction, what would be the right choice? 
We have to remember that we replaced the probability density function at each step of the renormalisation scheme by a Gaussian with the correct mean and dispersion. However, the real probability is not a Gaussian, and therefore our mean field estimator is not optimal. Reconstructions with different $T$ probe the non-Gaussian probability structure with a differently wide Gaussian kernel in phase space, and therefore result in a slightly different signal means due to the anharmonic nature of our Hamiltonian. 

\subsection{Uncertainty structure}

The remaining uncertainties at the end of the renormalization flow can
mainly be read of the renormalized propagator $D$, which we display in  
top part of Fig. \ref{fig:Dprop} in comparison to the original, un-renormalized
one $D_0$. The renormalised propagator is a much better approximation 
to the uncertainty-dispersion of the signal posterior distribution around the mean 
map than the original one. One can clearly see that the data imprinted a highly
non-uniform structure into the uncertainty pattern visible in the
renormalized propagator with small uncertainties where there were many
galaxy counts. Also the density estimator in Eq. \ref{eq:rhomap} benefits from the knowledge of the uncertainty structure contained in the renormalised propagator, as the lower panel of Fig. \ref{fig:reconstructionstatistics} shows.

The propagators also visualize the effect any additional data would
have at different locations. The height and width of the propagator
values define respectively the strength of the response to, and the
distance of information propagation from an information source.

The structure of  $D_0$ is imprinted by the prior
and the mask. At $D_0$'s widest locations the mask blocks any
information source and  the structure of the signal prior $S$ becomes
visible. At locations where the mask is transparent, the
reconstruction response per information source is lower, as plenty 
information can be expected there. Also the propagator width is smaller, since
the individual informations do not need to be propagated that far,
thanks to the richer information source density in such regions. 

The structure of  $D_m$ has additionally
imprinted the expected information source density structure given the
reconstruction $m$. The strongly non-linear signal response has lead
to regions with very high galaxy count rates, which have larger
information densities, and therefore lower and narrower information
propagators. This implies, that any additional galaxy detection in the 
regions with high galaxy counts will have little impact on the updated map, 
whereas any additional detected galaxies in low density regions will more strongly 
change it. However, the number of additional galaxies per invested observing 
time will be larger in high density regions, which may compensate the lower 
information-per-galaxy ratio there. It is therefore interesting to look at the observational
information content and how it depends on the actual data realization.

\subsection{Information gain}

In case of a free theory, the amount of information depends on the
experimental setup and on the prior, but is independent of the data
obtained as we have shown in Sect. \ref{sec:freeinfo}. 
This changes in case that one wants to harvest information in a
situation described by a non-linear IFT. There, the amount of information can
strongly depend on the actual data.

This is well illustrated by our LSS reconstruction problem. 
A perturbative calculation of the non-linear information gain 
is possible if either the bias-factor or the signal amplitude, which
both control the strength of the non-linear interactions, are small
compared to unity.\footnote{The signal amplitude can, for example, be made
small by defining the signal of interest to be the cosmic density
field, smoothed on a sufficiently large scale ($> 10$ Mpc) so that
$\langle s^2 \rangle_{(s)} <1$.} 

\begin{figure*}[tbh]
 \centering
\includegraphics[width=0.8\textwidth,bb = 160 264 480 523]{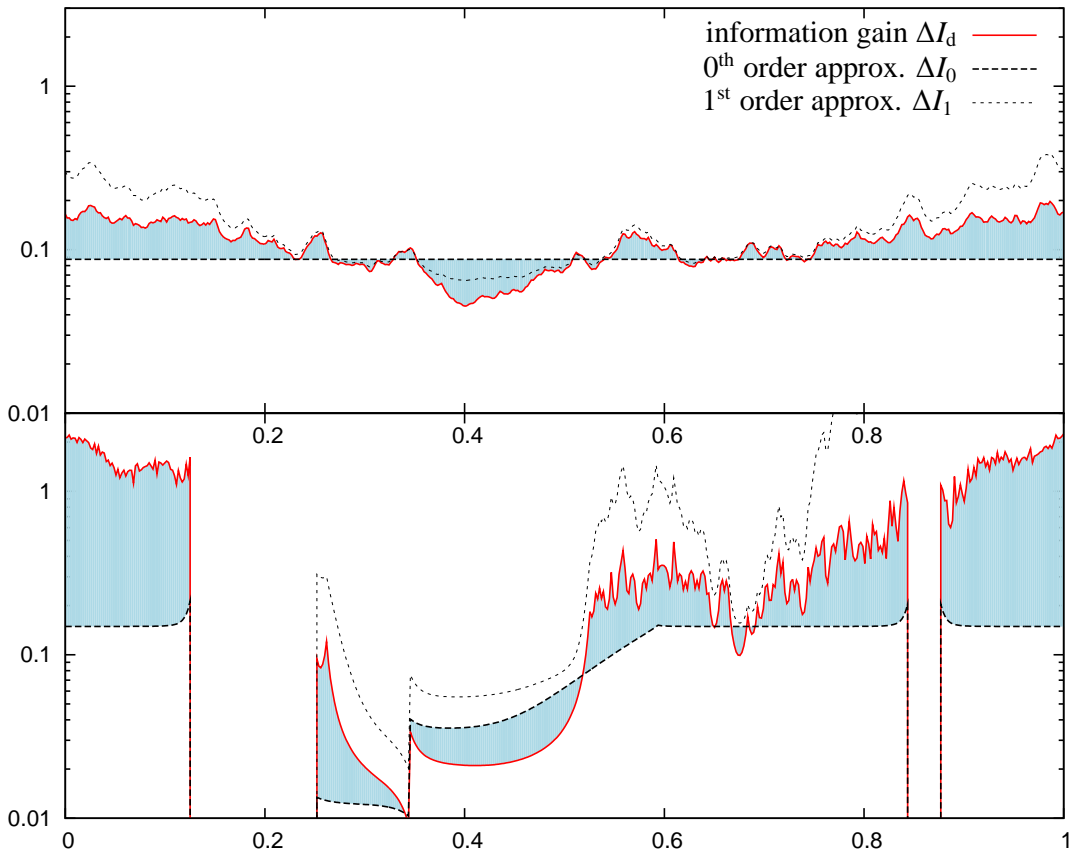}
 \caption{
Information gain density (the integrands of Eq. \ref{eq:DInfoapprox}
and \ref{eq:full Poissonian information gain}) for the two
reconstruction examples presented, the only weakly nonlinear one (top,
and  
Fig. \ref{fig:PoissonSmallBias}) and the strongly non-linear one
(bottom, and Fig. \ref{fig:PoissonLargeBias}). 
The renormalization result for $T=1$ (Eq. \ref{eq:full Poissonian
information gain}), the zero- and first-order perturbative results
(Eq. \ref{eq:DInfoapprox}) are shown.  
The information gain depends on the observational sensitivity as well
as the actual data. The latter influence is stronger in the non-linear
regime, and disappears in linear inference problems.} 
 \label{fig:info} 
\end{figure*}

\begin{figure*}[tbh]
 \centering
\includegraphics[width=0.8\textwidth,bb = 160 264 480 523]{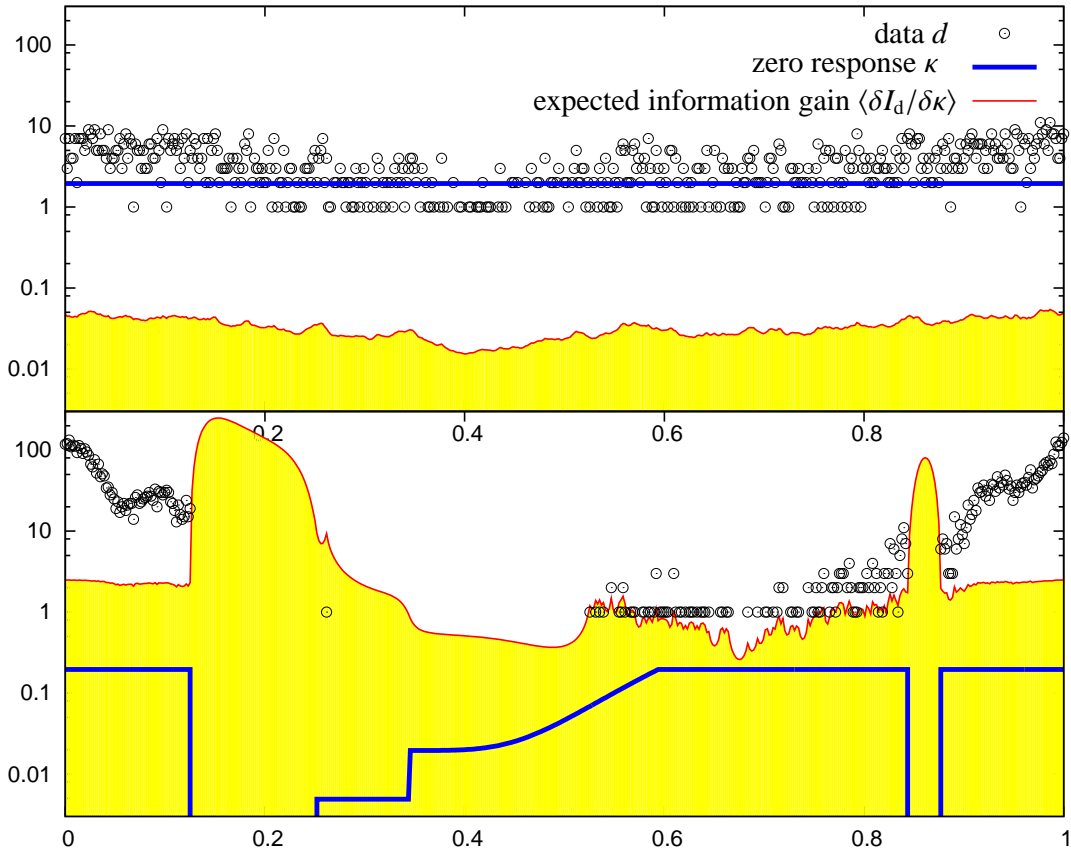}
 \caption{
Differential information gain density  for the two
reconstruction examples presented, the only weakly nonlinear one (top,
and  
Fig. \ref{fig:PoissonSmallBias}) and the strongly non-linear one
(bottom, and Fig. \ref{fig:PoissonLargeBias}). 
} 
 \label{fig:infogain} 
\end{figure*}

The information gain, as given by Eq. \ref{eq:I_explicit}, expanded to
the first few orders in $b$  
\begin{eqnarray}\label{eq:DInfoapprox}
  \Delta I_1   &=& \underbrace{\frac{1}{2} \,\mathrm{Tr} \log  \left( 1 + S\,
  \widehat{\kappa\, b^2} \right)}_{\Delta I_0} \\ 
&+& \frac{1}{2} \, \left(\kappa\, b^3 \widehat{D_0} \right)^\dagger \!\! \left(
m_0+ \frac{1}{2}\, b\, (\widehat{D}_0 +m_0^2) \right) +
\mathcal{O}(b^5),\nonumber
\end{eqnarray}
clearly depends on the actual realization of the data. The different
fluctuations in the Wiener map $m_0 = D_0\, j$, with $D_0=(S^{-1}+
\widehat{b^2\, \kappa})^{-1}$ and $j=b\,(d-\kappa)$, imply positive
and negative information density fluctuations. 

To conveniently calculate the information gain of the observation in case of
a large bias factor, we use the Gaussian approximation of the jointed 
probability function, as provided by the renormalization scheme. 
Due to the Gaussianity of this approximate solution, we can simply
use the formula for the information gain of a free theory, 
as given by Eq. \ref{eq:data induced information gain}. This yields
\begin{eqnarray}\label{eq:full Poissonian information gain}
\Delta I_d  &=&  \frac{1}{2} \, \mathrm{Tr} \left( \log\left( 1 +
S\,\widehat{b^2\kappa\eta} \right) \right)\!, 
\end{eqnarray}
with $\eta = e^{b\, m + \frac{1}{2}\,
b^2 \widehat{D}_{T=1}}$ being proportional to the expected number density of glaxies in this region (see Eq. \ref{eq:rhomap}).
It is also here obvious that the information gain depends on the data. 
In regions with higher observed galaxy numbers $\eta$ is larger, and  
more information is expected to be harvested by further observations.
This is illustrated in
Fig. \ref{fig:info}, where the information gain density, the
individual contributions to the trace in Eq. \ref{eq:full Poissonian
information gain}, as well as the first and and all terms of
Eq. \ref{eq:DInfoapprox} are shown for the cases displayed in
Figs. \ref{fig:PoissonSmallBias} and \ref{fig:PoissonLargeBias}. 
The approximate Eq. \ref{eq:DInfoapprox} seems to be adequate for $b\ll 1$, 
but not for our cases of $b=0.5$ and $2.5$.

The expected benefit of additional observations at location $x$ can also 
be calculated by differentiating Eq. \ref{eq:full Poissonian information gain} with respect to $\kappa(x)$. 
Using Eqs. \ref{eq:renormalizationwithT} and \ref{eq:muiifinal} we find
\begin{equation}
 \langle \frac{\delta I_d}{\delta \kappa_0}\rangle_{(\mathrm{new\,data}|d)} = \frac{1}{2}\, b^2 \, \eta \left(1+  \frac{1}{2}\, \widehat{\kappa_0 b^2  \eta} \, D^2 b^2\right)^{-1} \widehat{D}.
\end{equation}

The expected information gain is especially large for observations at locations where the uncertainty $\widehat{D}$ is large, where a large number density of galaxies ($\propto \eta$) can be expected, and where strong non-linearities are present($\propto b^2$). The inverse term caps the maximally available information gain at some level. For the two reconstruction examples given in Figs. \ref{fig:PoissonSmallBias} and \ref{fig:PoissonLargeBias} we display the expected information gain as a function of the observing postion in Fig. \ref{fig:infogain}.

It is apparent from the top panel, showing the case of uniform observation coverage, that additional observations are more advantageous at locations where already an increased matter density is identified. The bottom panel, showing the case of an very inhomogeneous observation of strongly nonlinear data, demonstrates that filling observational gaps should have the highest priority. But there again, regions where the extrapolated galaxy density seems to be larger should be preferred, as can be seen from the asymmetric shape of the expected information gain for observations in the gap around $x=0.2$. In this example, the information-harvest of high galaxy density regions can be so large, that further observations of the already well observed regions at the boundary of the domain seems to be more advantageous than improving the poorly observed regions around $x=0.4$, where a low galaxy density is already aparent from the existing data.

Of course, in order to plan observations in a real case, the dependence of observational costs as a function of location $x$ and already achieved zero-response there, $\kappa(x)$, have to be folded into the considerations.

\section{Non-Gaussian CMB fluctuations via $f_\mathrm{nl}$-theory}\label{sec:fnl}
\subsection{Data model}

As an IFT example on the sphere $\Omega = S^2$, involving two
interacting uncertainty fields, we investigate the so called $
f_\mathrm{nl} $-theory of local non-Gaussianities in the CMB
temperature fluctuations. This problem has currently a high scientific relevance due to the strongly
increasing availability of high fidelity CMB measurements, which
permit to constrain the physical conditions at very early epochs of
the Universe. The relevant references for this topic were provided in
Sect. \ref{sec:lit:cmb}. 

On top of the very uniform CMB sky with a mean temperature
$T_\mathrm{CMB} $, small temperature fluctuations on the level of
$\delta T^{\{\mathrm{I,E,B}\}}_\mathrm{obs}/T_\mathrm{CMB} \sim 10^{-\{5, 6, 7\}}$
are observed or expected in total Intensity (Stokes I) and in
polarization E- and B-modes, respectively. 
The weak B-modes are mainly due to lensing of E-modes and some unknown level of gravity waves. We will disregard them in the following.
These CMB temperature fluctuations are believed and observed to follow
mostly a Gaussian distribution. However, inflation predicts some level
of non-Gaussianity. Some of the secondary anisotropies
imprinted by the LSS of the Universe via 
CMB lensing, the Integrated Sachs-Wolfe and the Rees-Sciama effects should
also have imprinted non-Gaussian signatures \citep{1967ApJ...147...73S,
  1968Natur.217..511R}.  The primordial, as well as
some of the secondary CMB temperature fluctuations are a response to
the gravitational potential initially seeded during inflation. Since
we are interested in primordial fluctuations, we write 
\begin{equation}
d \equiv  \delta T^{\{\mathrm{I,E}\}}_\mathrm{obs}/T_\mathrm{CMB} =
R\,\varphi +n, 
\end{equation}
where $\varphi$ is the 3-dimensional, primordial
gravitational potential, and $R$ is the response on it of a
CMB-instrument, observing the induced CMB temperature fluctuations in
intensity and E-mode polarization. These are imprinted by a number of effects, 
like gravitational redshifting, the Doppler effect, and anisotropic Thomson scattering.
In case that the data of the instrument
are foreground-cleaned and deconvolved all-sky maps (assuming the data
processing to be part of the instrument) the response, which
translates the 3-d gravitational field into   temperature maps, is
well known from CMB-theory and can be calculated with publicly
available codes like \texttt{cmbfast, camb,} and \texttt{cmbeasy} (see
Sect. \ref{sec:lit:cmb}). 
The precise form of the response does not 
matter for a development of the basic concept, and can be inserted later. 

Finally, the noise $n$ subsumes all deviation of the measurement from the
signal response due to instrumental and physical effects, which are
not linearly correlated with the primordial gravitational potential,
such are detector noise, remnants of foreground signals, but also
primordial gravitational wave contributions to the CMB fluctuations. 

The small level of non-Gaussianity expected in the CMB temperature
fluctuations is a consequence of some non-Gaussianity in the
primordial gravitational potential. Despite the lack of a generic
non-Gaussian probability function, many of the inflationary
non-Gaussianities seem to be well described by a local process, which
taints an initially Gaussian random field, $\phi  \hookleftarrow
P(\phi) = \G(\phi, \Phi)$ (with  the $\phi$-covariance $\Phi = \langle
\phi\, \phi^\dagger \rangle_{(\phi)}$), with some level of
non-Gaussianity. A well controllable realization of such a tarnishing
operation is provided by a slightly non-linear transformation of
$\phi$ into the primordial gravitational potential $\varphi$ via
\begin{equation}\label{eq:fnldef}
\varphi(x) = \phi(x) + f_\mathrm{nl} \,(\phi^2(x) -
\langle \phi^2(x) \rangle_{(\phi)}) 
\end{equation}
for any $x$. The parameter
$f_\mathrm{nl}$ controls the level and nature of non-Gaussianity via
its absolute value and sign, respectively. This means that our data
model reads 
\begin{equation}
 d = R\, (\phi + f \,(\phi^2 - \widehat{\Phi})) + n,
\end{equation}
where we dropped the subscript of $f_\mathrm{nl}$.  In the following
we assume the noise $n$  to be Gaussian with covariance $N = \langle
n\, n^\dagger \rangle_{(n)}$ and define as usual $M = R^\dagger N^{-1}R$ 
for notational convenience.\footnote{
Non-Gaussian noise components are in
fact expected, and would need to be included into the construction of
an optimal $f_\mathrm{nl}$-reconstruction. However, currently we aim
only at outlining the principles and we are furthermore not aware of an
traditional  $f_\mathrm{nl}$-estimator constructed while taking such
noise into account. And finally, we show at the end how  to identify
some of such non-Gaussian noise sources by producing
$f_\mathrm{nl}$-maps on the sphere, which can morphologically be
compared to known foreground structures, like our Galaxy. }

\subsection{Spectrum, bispectrum, and trispectrum}

The nonlinearity of the relation between the hidden Gaussian random field $\phi$ and the 
observable gravitational potential $\varphi$ (Eq. \ref{eq:fnldef}) imprints non-Gaussianity into the latter. 
In order to be able to extract the value of the non-Gaussianity parameter $f$ 
from any data containing information on $\varphi$, we need to know
its statistic at least up to the four-point function, the trispectrum, which we briefly derive with IFT methods.

To that end, it is convenient to define a $\varphi$-moment generating function $Z[J]$ and its logarithm
\begin{eqnarray}
\log Z[J] &=& \log \int \!\! \mathcal{D}\phi\, P(\phi) \, e^{J^\dagger \varphi(\phi)}\\
&=&  \frac{1}{2} \, J^\dagger (\Phi^{-1} -2 \, \widehat{fJ})^{-1} J - (f\, J)^\dagger \widehat{\Phi} \nonumber\\
&-& \frac{1}{2} \,\mathrm{Tr} \left[\log \left(1-2\,\Phi\,\widehat{fJ}\right)\right]\nonumber
\end{eqnarray}
This permits to calculate via $J$-derivatives (see Eqs. \ref{eq:Es}-\ref{eq:34point})
the mean
\begin{equation}
 \bar{\varphi} = \langle \varphi \rangle_{(\phi)} = 0,
\end{equation}
the spectrum (or covariance)
\begin{eqnarray}
  C^{(\varphi)}_{xy} &=& \langle \varphi_x \, \varphi_y \rangle_{(\phi)}^{\mathrm{c}} 
=  \langle (\varphi  - \bar{\varphi})_x\, (\varphi  - \bar{\varphi})_y \rangle_{(\phi)} \nonumber\\
&=& \Phi_{xy} + 2\, f_x \Phi_{xy}^2 f_y,
\end{eqnarray}
the bispectrum%
\footnote{Since the bispectrum contains most of the non-Gaussianity signature, we also provide its 
Fourier-space version, which is well-known for the $f_\mathrm{nl}$-model \citep[e.g.][]{2008arXiv0812.3413F}.
The bispectrum for $f=\mathrm{const}$, expressed in terms of the $\varphi$-covariance reads
\begin{eqnarray}
 B^{(\varphi)}_{xyz} &=&  
2\,f\,[C^{(\varphi)}_{xy} C^{(\varphi)}_{yz} + C^{(\varphi)}_{xz} C^{(\varphi)}_{zy} +
C^{(\varphi)}_{yx} C^{(\varphi)}_{xz}]
+ \mathcal{O}(f^3).\nonumber
\end{eqnarray}
Fourier transforming this yields
\begin{eqnarray}
 B^{(\varphi)}_{k_1k_2k_3}\!\! \!\!&=&\!\!  2\,f\,(2\,\pi)^3\delta(k_1+k_2+k_3)\nonumber\\
\!\!&\times&\!\! [P(k_1) P(k_2) + P(k_2)P(k_3)+ P(k_3)P(k_1)] + \mathcal{O}(f^3),\nonumber
\end{eqnarray}
where $P(k)$ is the power spectrum of $\varphi$, which is identical to that of $\phi$ up to $\mathcal{O}(f^2)$.}
%
%
\begin{eqnarray}
 B^{(\varphi)}_{xyz} &=&   \langle (\varphi  - \bar{\varphi})_x\, (\varphi  - \bar{\varphi})_y \, (\varphi  - \bar{\varphi})_z\rangle_{(\phi)} = \langle \varphi_x \, \varphi_y \, \varphi_z \rangle_{(\phi)}^{\mathrm{c}} \nonumber\\
&=& 2\,[\Phi_{xy}f_y \Phi_{yz} + \Phi_{yz}f_z \Phi_{zx} +\Phi_{zx}f_x \Phi_{xy} ]\nonumber\\
&&+  8 \, \Phi_{xy} f_y \Phi_{yz} f_z \Phi_{zx} f_x 
%
\end{eqnarray}
and the trispectrum
\begin{eqnarray}
T^{(\varphi)}_{xyzu} &=& \langle (\varphi-\bar{\varphi})_x
(\varphi-\bar{\varphi})_y(\varphi-\bar{\varphi})_z (\varphi-\bar{\varphi})_u
\rangle_{(\phi)}\\
&=&  \Phi_{xy}\Phi_{zu} +\Phi_{xz}\Phi_{yu} +\Phi_{xu}\Phi_{yz}+
\langle \varphi_x\, \varphi_y \,\varphi_z\,\varphi_u\rangle_{(\phi)}^{\mathrm{c}}\!\!\!\!
 \nonumber\\
&=& \left[\frac{1}{8}\Phi_{xy}\Phi_{zu}+2\, \Phi_{xy} f_y \Phi_{yz} f_z \Phi_{zu}\right.
\nonumber\\
&+&  \left. \Phi_{xy} f_y \Phi_{yz} f_z \Phi_{zu} f_u \Phi_{ux} f_x \frac{^{}}{_{}}\right]
+\mathrm{23\, perm.}\nonumber
\end{eqnarray}
of the gravitational potential. Since we will investigate the possibility of a  spatially varying non-Gaussianity
parameter at the end of this section, we keep track of the spatial coordinate of $f$, but for the time being 
read $f_x = f$.

The spectrum, bispectrum and trispectrum of our CMB-measurement can easily 
be calculated from the gravitational spectrum and bispectrum, respectively:
\begin{eqnarray}\label{eq:CBcmb}
 C^{(d)} &=& R\, C^{(\varphi)} R^\dagger + N,\\
B^{(d)}_{\hat{n}_1\hat{n}_2\hat{n}_3} &=& R_{\hat{n}_1x}\,R_{\hat{n}_2y}\,R_{\hat{n}_3z}\, B^{(\varphi)}_{xyz},\nonumber\\
T^{(d)}_{\hat{n}_1\hat{n}_2\hat{n}_3\hat{n}_4} &=& R_{\hat{n}_1x}\,R_{\hat{n}_2y}\,R_{\hat{n}_3z}\,
R_{\hat{n}_3u}\, T^{(\varphi)}_{xyzu} \nonumber\\
&+& \left[\left(R\, C^{(\varphi)} R^\dagger + \frac{1}{8}\,N\right)_{\hat{n}_1\hat{n}_2} N_{\hat{n}_3\hat{n}_4}
\right.
\nonumber\\
&&+
\left. \mathrm{23\; permutations} \frac{^{}}{_{}}\right],\nonumber
\end{eqnarray}
where $ \hat{n}$ denotes the unit vector on the sphere, and we have made use of the assumption of the noise being Gaussian and independent of the signal. In case the noise itself has a bi- or trispectrum, or there is a signal dependent noise, e.g. due to an incorrect instrument calibration, then more terms have to be added to the expressions. The usually quoted formulae
\citep[e.g.][]{2001PhRvD..63f3002K, 2005ApJ...634...14K, 2008arXiv0812.3413F,2006PhRvD..73h3007K} 
can be obtained from Eq. \ref{eq:CBcmb} by applying spherical harmonic transformations.

\subsection{CMB-Hamiltonian}

Although we are not interested in the auxiliary field $\phi$, it is
nevertheless very useful for its marginalization to define its
Hamiltonian, which is 
\begin{eqnarray}
 H_f[d,\phi] \!&=&\! - \log( \G(\phi, \Phi)\, \G(d - R\,(\phi + f \,(\phi^2 -
 \widehat{\Phi})), N))\nonumber \\ 
\!&=&\!\! \frac{1}{2}  \phi^\dagger D^{-1} \phi + H_0 - j^\dagger \phi  +
\sum_{n=0}^4 \frac{1}{n!}\, \Lambda^{(n)}[\phi, \ldots, \phi],\nonumber\\ 
\mbox{with}\!&&\nonumber\\
D^{-1}  \!&=&\! \Phi^{-1} + R^\dagger N^{-1} R \equiv  \Phi^{-1} +
M,\nonumber\\ 
j \!&=&\! R^\dagger N^{-1} d, \nonumber\\
\Lambda^{(0)} \!&=& \! j^\dagger ( f\, \widehat{\Phi}) +  \frac{1}{2} \, (f\,
\widehat{\Phi})^\dagger M\, (f\, \widehat{\Phi}),\\ 
\Lambda^{(1)} \!&=& \! -  ( f\, \widehat{\Phi})^\dagger M \; \mbox{and}\;
j' = j - {\Lambda^{(1)}}^\dagger,\nonumber\\ 
\Lambda^{(2)} \!&=&\! -2\,\widehat{fj'},\nonumber\\
\Lambda^{(3)}_{xyz} \!&=&\! (M_{x y} \,f_y\, \delta_{yz} +
5\;\mbox{permutations}),\nonumber\\ 
\Lambda^{(4)}_{xyzu} \!&=&\! \frac{1}{2}\, (f_x\, \delta_{xy} \,M_{y
  z}\,\delta_{zu}\,f_u+ 23\;\mbox{permutations}),\nonumber 
\end{eqnarray} 
and $H_0$ collects all terms independent of $\phi$ and $f$. The last
two tensors should be read without the Einstein sum-convention, but 
with all possible index-permutations. Note, that this is a
non-local theory for $\phi$ in case that either the noise covariance or the
response matrix is non-diagonal, yielding a non-local $M$ and
therefore non-local interactions $\Lambda^{(3)}$ and $\Lambda^{(4)}$.

We should note, that  \citet{2005PhRvD..72d3003B} derived the now traditional 
$f_\mathrm{nl}$-estimator from a very similar starting point, 
the log-probability for $\varphi$. The difference of the resulting estimators is not due 
to the slightly different approaches ($H_f[d,\varphi]$ versus $H_f[d,\phi]$), 
but because of the frequentist and Bayes statistics he and we use, respectively.

In case that the noise as well as the response is diagonal in position
space, as it is often assumed for the instrument response of properly
cleaned CMB maps, and is also approximately valid on large angular
scales, where the Sachs-Wolfe effect dominates, 
we have $N_{x y}=\sigma_n^2(x)\, \delta(x-y)$,  $R = -3$
\citep{1967ApJ...147...73S} for the total intensity fluctuations, and
thus  $M_{x y}=9\, \sigma_n^{-2}(x)\, \delta(x-y)$, if we restrict the
signal space to the last-scattering surface, which we identify with
$S^2$. This permits to simplify the Hamiltonian to 
\begin{eqnarray}
 H_f[d,\phi] 
\!&=&\! \frac{1}{2}  \phi^\dagger D^{-1} \phi + H_0 - j^\dagger \phi  +
\sum_{n=0}^4 \frac{1}{n!}\, \lambda_{n}^\dagger \,\phi^n,\ 
\mbox{with}\nonumber\\ 
D^{-1}  \!&=&\! \Phi^{-1} + 9\,\widehat{\sigma_n^{-2}},\; 
j' = j - \lambda_{1} = 3 \,( 3 \widehat{\Phi}\, f-d)/\sigma_n^2, \nonumber\\
\lambda_{0} \!&=& \!   3\,(\widehat{\Phi}/\sigma_n^2)^\dagger ( \frac{3}{2} f^2
\widehat{\Phi} - f\, d ),\; 
\lambda_{2} = -2 \, f\,j', \nonumber\\
\lambda_{3} \!&=&\!  54\, f/\sigma_n^2,\; \mbox{and} \;
\lambda_{4} = 108\,f^2/\sigma_n^2.
\end{eqnarray} 
The numerical coefficients of the last two terms may look large, 
however, these coefficients stand in front of terms of typically $\phi^3\sim
10^{-15}$, and $\phi^4\sim 10^{-20}$, which ensures 
their well-behavedness in any diagrammatic expansion series. 

For later usage, we define the Wiener-filter reconstruction of the 
gravitational potential as $m_0 = D\, j$.

\subsection{$f_\mathrm{nl}$-evidence and map making}

Since we are momentarily not interested in reconstructing the
primordial fluctuations, but to extract knowledge on $f_\mathrm{nl}$,
we marginalize the former by calculating the log-evidence $\log
P(d|f) $ up to 
quadratic order in $f$: 
\begin{eqnarray} \label{eq:fnlpartionsum}
  \log Z_f[d] &=& 
\log \int \mathcal{D}\phi\, P(d,\phi|f) \nonumber\\
&=&\log \int \mathcal{D}\phi\, e^{-H_f[d,\phi]} \nonumber\\
&&
\!\!\!\!\!\!\!\!\!\!\!\!\!\!\!\!\!\!\!\!\!\!\!\!\!\!\!\!\!\!\!\!\!\!\!\!\!\!\!\!
= - H_0 - \Lambda_0 + 
\includegraphics[width=0.6\fgwidth]{fg/fg_d1.eps} +
\includegraphics[width=0.6\fgwidth]{fg/fg_d2.eps} +
\includegraphics[width=0.6\fgwidth]{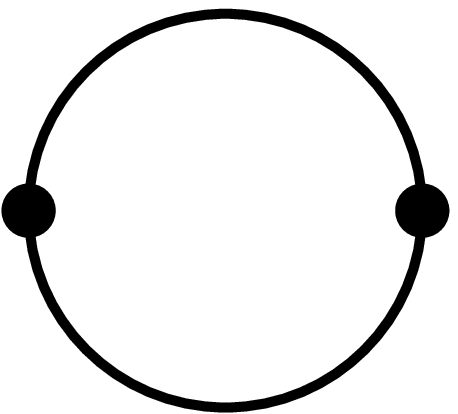} +
\includegraphics[width=0.6\fgwidth]{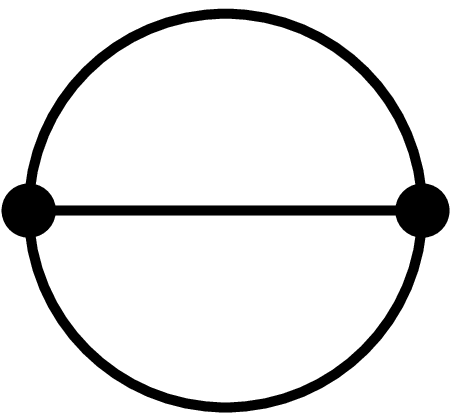} +
\includegraphics[width=1.4\fgwidth]{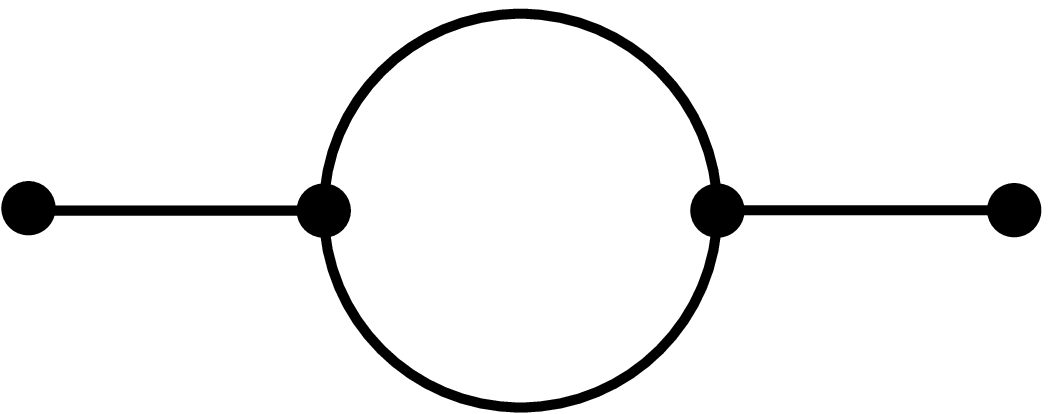} 
\nonumber \\ &&
\!\!\!\!\!\!\!\!\!\!\!\!\!\!\!\!\!\!\!\!\!\!\!\!\!\!\!\!\!\!\!\!\!\!\!\!\!\!\!\!
+\includegraphics[width=\fgwidth]{fg/fg_d9.eps} +
\includegraphics[width=\fgwidth]{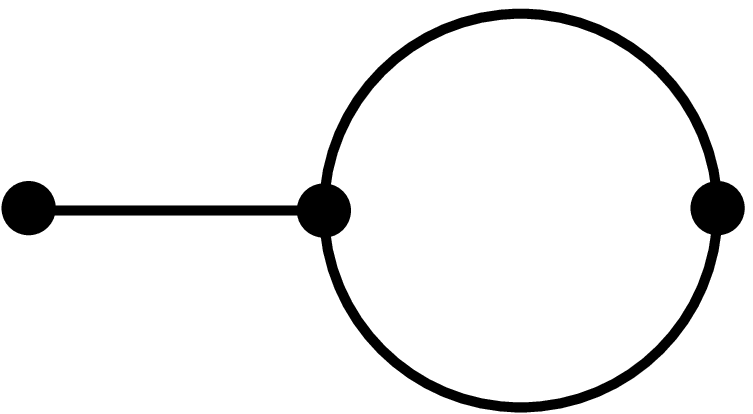} + 
\includegraphics[width=\fgwidth]{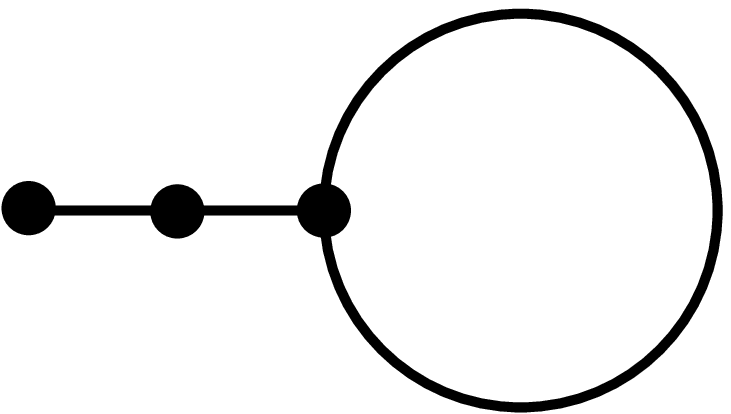} +
\includegraphics[width=\fgwidth]{fg/fg_d8.eps} +
\includegraphics[width=0.8\fgwidth]{fg/fg_d10.eps} +
\includegraphics[width=1.5\fgwidth]{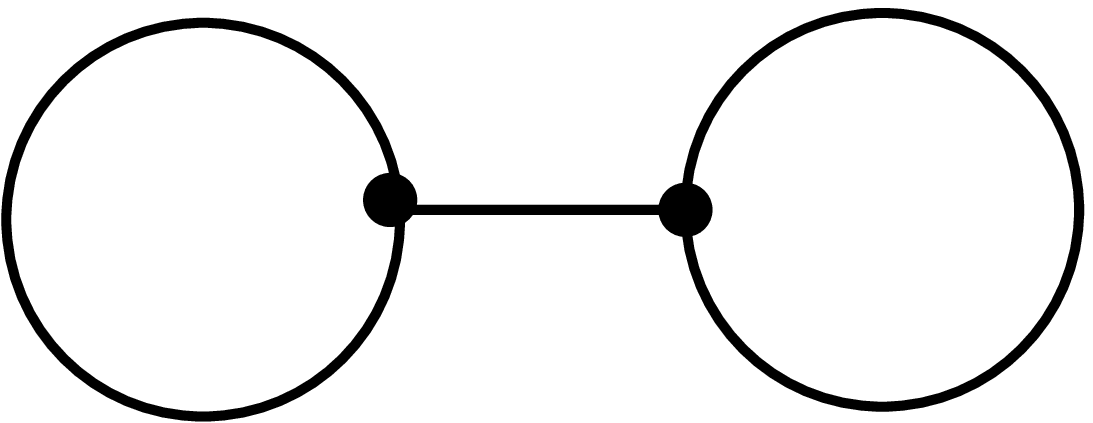}
\nonumber \\ &&
\!\!\!\!\!\!\!\!\!\!\!\!\!\!\!\!\!\!\!\!\!\!\!\!\!\!\!\!\!\!\!\!\!\!\!\!\!\!\!\!
 + 
\includegraphics[width=1.25\fgwidth]{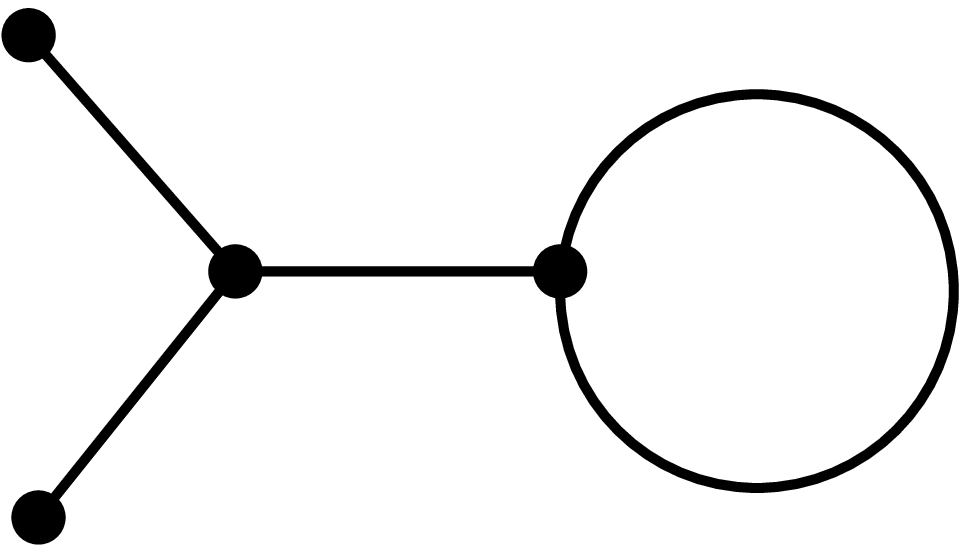} + 
\includegraphics[width=\fgwidth]{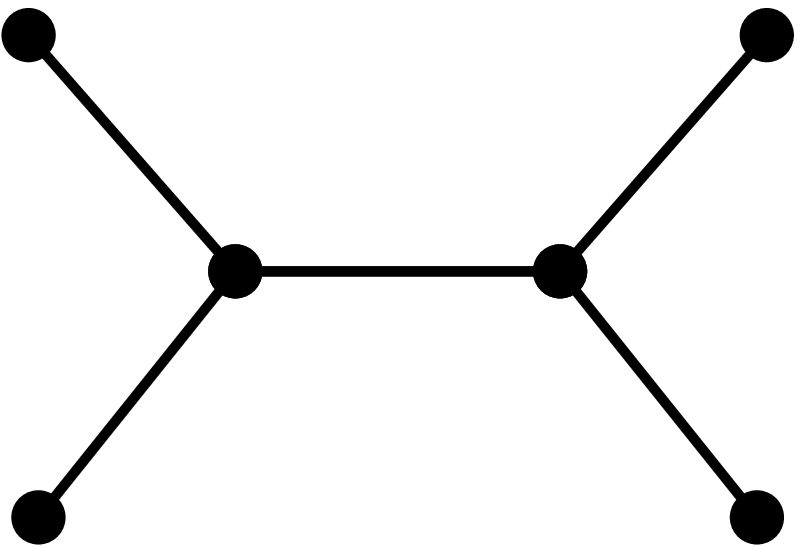} + 
\includegraphics[width=0.8\fgwidth]{fg/fg_d6.eps} + 
\includegraphics[width=0.8\fgwidth]{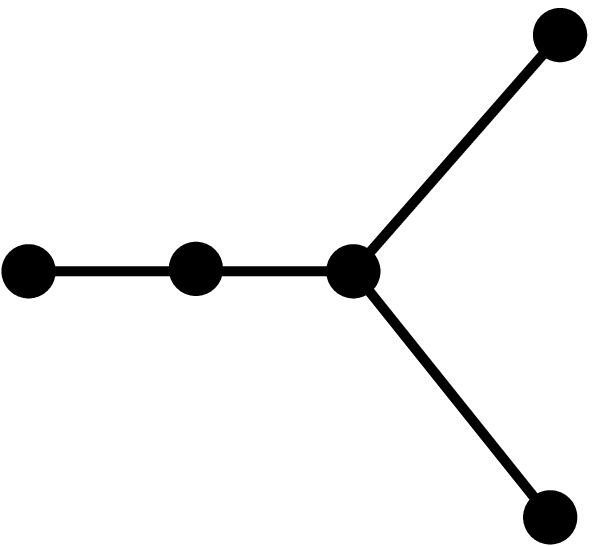} + 
\includegraphics[width=0.8\fgwidth]{fg/fg_d7.eps} 
\nonumber \\ &&
\!\!\!\!\!\!\!\!\!\!\!\!\!\!\!\!\!\!\!\!\!\!\!\!\!\!\!\!\!\!\!\!\!\!\!\!\!\!\!\!
+
\includegraphics[width=\fgwidth]{fg/fg_d5.eps} +
\includegraphics[width=\fgwidth]{fg/fg_d4.eps} +
\includegraphics[width=\fgwidth]{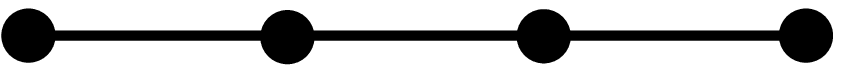} +
 \mathcal{O}(f^3).
\end{eqnarray}
We have made use of the fact that the logarithm of the partition sum
is provided by all connected diagrams, and that  
$j'$ contains a term of the order $\mathcal{O}(f^0)$,
 $\Lambda^{(2)}$ and $\Lambda^{(3)}$ contain terms of the order
$\mathcal{O}(f^1)$, and $\Lambda^{(4)}$ one of the order
$\mathcal{O}(f^2)$, so that they can appear an unrestricted number of
times, twice 
and once in diagrams of order up to $\mathcal{O}(f^2)$, respectively. 
Since only $4^{\mathrm{th}}$ order interactions are involved, an implementation 
in spherical harmonics space may be feasible using the  only $4^{\mathrm{th}}$ 
order $C$-coefficients (Eq. \ref{eq:C}), which can be calculated
computer algebraically. Finally, we recall 
\begin{equation}
  \includegraphics[width=0.5\fgwidth]{fg/fg_d1.eps}  = \frac{1}{2}\, \log
  |2\pi\, D| =\frac{1}{2}\, \mathrm{Tr} (\log (2\pi\, D)). 
\end{equation}

Although $f$ is not known, the expressions in
Eq. \ref{eq:fnlpartionsum} proportional to $f$ and $f^2$ can be
calculated separately, permitting to write down the Hamiltonian of
$f$ if a suitable prior $P(f)$ is chosen,
\begin{eqnarray}
 H_d[f] &\equiv& -\log  (P(d|f)\,P(f)) \nonumber\\
&=& \tilde{H}_0 + \frac{1}{2}\,f^\dagger
 \tilde{D}^{-1} f +  \tilde{j}^{\dagger}f +\mathcal{O}(f^3), 
\end{eqnarray}
where we collected the linear and quadratic coefficients into
$\tilde{j}$ and $\tilde{D}^{-1}$. It is obvious that the optimal
$f$-estimator to lowest order is therefore 
\begin{equation}
 m_f = \langle f\rangle_{(s,f|d)} =  \tilde{D}\,\tilde{j},
\end{equation}
and its uncertainty variance is just
\begin{equation}
 \langle (f - m_f)\, (f - m_f)^\dagger\rangle_{(s,f|d)} = \tilde{D}.
\end{equation}

So far, we have assumed $f$ to have a single universal value. However,
we can also permit $f$ to to vary spatially, or on the sphere of the
sky. In the latter case one would expand $f$ as 
\begin{equation}
 f(x) = \sum_{l=0}^{l_{\mathrm{max}}}\, \sum_{m=-l}^{l}\, f_{lm}\, Y_{lm}(\hat{x})
\end{equation}
up to some finite $l_{\mathrm{max}}$. Then one would recalculate the
partition sum, now separately for terms proportional to $ f_{lm}$ and
$f_{lm}\, f_{l'm'}$, which are then sorted into the vector and matrix
coefficients of $\tilde{j}$ and $\tilde{D}^{-1}$,
respectively and according to 
\begin{eqnarray}
 \tilde{j}_{(lm)} &=& \left. \frac{dH_d[f] }{df_{lm}}\right|_{f=0},\; \mbox{and}\\
 \tilde{D}^{-1}_{(lm)\,(l'm')} &=& \left. \frac{d^2H_d[f]
 }{df_{lm}\,df_{l'm'}}\right|_{f=0}.\nonumber 
\end{eqnarray}

$f$-map making can then proceed as described above in
spherical harmonics space. Comparing the resulting map in angular
space to known foreground sources, as our Galaxy, the level of
non-Gaussian contamination due to their imperfect removal from the
data may be assessed. 

\subsection{Comparison to traditional estimator}

We conclude this chapter with a short comparison to traditional
$f_\mathrm{nl}$-estimators. To our knowledge, the most developed estimator in
the literature is based on the CMB-bispectrum, which is the third order
correlation functions of the data \citep[e.g.][and references in
Sect. \ref{sec:lit:cmb}]{1998MNRAS.299..805H, 2005PhRvD..72d3003B}.  
The IFT based filter presented here 
contains terms which are up to fourth order in the data, and therefore can be 
expected to be of higher accuracy since both methods are supposed to be optimal. 
\citet{2006PhRvD..73h3007K} note that the CMB trispectrum should contain significant 
information on $f_\mathrm{nl}^2$, and may be superior to non-Gaussianity detection 
compared to the bi\-spectrum on small angular scales. However, since the trispectrum is insensitive 
to the sign of $f_\mathrm{nl}$, its actual usage as a proxy is a it more subtle. In the IFT estimator, 
any term proportional to $f_\mathrm{nl}^2$ enters the inverse of the propagator $\tilde{D}$, and therefore
the trispectrum seems to unfold its  $f_\mathrm{nl}$-estimation power mostly in combination with the bi\-spectrum, 
which drives $\tilde{j}$.

Under which conditions
does the traditional estimator emerge from the IFT one? There are 
three conceptual differences between the estimators, in that the IFT filter can handle inhomogeneous non-Gaussianity, correct for CMB sky and exposure chance coupling, and is unbiased with respect to the 
posterior.

The traditional estimator is usually written as
\begin{equation}\label{eq:fnleps}
 \eps = \frac{1}{\mathcal{N}} \, \int dx\, A(x)\,B^2(x) 
= \frac{1}{\mathcal{N}} \, m_0^\dagger \Phi^{-1} m_0^2,
\end{equation}
where $B = D\, j = m_0$ is the Wiener-filter reconstruction of the gravitational potential, $A = \Phi^{-1} B$ is the same, just additionally filtered by the inverse power spectrum, and $\mathcal{N}$ is a normalization constant
\citep[e.g.][]{2002ApJ...566...19K}. This is fixed by the condition that the estimator should be unbiased with respect to all signal and noise realizations,
\begin{eqnarray}
 \mathcal{N} &=& \langle m_0^\dagger \Phi^{-1} m_0^2 \rangle_{(d,s|f=1)}\nonumber\\
&=& B^{(\varphi)}_{xyz}|_{f=1}\, \left[ (M\,D)_{xu} \,  \Phi^{-1}_{uv}   (D\,M)_{vy} \,  (D\,M)_{vz} \, \right]\nonumber\\
 &=& 2\, \left[ \Phi_{xy}\Phi_{yz}+ \Phi_{yz}\Phi_{zx}+\Phi_{zx}\Phi_{xy}\right]\,\nonumber\\
&\times& \left[ (M\,D)_{xu} \,  \Phi^{-1}_{uv}   (D\,M)_{vy} \,  (D\,M)_{vz} \, \right]
\end{eqnarray}

The first difference between the estimators is obvious, in that the IFT estimator can handle a spatially varying $f(x)$.
Therefore, we will only regard spatially constant non-linearity parameters in the following. Since no CMB experiment
is able to measure the monopole temperature fluctuation, the response to any spatially homogeneous signal is zero.
This means, in Fourier basis, that $R_{\hat{n},k=0}=0$ and therefore $j_{k=0}=M_{k=0,k'}=0$. 
Thus, we find for a Universe with homogeneous statistics ($\widehat{\Phi}_{k\ne 0}=0 $) that 
$\Lambda^{(0)} = \Lambda^{(1)} = 0$, 
$j' = j$, and $\Lambda^{(2)} = -2f \,\widehat{j}$, which reduces the number of diagrams we have to calculate.

The IFT estimator is driven by the $f$-information source $\tilde{j}$, which is given by all diagrams which contain terms linear in $f$. There are four of them, yielding
\begin{eqnarray}
\tilde{j} &=& 
\frac{1}{f}\,\left[
\includegraphics[width=\fgwidth]{fg/fg_d4.eps} +
\includegraphics[width=0.8\fgwidth]{fg/fg_d6.eps} +
\includegraphics[width=0.6\fgwidth]{fg/fg_d2.eps} +
\includegraphics[width=\fgwidth]{fg/fg_d9.eps}
\right]\nonumber\\
&=& m_0^\dagger \Phi^{-1} m_0^2 + m_0^\dagger \, \left[ \Phi^{-1} \widehat{D} - 2\, \widehat{MD} \right],
\end{eqnarray}
where we used $M=D^{-1}-\Phi^{-1}$ in order to combine the two tree and 
the two loop diagrams into the first and second term, respectively. 
The term resulting from the tree diagrams is actually identically to 
the unnormalised traditional estimator $\eps$ (Eq. \ref{eq:fnleps}). 

The terms resulting from the loop diagrams vanish for an homogeneous $M$, 
which a CMB experiment with uniform exposure and constant noise could produce. 
In case of an inhomogeneous $M$, which is the more realistic case, 
the loop term does not vanish and corrects for chance correlations between the 
CMB-realization (as seen through $j$) and the noise and response structure of the experiment 
(as encoded in $M$ and $D$). \citet{2006JCAP...05..004C} already pointed out that such a linear correction term is 
necessary in case of an inhomogeneous sky coverage.

Anyhow, the second difference between the estimators is that the IFT based one applies a 
correction for chance correlations of CMB sky and sky exposure and the traditional one does not. 
This term is absent in the traditional estimator since the latter 
was constructed as the optimal estimator which is third order in the data. 
This excluded the loop term, which is linear in the data.

An inclusion of this term into the traditional estimator is straightforward and actually done by
the more recent $f_\mathrm{nl}$ measurements \cite[e.g.][]{2005PhRvD..71l3004Y}. 
The normalization constant 
$\mathcal{N}$ is unaffected by this, since the expectation value of the loop term averaged over 
all possible signal realization is zero.

This brings us to the third difference between the estimators, the different normalization.
The traditional estimator is normalized by a data independent constant $\mathcal{N}$, where the IFT 
estimator is normalized by a data dependent term
\begin{eqnarray}  \label{eq:Dfnl}
 \tilde{D}^{-1} &=& \frac{1}{\sigma_f^{2}}+ \frac{2}{f^{2}} \left[  
\includegraphics[width=\fgwidth]{fg/fg_d8.eps} +
\includegraphics[width=1.5\fgwidth]{fg/fg_ee11.eps}+ 
\includegraphics[width=0.6\fgwidth]{fg/fg_ee2.eps} +
\includegraphics[width=0.6\fgwidth]{fg/fg_ee1.eps} +
\right.\nonumber
\\ &&
\includegraphics[width=\fgwidth]{fg/fg_ee4.eps} +
\includegraphics[width=1.4\fgwidth]{fg/fg_ee12.eps} + 
\includegraphics[width=\fgwidth]{fg/fg_ee7.eps} +
\includegraphics[width=0.8\fgwidth]{fg/fg_d10.eps} +
\includegraphics[width=1.25\fgwidth]{fg/fg_ee10.eps}\!\!\!\! \nonumber
\\ &&
+ \left. 
\includegraphics[width=\fgwidth]{fg/fg_ee14.eps} + 
\includegraphics[width=0.8\fgwidth]{fg/fg_ee9.eps} + 
\includegraphics[width=0.8\fgwidth]{fg/fg_d7.eps}  +
\includegraphics[width=\fgwidth]{fg/fg_ee8.eps} 
\right] ,
\end{eqnarray}
where only the first three diagrams are data independent and $\sigma_f$ is the variance of the prior, which we assume to be  $P(f) = \G(f, \sigma_f^2)$. The detailed expressions for the different diagrams can be found in Appendix \ref{sect:fnlD}. For both estimators, the traditional and the IFT one, the normalization 
is supposed to guarantee unbiasedness, however, with respect to different 
probability distributions. 

The traditional estimator is unbiased in the frequentist sense, 
for an average over all signal $f$ and data realizations.
However, the IFT estimator is unbiased in the Bayesian sense, with respect to the posterior, 
the probability distribution of all signals given the data. Since the data are given, and not
assumed to vary any more after the observation is performed, it can and should affect 
the normalization constant, which encodes the sensitivity of our non-Gaussianity estimation. 

The reason for the IFT normalization constant (or $f$-propagator) to be data dependent can be 
understood as follows. There are data realizations which are better suited to reveal the presence 
of  a non-Gaussianities than others, even if they have identical $\tilde{j}$.  Such a dependence 
of the detectability of a effect on the concrete data realization is common in non-linear Baysian inference,
and was even more prominent in the example of the reconstruction of a log-normal density field in Sect. \ref{sec:poisson}.

\section{Summary and Outlook}\label{sec:summary}

Starting with fundamental information theoretical considerations
about the nature of measurements, signals, noise and their relation to
a physical reality given a model of the Universe or the system under
consideration, we reformulated the inference problem in the language
of  \textit{information field theory} (IFT). IFT is actually a
statistical field theory. The information field is identified with
a spatially distributed signal, which can freely be chosen by the
scientist according to needs and technical constraints. The
mathematical apparatus of field theory permits to deal with the
ensemble of all possible field configurations given the data and prior
information in a consistent way. 

With this conceptual framework, we derived the Hamiltonian of the theory,
showed that the free theory reproduces the well known results of Wiener-filter
theory, and presented the Feynman-rules for non-linear, interacting
Hamiltonians in general, and in particular cases. The latter are
information fields over Fourier- and spherical harmonics-spaces for
inference problems in $R^n$ and $S^2$, respectively. Our ``philosophical''
considerations permitted to argue why the resulting IFTs are usually
well normalized, but often non-local. Since the propagator of the
theory is closely related to the Wiener-filter, for which nowadays
efficient numerical algorithms exist as image reconstruction and
map-making codes, and the information source term is usually a noise
weighted version of the data, the necessary computational tools are at
hand to convert the diagrammatic expressions into well performing
algorithms.

Furthermore, we provided the Boltzmann-Shannon information measure of IFT
based on the Helmholtz free energy, thereby highlighting the embedding 
of IFT in the framework of statistical mechanics.

As examples of the IFT recipe, two concrete IFT problems with cosmological
motivation were discussed, which are also thought as blueprints for other inference problems.
The first was targeting at the problem of
reconstructing the spatially continuous cosmic LSS matter
distribution from discrete galaxy counts in incomplete galaxy surveys. 
The resulting algorithm can also be used for image
reconstruction with low-number photon statistics, e.g in low-dose X-ray
imaging.
 
The second example was the design of an optimal method to measure or
constrain any possible local non-linearities in the CMB temperature
fluctuations. This may serve as
a blueprint for statistical monitoring of the linearity of a signal amplifier.

We conclude here with a short outlook on some problems that
are accessible to the presented theory. 

Many signal inference problems involve the reconstruction of fields without
precisely known statistics.  Some coefficients in the IFT-Hamiltonians may
only be phenomenological in nature, and therefore have to be derived from the
same data used for the reconstruction itself. This more intricate interplay of
parameter and information field can also be incorporated into the IFT
framework, as we will show with a subsequent work.

For cosmological applications, along the lines started in this work, clearly
more realistic data models need to be investigated. For example, to understand
the response in galaxy formation to the underlying dark matter distribution in
terms of a realistic, statistical model, to be used in constructing the
corresponding IFT Hamiltonian for a dark-matter information field, detailed
higher-order correlation coefficients have to be distilled from numerical
simulations or semi-analytic descriptions. Also the CMB Hamiltonian may benefit
from the inclusion of remnants from the CMB foreground subtraction process,
permitting to gather more solid evidence on fundamental parameters which are
hidden in the CMB fluctuations, like the amplitude of non-Gaussianities.

Furthermore, there exist a number of more or less heuristic algorithms
for inverse problems, which have proven to serve well under certain
circumstances. Reverse engineering of their implicitly assumed priors
and data models may permit to understand better for which conditions
they are best suited, as well how to improve them in case these
conditions are not exactly met.

Finally, we are very curious to see whether and how the presented framework may
be suitable to inference problems in other scientific fields.

\section*{Acknowledgements}
It is a pleasure to thank the following people for helpful scientific
discussions on various aspects of this work: Simon White on the dangers of
perturbation theory, Benjamin Wandelt on the prospects of large-scale structure
reconstruction, Jens Jasche on the pleasures and pains of signal processing,
J{\"o}rg Rachen on the philosophy of science, and Andr\'e Waelkens on the
invariant, but vertiginous theory of isotropic tensors. We thank
Cornelius Weig and Henrik Junklewitz  
for debates on the connection between IFT and QFT. We gratefully
acknowledge helpful comments on the manuscript by Marcus Br\"uggen and Thomas
Riller and by three very constructive referees.

\appendix
\section{Notation}\label{sec:notation}

We briefly summarize our notation of functions in position and Fourier space.

A here usually real, but in principle also complex function $f(x)$ over the
$n$-dimensional space is 
regarded as a vector $f$ in a discrete and finite-dimensional, or
continuous and infinite-dimensional Hilbert space. $f$ will denote
this vector, independently of the momentarily chosen function basis,
be it the real space $f(x) =\langle x|f\rangle$ or the Fourier basis 
\begin{equation}
f(k) = \langle k|f\rangle =\int \! dx\, f(x) \, e^{i\, k\cdot x}.
\end{equation}
Here, the volume integration usually is performed only over an finite
domain with volume $V$. This leads to the convention for the origin of the
delta function in $k$-space, 
\begin{equation}
\delta(0) = \frac{V}{(2\,\pi)^n},
\end{equation}
and also to a Fourier transformation operator $F=|k\rangle \langle x|$, with
$F_{kx}=e^{i\,k\,x}$, and its inverse $F^\dagger=|x\rangle \langle k|$, with 
$F_{xk}^\dagger=e^{- i\,k\,x}$. The dagger
is used to denote transposed and complex conjugated objects. 
We have $(F^\dagger F)_{xy} = 1_{xy}$ as well as 
$(F\, F^\dagger )_{kk'} = 1_{kk'}$ for the following definition of the 
scalar product of two functions $f$ and $g$ in real and Fourier space:
\begin{equation}
f^{\dagger}g = \langle f |g\rangle = \int \! dx\, {f^*(x)} \, g(x) =
\int \! \frac{dk}{(2\,\pi)^n}\, {f^*(k)} \, g(k), 
\end{equation}
where the asterix denotes complex conjugation.
The statistical power-spectrum of $f$ is denoted by $P_f(k) = \langle |f(k)|^2 \rangle_{(f)}/V$. 

We also introduce for convenience the position-space component-wise product of two functions
\begin{equation}
(f\,g)(x) \equiv f(x)\, g(x),
\end{equation}
which also permits compact notations like
\begin{equation}
(\log f)(x) = \log(f(x)),\, (f/g)(x) = f(x)/g(x),
\end{equation}
and alike. The component-wise product should not be confused with the
tensor product of two vectors $(f\,g^{\dagger})(x,y) = f(x)\,{g^*(y)}$.  

The diagonal components of a matrix $M$ in position-space
representation form a vector which we denote by 
\begin{equation}
\widehat{M} = \mathrm{diag}_x M,\; \mbox{with}\;\widehat{M}_x = M_{xx}.
\end{equation}
Similarly, a diagonal matrix in position-space representation, whose
diagonal components are given by a vector $f$, will be denoted by  
\begin{equation}
\widehat{f} = \mathrm{diag}_x f\; \mbox{with}\;\widehat{f}_{xy} = f_x\, 1_{xy}.
\end{equation}
Thus, $\widehat{\widehat{M}} = M$ if and only if $M$ diagonal, and
$\widehat{\widehat{f}} = f$ always.

In our notation a multivariate Gaussian reads:
\begin{equation}
\G(s, S) = \frac{1}{|2\pi S|^\frac{1}{2}} \exp\left( -\frac{1}{2}
s^{\dagger} S^{-1} s\right)  
\end{equation}
Here, $S = \langle s\,s^{\dagger} \rangle_{(s)}$ denotes the covariance
tensor of the Gaussian field $s$, which is drawn from $P(s) = \G(s,S)$. 
If $s$ is statistically homogeneous, $S$ is fully described by the
power-spectrum $P_s(k)$: 
\begin{eqnarray}
\label{powspec}
S_{k\,k'} &=& (2\,\pi)^n\, \delta(k-k') \, P_s(k),\nonumber\\ 
S_{k\,k'}^{-1} &=& (2\,\pi)^n\, \delta(k-k') \, \left(P_s(k)\right)^{-1}.
\end{eqnarray}
The Fourier representation of the trace of a Fourier-diagonal operator,
\begin{equation}
\mathrm{Tr} (A) = \int \!dx\, A_{x\,x} = V \, \int\! \frac{dk}{(2\,\pi)^n}\, P_A(k),
\end{equation}
is very useful in combination with the following expression for the determinant
of a Hermitian matrix, 
\begin{equation}
 \log |A| = \mathrm{Tr} (\log A).
\end{equation}

Furthermore, we usually suppress the dependency of probabilities on the
underlying model $I$ and its parameters $\theta$ in our notation. I.e. instead of
$P(s|\theta, I)$ we just write $P(s)$ or $P(s|\theta)$ depending on our focus. 
Here $\theta = (S,N,R, ...)$
contains all the parameters of the model, which are assumed to be
known within this work. 

\section{Feynman rules on the sphere}
\label{sec:sphere}

Here, we provide the Feynman rules on the sphere. The
real-space rules are identical to those of flat spaces, with just the
scalar product replaced by the integral over the sphere, etc. In case
the problem at hand has an isotropic propagator, which only depends on
the distance of two points on the sphere, but not on their location or
orientation, the propagator is diagonal if expressed in spherical
harmonics $Y_{lm}(x)$. Thanks to the orthogonality relation of
spherical harmonics, we have for $x,y \in S^2$ 
\begin{equation}
\label{eq:YY+}
 (Y\, Y^\dagger)_{xy} = \sum_{lm} Y_{lm}(x) \, {Y^*_{lm}(y)} =  \delta(x-y) = (1)_{xy}
\end{equation}
and 
\begin{eqnarray}
\label{eq:Y+Y}
 (Y^\dagger Y)_{(l,m)(l',m')} &=& \int\! dx\, {Y^*_{lm}(x)} \, Y_{l'm'}(x) \nonumber\\
&=&  \delta_{l l'}\, \delta_{m m'} = (1)_{(l,m)(l',m')}.
\end{eqnarray}
Therefore, we can just insert real-space identity matrices $1=Y\,
Y^\dagger$ in between any expression in real-space diagrammatic
expression and assign $ Y^\dagger$ to the right, and $Y$ to the left
term of it. This way we find the spherical-harmonics Feynman rules,
which are very similar to the Fourier-space ones, in that they also
require directed propagators-lines for proper angular-momentum
conservation. For a theory with only local interactions, these read: 
\begin{enumerate}
\item An open end of a line has external (not summed) angular-momentum
quantum numbers $(l,m)$. 
\item A line connecting momentum $(l,m)$ with momentum $(l',m')$
corresponds to a 
    propagator between these momenta: $D_{(l,m)(l',m')}=C_D(l)\,
\delta_{l l'}\, \delta_{m m'}$, where $C_D(l)$ is the angular power
spectrum of the propagator. 
\item A data source vertex is  $(j+J-\lambda_1)(l,m)$, where $(l,m)$ is the angular
momentum at the data-end of the line.
\item A vertex with quantum number $(l_0,m_0)$ with $n_\mathrm{in}$
incoming and $n_\mathrm{out}$ outgoing lines ($n_\mathrm{in} +
n_\mathrm{out}>1$) with momentum labels $(l_1,m_1) \ldots
(l_{n_\mathrm{in}},m_{n_\mathrm{in}}) $  and $(l'_1,m'_1) \ldots
(l'_{n_\mathrm{out}},m'_{n_\mathrm{out}}) $, respectively, is given by 
$ -\lambda_m(l_0,m_0) \, C_{(l_0,m_0) \ldots
  (l_{n_\mathrm{in}},m_{n_\mathrm{in}}) }^{ (l'_1,m'_1)
  \ldots(l'_{n_\mathrm{out}},m'_{n_\mathrm{out}})}$, where $C$ will be defined
in Eq. \ref{eq:C}.
\item An internal vertex has  internal (summed) angular-momentum
quantum numbers $(l',m')$. Summation means a term
$\sum_{l'=0}^{\infty}\sum_{m=-l'}^{l'}$ in 
front of the expression. 
\item The expression gets divided by the symmetry factor of its diagram. 
\end{enumerate}
The interaction structure in spherical harmonics-space is complicated
due to the non-orthogonality of powers and products of the spherical
harmonic functions, compared to the Fourier-space case, where any
power or product of Fourier-basis functions is again a single
Fourier-basis function.  

The spherical structure is encapsulated in the coefficients
\begin{equation} \label{eq:C}
 C_{(l_0,m_0) \ldots (l_{n_\mathrm{in}},m_{n_\mathrm{in}}) }^{(l'_1,m'_1)
 \ldots (l'_{n_\mathrm{out}},m'_{n_\mathrm{out}})} \!\equiv \!\int \!\!dx  \left(
 \prod_{i= 0}^{n_\mathrm{in}} Y_{l_i m_i}(x) \right) \! \left( \prod_{i=
 1}^{n_\mathrm{out}}\,{Y^*_{l'_i m'_i}(x)} \right)\!\!, 
\end{equation}
which can be expressed in terms of sums and products of Wigner
coefficients, thanks to the relations ${Y^*_{l m}(x)} = Y_{l\,,-m}(x)$,  
\begin{eqnarray}
 Y_{l_1m_1}(x)\,Y_{l_2\,m_2}(x) \!&=&\! \sum_{l
 m}\,\sqrt{\frac{(2\,l_1+1)\,(2\,l_2+1)\,(2\,l+1)}{4\,\pi}}  \nonumber\\ 
&& \!\!\!\!\!\!\!\!\!\!\!\!\!\!\!\!\!\!\!\!\!\!\!\!\!\!\!\!\!\!\!\!\!\!\!\!
\times \left( 
\begin{array}{ccc}
 l_1 & l_2 & l\\
m_1 & m_2 & m
\end{array}
\right)\,
Y_{lm}(x)\,
\left(
\begin{array}{ccc}
 l_1 & l_2 & l\\
0 & 0 & 0
\end{array}
\right),
\end{eqnarray}
and the orthogonality relation in Eq. \ref{eq:Y+Y}, to be applied
successively in this order. 
Due to this complication, it is probably most efficient to calculate
propagation in spherical harmonics space, but to change back to
real space for any interaction vertex of high order.

\section{$f_\mathrm{nl}$-Propagator}\label{sect:fnlD}
We provide in the following the individual terms of the $f_\mathrm{nl}$-Propagator in Eq. \ref{eq:Dfnl}.
The individual diagrams are all $\mathcal{O}(f^2)$ and are given here for the case $f=1$:
\begin{eqnarray} \label{eq:Dfnldetails}
\includegraphics[width=\fgwidth]{fg/fg_d8.eps} &=& 
- \mathrm{Tr}\left[ D^2\, M \right] -\frac{1}{2} \widehat{D}^\dagger M \,  \widehat{D}
\\
\includegraphics[width=1.5\fgwidth]{fg/fg_ee11.eps} &=&
\frac{1}{2} \left[2\widehat{DM}  + \widehat{D}M  \right]^\dagger  \!\!D 
\left[2 \widehat{MD}  + M\widehat{D} \right]
\\
\includegraphics[width=0.6\fgwidth]{fg/fg_ee2.eps} &=&
\mathrm{Tr}\left[ D^2\, M\, D\, M \right] 
\nonumber\\
&&+  2 M_{xy} D_{yy'} M_{y'x'} D_{x'y} D_{xx'} 
\\
\includegraphics[width=0.6\fgwidth]{fg/fg_ee1.eps} &=&
j^\dagger D^2 j
\\ 
\includegraphics[width=\fgwidth]{fg/fg_ee4.eps} &=&
-2\,m^\dagger M\,D^2 j-4\,\mathrm{Tr}\left[ \widehat{m} \, D\, \widehat{j} \,D\, M \right]
\\
\includegraphics[width=1.4\fgwidth]{fg/fg_ee12.eps} &=&
m^\dagger M\,D^2M\, m + 4\,\mathrm{Tr}\left[ \widehat{m} \, D\, \widehat{Mm} \,D\, M \right]
\nonumber\\
&&+ 2\,\mathrm{Tr}\left[ \widehat{m} \, D\, (\widehat{m} \,M +M \, \widehat{m})\,D\, M\right]
\\
\includegraphics[width=\fgwidth]{fg/fg_ee7.eps} &=&
-2  \left[2\widehat{DM}  + \widehat{D}M  \right]^\dagger \!\! D \,\widehat{j}\,m
\\
\includegraphics[width=0.8\fgwidth]{fg/fg_d10.eps} &=&
-{m^2}^\dagger M\, \widehat{D} 
- 2\, \mathrm{Tr}\left[ \widehat{m}\, M\, \widehat{m}\, D \right]
\\
\includegraphics[width=1.25\fgwidth]{fg/fg_ee10.eps}&=&
\left[2\widehat{DM}  + \widehat{D}M  \right]^\dagger \!\! D \left[2\widehat{m}Mm  + M m^2 \right]\,\,\,\,
\\
\includegraphics[width=\fgwidth]{fg/fg_ee14.eps} &=&
\frac{1}{2} 
\left[2\widehat{m}Mm  + M m^2 \right]^\dagger \!\! D \left[2\widehat{m}Mm  + M m^2 \right]\nonumber
\\
&&\\
\includegraphics[width=0.8\fgwidth]{fg/fg_ee9.eps} &=&
-2\,(m\,j)^\dagger  D\,(M\,m^2  +2 \, \widehat{m}\,   M \, m  )
\\
\includegraphics[width=0.8\fgwidth]{fg/fg_d7.eps} &=&
-\frac{1}{2}\,{m^2}^\dagger M\,m^2
\\
\includegraphics[width=\fgwidth]{fg/fg_ee8.eps} &=&
2\,(m\,j)^\dagger D \, (j\,m)
\end{eqnarray}
We used here the conventions $m = D\, j$ and $(D^2)_{xy}= (D_{xy})^2$
and remind  that $\Lambda^{(0)}=\Lambda^{(1)}=0$, $j'=j$, $\Lambda^{(2)}=-2\,f\,\widehat{j}$, $\Lambda^{(3)}_{xyz}=[M_{xy} \delta_{yz} + 5 \;\mathrm{perm.}]$,
$\Lambda^{(4)}_{xyzu}=\frac{1}{2}\,[\delta_{xy}M_{yz} \delta_{zu} + 23 \;\mathrm{perm.}]$.
\bibliographystyle{apsrev}
\bibliography{bibtex/ift}

\begin{thebibliography}{223}
\expandafter\ifx\csname natexlab\endcsname\relax\def\natexlab#1{#1}\fi
\expandafter\ifx\csname bibnamefont\endcsname\relax
  \def\bibnamefont#1{#1}\fi
\expandafter\ifx\csname bibfnamefont\endcsname\relax
  \def\bibfnamefont#1{#1}\fi
\expandafter\ifx\csname citenamefont\endcsname\relax
  \def\citenamefont#1{#1}\fi
\expandafter\ifx\csname url\endcsname\relax
  \def\url#1{\texttt{#1}}\fi
\expandafter\ifx\csname urlprefix\endcsname\relax\def\urlprefix{URL }\fi
\providecommand{\bibinfo}[2]{#2}
\providecommand{\eprint}[2][]{\url{#2}}

\bibitem[{\citenamefont{{Bayes}}(1763)}]{Bayes}
\bibinfo{author}{\bibfnamefont{T.}~\bibnamefont{{Bayes}}},
  \bibinfo{journal}{Phil. Trans. Roy. Soc.} \textbf{\bibinfo{volume}{53}},
  \bibinfo{pages}{370} (\bibinfo{year}{1763}).

\bibitem[{\citenamefont{{Shannon}}(1948)}]{Shannon1948}
\bibinfo{author}{\bibfnamefont{C.~E.} \bibnamefont{{Shannon}}},
  \bibinfo{journal}{Bell System Technical Journal}
  \textbf{\bibinfo{volume}{27}}, \bibinfo{pages}{379} (\bibinfo{year}{1948}).

\bibitem[{\citenamefont{{Shannon} and {Weaver}}(1949)}]{1949mtc..book.....S}
\bibinfo{author}{\bibfnamefont{C.~E.} \bibnamefont{{Shannon}}}
  \bibnamefont{and} \bibinfo{author}{\bibfnamefont{W.}~\bibnamefont{{Weaver}}},
  \emph{\bibinfo{title}{{The mathematical theory of communication}}}
  (\bibinfo{publisher}{Urbana: University of Illinois Press, 1949},
  \bibinfo{year}{1949}).

\bibitem[{\citenamefont{{Jaynes}}(1957{\natexlab{a}})}]{1957PhRv..106..620J}
\bibinfo{author}{\bibfnamefont{E.~T.} \bibnamefont{{Jaynes}}},
  \bibinfo{journal}{Physical Review} \textbf{\bibinfo{volume}{106}},
  \bibinfo{pages}{620} (\bibinfo{year}{1957}{\natexlab{a}}).

\bibitem[{\citenamefont{{Jaynes}}(1957{\natexlab{b}})}]{1957PhRv..108..171J}
\bibinfo{author}{\bibfnamefont{E.~T.} \bibnamefont{{Jaynes}}},
  \bibinfo{journal}{Physical Review} \textbf{\bibinfo{volume}{108}},
  \bibinfo{pages}{171} (\bibinfo{year}{1957}{\natexlab{b}}).

\bibitem[{\citenamefont{{Jaynes}}(1963)}]{jaynes1963}
\bibinfo{author}{\bibfnamefont{E.~T.} \bibnamefont{{Jaynes}}}, in
  \emph{\bibinfo{booktitle}{Statistical Physics 3}} (\bibinfo{year}{1963}), p.
  \bibinfo{pages}{181}.

\bibitem[{\citenamefont{{Jaynes}}(1965)}]{1965AmJPh..33..391J}
\bibinfo{author}{\bibfnamefont{E.~T.} \bibnamefont{{Jaynes}}},
  \bibinfo{journal}{American Journal of Physics} \textbf{\bibinfo{volume}{33}},
  \bibinfo{pages}{391} (\bibinfo{year}{1965}).

\bibitem[{\citenamefont{{Jaynes}}(1968)}]{jaynes1968}
\bibinfo{author}{\bibfnamefont{E.~T.} \bibnamefont{{Jaynes}}},
  \bibinfo{journal}{IEEE Trans. on Systems Science and Cybernetics}
  \textbf{\bibinfo{volume}{SSC-4}}, \bibinfo{pages}{227}
  (\bibinfo{year}{1968}).

\bibitem[{\citenamefont{{Jaynes}}(1982)}]{1982ieee...70..939J}
\bibinfo{author}{\bibfnamefont{E.~T.} \bibnamefont{{Jaynes}}}, in
  \emph{\bibinfo{booktitle}{Proc. IEEE, Volume 70, p. 939-952}}
  (\bibinfo{year}{1982}), pp. \bibinfo{pages}{939--952}.

\bibitem[{\citenamefont{{Jaynes} and {Baierlein}}(2004)}]{2004PhT....57j..76J}
\bibinfo{author}{\bibfnamefont{E.~T.} \bibnamefont{{Jaynes}}} \bibnamefont{and}
  \bibinfo{author}{\bibfnamefont{R.}~\bibnamefont{{Baierlein}}},
  \bibinfo{journal}{Physics Today} \textbf{\bibinfo{volume}{57}},
  \bibinfo{pages}{76} (\bibinfo{year}{2004}).

\bibitem[{\citenamefont{{Metropolis} et~al.}(1953)\citenamefont{{Metropolis},
  {Rosenbluth}, {Rosenbluth}, {Teller}, and {Teller}}}]{metroplis}
\bibinfo{author}{\bibfnamefont{N.}~\bibnamefont{{Metropolis}}},
  \bibinfo{author}{\bibfnamefont{A.~W.} \bibnamefont{{Rosenbluth}}},
  \bibinfo{author}{\bibfnamefont{M.~N.} \bibnamefont{{Rosenbluth}}},
  \bibinfo{author}{\bibfnamefont{A.~H.} \bibnamefont{{Teller}}},
  \bibnamefont{and} \bibinfo{author}{\bibfnamefont{E.~T.}
  \bibnamefont{{Teller}}}, \bibinfo{journal}{Journal of Chemical Physics}
  \textbf{\bibinfo{volume}{21}}, \bibinfo{pages}{1087} (\bibinfo{year}{1953}).

\bibitem[{\citenamefont{{Hastings}}(1970)}]{hastings}
\bibinfo{author}{\bibfnamefont{W.~K.} \bibnamefont{{Hastings}}},
  \bibinfo{journal}{Biometrika} \textbf{\bibinfo{volume}{57}},
  \bibinfo{pages}{97} (\bibinfo{year}{1970}).

\bibitem[{\citenamefont{{Geman} and {Geman}}(1984)}]{gibbsamp}
\bibinfo{author}{\bibfnamefont{S.}~\bibnamefont{{Geman}}} \bibnamefont{and}
  \bibinfo{author}{\bibfnamefont{D.}~\bibnamefont{{Geman}}},
  \bibinfo{journal}{IEEE Transactions on Pattern Analysis and Machine
  Intelligence} \textbf{\bibinfo{volume}{6}}, \bibinfo{pages}{721}
  (\bibinfo{year}{1984}).

\bibitem[{\citenamefont{{Duan} et~al.}(1987)\citenamefont{{Duan}, {Kennedy},
  {Pendleton}, and {Roweth}}}]{HMCMC}
\bibinfo{author}{\bibfnamefont{S.}~\bibnamefont{{Duan}}},
  \bibinfo{author}{\bibfnamefont{A.}~\bibnamefont{{Kennedy}}},
  \bibinfo{author}{\bibfnamefont{B.}~\bibnamefont{{Pendleton}}},
  \bibnamefont{and} \bibinfo{author}{\bibfnamefont{D.}~\bibnamefont{{Roweth}}},
  \bibinfo{journal}{Phys. Lett. B} \textbf{\bibinfo{volume}{195}},
  \bibinfo{pages}{216} (\bibinfo{year}{1987}).

\bibitem[{\citenamefont{{Murthy} et~al.}(2005)\citenamefont{{Murthy}, {Janani},
  and {Shenbga Priya}}}]{2005cs........4037M}
\bibinfo{author}{\bibfnamefont{K.~P.~N.} \bibnamefont{{Murthy}}},
  \bibinfo{author}{\bibfnamefont{M.}~\bibnamefont{{Janani}}}, \bibnamefont{and}
  \bibinfo{author}{\bibfnamefont{B.}~\bibnamefont{{Shenbga Priya}}},
  \bibinfo{journal}{ArXiv Computer Science e-prints}  (\bibinfo{year}{2005}),
  \eprint{arXiv:cs/0504037}.

\bibitem[{\citenamefont{{Tanner}}(1996)}]{toolsstatinf}
\bibinfo{author}{\bibfnamefont{M.~A.} \bibnamefont{{Tanner}}},
  \emph{\bibinfo{title}{Tools for statistical inference}}
  (\bibinfo{publisher}{Springer-Verlag}, \bibinfo{address}{New York},
  \bibinfo{year}{1996}).

\bibitem[{\citenamefont{{Neal}}(1993)}]{neal1993}
\bibinfo{author}{\bibfnamefont{R.~M.} \bibnamefont{{Neal}}}, in
  \emph{\bibinfo{booktitle}{Technical Report CRG-TR-93-1}}
  (\bibinfo{publisher}{Dept. of Computer Science}, \bibinfo{address}{University
  of Toronto}, \bibinfo{year}{1993}).

\bibitem[{\citenamefont{{Robert}}(2001)}]{bayeschoice}
\bibinfo{author}{\bibfnamefont{C.~P.} \bibnamefont{{Robert}}},
  \emph{\bibinfo{title}{The Bayesian choice}}
  (\bibinfo{publisher}{Springer-Verlag}, \bibinfo{address}{New York},
  \bibinfo{year}{2001}).

\bibitem[{\citenamefont{{Gelman} et~al.}(2004)\citenamefont{{Gelman}, {Carlin},
  {Stern}, and {Rubin}}}]{bayesdataanal}
\bibinfo{author}{\bibfnamefont{A.}~\bibnamefont{{Gelman}}},
  \bibinfo{author}{\bibfnamefont{J.~B.} \bibnamefont{{Carlin}}},
  \bibinfo{author}{\bibfnamefont{H.~S.} \bibnamefont{{Stern}}},
  \bibnamefont{and} \bibinfo{author}{\bibfnamefont{D.}~\bibnamefont{{Rubin}}},
  \emph{\bibinfo{title}{Bayesian data analysis}} (\bibinfo{publisher}{Chapman
  \& Hall/CRC}, \bibinfo{address}{Boca Raton, Florida}, \bibinfo{year}{2004}).

\bibitem[{\citenamefont{{Aster} et~al.}(2005)\citenamefont{{Aster}, {Brochers},
  and {Thurber}}}]{paramest}
\bibinfo{author}{\bibfnamefont{R.~A.} \bibnamefont{{Aster}}},
  \bibinfo{author}{\bibfnamefont{B.}~\bibnamefont{{Brochers}}},
  \bibnamefont{and} \bibinfo{author}{\bibfnamefont{C.~H.}
  \bibnamefont{{Thurber}}}, \emph{\bibinfo{title}{Parameter estimation and
  inverse problems}} (\bibinfo{publisher}{Elsevier Academic Press},
  \bibinfo{address}{London}, \bibinfo{year}{2005}).

\bibitem[{\citenamefont{{Trotta}}(2008)}]{2008arXiv0803.4089T}
\bibinfo{author}{\bibfnamefont{R.}~\bibnamefont{{Trotta}}},
  \bibinfo{journal}{ArXiv e-prints} \textbf{\bibinfo{volume}{0803.4089}}
  (\bibinfo{year}{2008}), \eprint{0803.4089}.

\bibitem[{\citenamefont{{Wiener}}(1949)}]{1949wiener}
\bibinfo{author}{\bibfnamefont{N.}~\bibnamefont{{Wiener}}},
  \emph{\bibinfo{title}{{Extrapolation, Interpolation, and Smoothing of
  Stationary Time Series}}} (\bibinfo{publisher}{New York: Wiley},
  \bibinfo{year}{1949}).

\bibitem[{\citenamefont{{Richardson}}(1972)}]{1972JOSA...62...55R}
\bibinfo{author}{\bibfnamefont{W.~H.} \bibnamefont{{Richardson}}},
  \bibinfo{journal}{Journal of the Optical Society of America (1917-1983)}
  \textbf{\bibinfo{volume}{62}}, \bibinfo{pages}{55} (\bibinfo{year}{1972}).

\bibitem[{\citenamefont{{Lucy}}(1974)}]{1974AJ.....79..745L}
\bibinfo{author}{\bibfnamefont{L.~B.} \bibnamefont{{Lucy}}},
  \bibinfo{journal}{\aj} \textbf{\bibinfo{volume}{79}}, \bibinfo{pages}{745}
  (\bibinfo{year}{1974}).

\bibitem[{\citenamefont{{Frieden}}(1972)}]{1972JOSA...62..511F}
\bibinfo{author}{\bibfnamefont{B.~R.} \bibnamefont{{Frieden}}},
  \bibinfo{journal}{Journal of the Optical Society of America (1917-1983)}
  \textbf{\bibinfo{volume}{62}}, \bibinfo{pages}{511} (\bibinfo{year}{1972}).

\bibitem[{\citenamefont{{Gull} and {Daniell}}(1978)}]{1978Natur.272..686G}
\bibinfo{author}{\bibfnamefont{S.~F.} \bibnamefont{{Gull}}} \bibnamefont{and}
  \bibinfo{author}{\bibfnamefont{G.~J.} \bibnamefont{{Daniell}}},
  \bibinfo{journal}{\nat} \textbf{\bibinfo{volume}{272}}, \bibinfo{pages}{686}
  (\bibinfo{year}{1978}).

\bibitem[{\citenamefont{{Skilling} et~al.}(1979)\citenamefont{{Skilling},
  {Strong}, and {Bennett}}}]{1979MNRAS.187..145S}
\bibinfo{author}{\bibfnamefont{J.}~\bibnamefont{{Skilling}}},
  \bibinfo{author}{\bibfnamefont{A.~W.} \bibnamefont{{Strong}}},
  \bibnamefont{and}
  \bibinfo{author}{\bibfnamefont{K.}~\bibnamefont{{Bennett}}},
  \bibinfo{journal}{\mnras} \textbf{\bibinfo{volume}{187}},
  \bibinfo{pages}{145} (\bibinfo{year}{1979}).

\bibitem[{\citenamefont{{Bryan} and {Skilling}}(1980)}]{1980MNRAS.191...69B}
\bibinfo{author}{\bibfnamefont{R.~K.} \bibnamefont{{Bryan}}} \bibnamefont{and}
  \bibinfo{author}{\bibfnamefont{J.}~\bibnamefont{{Skilling}}},
  \bibinfo{journal}{\mnras} \textbf{\bibinfo{volume}{191}}, \bibinfo{pages}{69}
  (\bibinfo{year}{1980}).

\bibitem[{\citenamefont{{Burch} et~al.}(1983)\citenamefont{{Burch}, {Gull}, and
  {Skilling}}}]{1983CVGIP..23..113B}
\bibinfo{author}{\bibfnamefont{S.~F.} \bibnamefont{{Burch}}},
  \bibinfo{author}{\bibfnamefont{S.~F.} \bibnamefont{{Gull}}},
  \bibnamefont{and}
  \bibinfo{author}{\bibfnamefont{J.}~\bibnamefont{{Skilling}}},
  \bibinfo{journal}{Computer Vision Graphics and Image Processing}
  \textbf{\bibinfo{volume}{23}}, \bibinfo{pages}{113} (\bibinfo{year}{1983}).

\bibitem[{\citenamefont{{Gull} and {Skilling}}(1983)}]{1983iimp.conf..267G}
\bibinfo{author}{\bibfnamefont{S.~F.} \bibnamefont{{Gull}}} \bibnamefont{and}
  \bibinfo{author}{\bibfnamefont{J.}~\bibnamefont{{Skilling}}}, in
  \emph{\bibinfo{booktitle}{Indirect Imaging. Measurement and Processing for
  Indirect Imaging. Proceedings of an International Symposium held in Sydney,
  Australia, August 30-September 2, 1983. Editor, J.A. Roberts; Publisher,
  Cambridge University Press, Cambridge, England, New York, NY, 1984. LC \#
  QB51.3.E43 I53 1984. ISBN \# 0-521-26282-8. P.267, 1983}}
  (\bibinfo{year}{1983}), p. \bibinfo{pages}{267}.

\bibitem[{\citenamefont{{Sibisi} et~al.}(1984)\citenamefont{{Sibisi},
  {Skilling}, {Brereton}, {Laue}, and {Staunton}}}]{1984Natur.311..446S}
\bibinfo{author}{\bibfnamefont{S.}~\bibnamefont{{Sibisi}}},
  \bibinfo{author}{\bibfnamefont{J.}~\bibnamefont{{Skilling}}},
  \bibinfo{author}{\bibfnamefont{R.~G.} \bibnamefont{{Brereton}}},
  \bibinfo{author}{\bibfnamefont{E.~D.} \bibnamefont{{Laue}}},
  \bibnamefont{and}
  \bibinfo{author}{\bibfnamefont{J.}~\bibnamefont{{Staunton}}},
  \bibinfo{journal}{\nat} \textbf{\bibinfo{volume}{311}}, \bibinfo{pages}{446}
  (\bibinfo{year}{1984}).

\bibitem[{\citenamefont{{Titterington} and
  {Skilling}}(1984)}]{1984Natur.312..381T}
\bibinfo{author}{\bibfnamefont{D.~M.} \bibnamefont{{Titterington}}}
  \bibnamefont{and}
  \bibinfo{author}{\bibfnamefont{J.}~\bibnamefont{{Skilling}}},
  \bibinfo{journal}{\nat} \textbf{\bibinfo{volume}{312}}, \bibinfo{pages}{381}
  (\bibinfo{year}{1984}).

\bibitem[{\citenamefont{{Skilling} and {Bryan}}(1984)}]{1984MNRAS.211..111S}
\bibinfo{author}{\bibfnamefont{J.}~\bibnamefont{{Skilling}}} \bibnamefont{and}
  \bibinfo{author}{\bibfnamefont{R.~K.} \bibnamefont{{Bryan}}},
  \bibinfo{journal}{\mnras} \textbf{\bibinfo{volume}{211}},
  \bibinfo{pages}{111} (\bibinfo{year}{1984}).

\bibitem[{\citenamefont{{Bryan} and {Skilling}}(1986)}]{1986JMOp...33..287B}
\bibinfo{author}{\bibfnamefont{R.~K.} \bibnamefont{{Bryan}}} \bibnamefont{and}
  \bibinfo{author}{\bibfnamefont{J.}~\bibnamefont{{Skilling}}},
  \bibinfo{journal}{Journal of Modern Optics} \textbf{\bibinfo{volume}{33}},
  \bibinfo{pages}{287} (\bibinfo{year}{1986}).

\bibitem[{\citenamefont{{Gull}}(1989)}]{gull1989}
\bibinfo{author}{\bibfnamefont{S.~F.} \bibnamefont{{Gull}}}, in
  \emph{\bibinfo{booktitle}{Maximum Entropy and Bayesian Methods}}, edited by
  \bibinfo{editor}{\bibfnamefont{J.}~\bibnamefont{{Skilling}}}
  (\bibinfo{publisher}{Kluwer Academic Publishers},
  \bibinfo{address}{Dordtrecht}, \bibinfo{year}{1989}), pp.
  \bibinfo{pages}{53--71}.

\bibitem[{\citenamefont{{Gull} and {Skilling}}(1990)}]{gullskilling}
\bibinfo{author}{\bibfnamefont{S.~F.} \bibnamefont{{Gull}}} \bibnamefont{and}
  \bibinfo{author}{\bibfnamefont{J.}~\bibnamefont{{Skilling}}},
  \emph{\bibinfo{title}{The MEMSYS5 User's Manual}}
  (\bibinfo{publisher}{Maximum Entropy Data Consultants Ltd},
  \bibinfo{address}{Royston}, \bibinfo{year}{1990}).

\bibitem[{\citenamefont{{Skilling}}(1998)}]{1998mebm.conf....1S}
\bibinfo{author}{\bibfnamefont{J.}~\bibnamefont{{Skilling}}}, in
  \emph{\bibinfo{booktitle}{Maximum Entropy and Bayesian Methods}}, edited by
  \bibinfo{editor}{\bibfnamefont{G.~J.} \bibnamefont{{Erickson}}},
  \bibinfo{editor}{\bibfnamefont{J.~T.} \bibnamefont{{Rychert}}},
  \bibnamefont{and} \bibinfo{editor}{\bibfnamefont{C.~R.}
  \bibnamefont{{Smith}}} (\bibinfo{year}{1998}), p.~\bibinfo{pages}{1}.

\bibitem[{\citenamefont{{Kitaura} and
  {En{\ss}lin}}(2008)}]{2008MNRAS.389..497K}
\bibinfo{author}{\bibfnamefont{F.~S.} \bibnamefont{{Kitaura}}}
  \bibnamefont{and} \bibinfo{author}{\bibfnamefont{T.~A.}
  \bibnamefont{{En{\ss}lin}}}, \bibinfo{journal}{\mnras}
  \textbf{\bibinfo{volume}{389}}, \bibinfo{pages}{497} (\bibinfo{year}{2008}),
  \eprint{0705.0429}.

\bibitem[{\citenamefont{{Narayan} and
  {Nityananda}}(1986)}]{1986ARA&A..24..127N}
\bibinfo{author}{\bibfnamefont{R.}~\bibnamefont{{Narayan}}} \bibnamefont{and}
  \bibinfo{author}{\bibfnamefont{R.}~\bibnamefont{{Nityananda}}},
  \bibinfo{journal}{\araa} \textbf{\bibinfo{volume}{24}}, \bibinfo{pages}{127}
  (\bibinfo{year}{1986}).

\bibitem[{\citenamefont{{Molina} et~al.}(2001)\citenamefont{{Molina}, {Nunez},
  {Cortijo}, and {Mateos}}}]{imagerest}
\bibinfo{author}{\bibfnamefont{R.}~\bibnamefont{{Molina}}},
  \bibinfo{author}{\bibfnamefont{J.}~\bibnamefont{{Nunez}}},
  \bibinfo{author}{\bibfnamefont{F.~J.} \bibnamefont{{Cortijo}}},
  \bibnamefont{and} \bibinfo{author}{\bibfnamefont{J.}~\bibnamefont{{Mateos}}},
  \bibinfo{journal}{Signal Processing Magazine, IEEE}
  \textbf{\bibinfo{volume}{18}}, \bibinfo{pages}{11} (\bibinfo{year}{2001}).

\bibitem[{\citenamefont{{Bertschinger}}(1987)}]{1987ApJ...323L.103B}
\bibinfo{author}{\bibfnamefont{E.}~\bibnamefont{{Bertschinger}}},
  \bibinfo{journal}{\apjl} \textbf{\bibinfo{volume}{323}},
  \bibinfo{pages}{L103} (\bibinfo{year}{1987}).

\bibitem[{\citenamefont{{Fry}}(1985)}]{1985ApJ...289...10F}
\bibinfo{author}{\bibfnamefont{J.~N.} \bibnamefont{{Fry}}},
  \bibinfo{journal}{\apj} \textbf{\bibinfo{volume}{289}}, \bibinfo{pages}{10}
  (\bibinfo{year}{1985}).

\bibitem[{\citenamefont{{Bialek} and {Zee}}(1987)}]{1987PhRvL..58..741B}
\bibinfo{author}{\bibfnamefont{W.}~\bibnamefont{{Bialek}}} \bibnamefont{and}
  \bibinfo{author}{\bibfnamefont{A.}~\bibnamefont{{Zee}}},
  \bibinfo{journal}{Physical Review Letters} \textbf{\bibinfo{volume}{58}},
  \bibinfo{pages}{741} (\bibinfo{year}{1987}).

\bibitem[{\citenamefont{{Bialek} and {Zee}}(1988)}]{1988PhRvL..61.1512B}
\bibinfo{author}{\bibfnamefont{W.}~\bibnamefont{{Bialek}}} \bibnamefont{and}
  \bibinfo{author}{\bibfnamefont{A.}~\bibnamefont{{Zee}}},
  \bibinfo{journal}{Physical Review Letters} \textbf{\bibinfo{volume}{61}},
  \bibinfo{pages}{1512} (\bibinfo{year}{1988}).

\bibitem[{\citenamefont{{Bialek} et~al.}(1996)\citenamefont{{Bialek}, {Callan},
  and {Strong}}}]{1996PhRvL..77.4693B}
\bibinfo{author}{\bibfnamefont{W.}~\bibnamefont{{Bialek}}},
  \bibinfo{author}{\bibfnamefont{C.~G.} \bibnamefont{{Callan}}},
  \bibnamefont{and} \bibinfo{author}{\bibfnamefont{S.~P.}
  \bibnamefont{{Strong}}}, \bibinfo{journal}{Physical Review Letters}
  \textbf{\bibinfo{volume}{77}}, \bibinfo{pages}{4693} (\bibinfo{year}{1996}),
  \eprint{arXiv:cond-mat/9607180}.

\bibitem[{\citenamefont{{Stoica} et~al.}(2000)\citenamefont{{Stoica},
  {Larsson}, and {Li}}}]{2000AJ....120.2163S}
\bibinfo{author}{\bibfnamefont{P.}~\bibnamefont{{Stoica}}},
  \bibinfo{author}{\bibfnamefont{E.~G.} \bibnamefont{{Larsson}}},
  \bibnamefont{and} \bibinfo{author}{\bibfnamefont{J.}~\bibnamefont{{Li}}},
  \bibinfo{journal}{\aj} \textbf{\bibinfo{volume}{120}}, \bibinfo{pages}{2163}
  (\bibinfo{year}{2000}).

\bibitem[{\citenamefont{{En{\ss}lin} and {Frommert}}(2009)}]{PURE}
\bibinfo{author}{\bibfnamefont{T.}~\bibnamefont{{En{\ss}lin}}}
  \bibnamefont{and}
  \bibinfo{author}{\bibfnamefont{M.}~\bibnamefont{{Frommert}}},
  \bibinfo{journal}{in preparation}  (\bibinfo{year}{2009}).

\bibitem[{\citenamefont{{Lemm}}(1999)}]{1999physics..12005L}
\bibinfo{author}{\bibfnamefont{J.~C.} \bibnamefont{{Lemm}}},
  \bibinfo{journal}{ArXiv Physics e-prints}  (\bibinfo{year}{1999}),
  \eprint{physics/9912005}.

\bibitem[{\citenamefont{{Lemm} and
  {Uhlig}}(2000{\natexlab{a}})}]{2000FBS....29...25L}
\bibinfo{author}{\bibfnamefont{J.~C.} \bibnamefont{{Lemm}}} \bibnamefont{and}
  \bibinfo{author}{\bibfnamefont{J.}~\bibnamefont{{Uhlig}}},
  \bibinfo{journal}{Few-Body Systems} \textbf{\bibinfo{volume}{29}},
  \bibinfo{pages}{25} (\bibinfo{year}{2000}{\natexlab{a}}),
  \eprint{arXiv:quant-ph/0006027}.

\bibitem[{\citenamefont{{Lemm} et~al.}(2000)\citenamefont{{Lemm}, {Uhlig}, and
  {Weiguny}}}]{2000PhRvL..84.2068L}
\bibinfo{author}{\bibfnamefont{J.~C.} \bibnamefont{{Lemm}}},
  \bibinfo{author}{\bibfnamefont{J.}~\bibnamefont{{Uhlig}}}, \bibnamefont{and}
  \bibinfo{author}{\bibfnamefont{A.}~\bibnamefont{{Weiguny}}},
  \bibinfo{journal}{Physical Review Letters} \textbf{\bibinfo{volume}{84}},
  \bibinfo{pages}{2068} (\bibinfo{year}{2000}),
  \eprint{arXiv:cond-mat/9907013}.

\bibitem[{\citenamefont{{Lemm} and
  {Uhlig}}(2000{\natexlab{b}})}]{2000PhRvL..84.4517L}
\bibinfo{author}{\bibfnamefont{J.~C.} \bibnamefont{{Lemm}}} \bibnamefont{and}
  \bibinfo{author}{\bibfnamefont{J.}~\bibnamefont{{Uhlig}}},
  \bibinfo{journal}{Physical Review Letters} \textbf{\bibinfo{volume}{84}},
  \bibinfo{pages}{4517} (\bibinfo{year}{2000}{\natexlab{b}}),
  \eprint{arXiv:nucl-th/9908056}.

\bibitem[{\citenamefont{{Lemm}}(2000)}]{2000PhLA..276...19L}
\bibinfo{author}{\bibfnamefont{J.~C.} \bibnamefont{{Lemm}}},
  \bibinfo{journal}{Physics Letters A} \textbf{\bibinfo{volume}{276}},
  \bibinfo{pages}{19} (\bibinfo{year}{2000}).

\bibitem[{\citenamefont{{Lemm}}(2001)}]{2001AIPC..568..425L}
\bibinfo{author}{\bibfnamefont{J.~C.} \bibnamefont{{Lemm}}}, in
  \emph{\bibinfo{booktitle}{Bayesian Inference and Maximum Entropy Methods in
  Science and Engineering}}, edited by
  \bibinfo{editor}{\bibfnamefont{A.}~\bibnamefont{{Mohammad-Djafari}}}
  (\bibinfo{year}{2001}), vol. \bibinfo{volume}{568} of
  \emph{\bibinfo{series}{American Institute of Physics Conference Series}}, pp.
  \bibinfo{pages}{425--436}.

\bibitem[{\citenamefont{{Lemm} et~al.}(2001)\citenamefont{{Lemm}, {Uhlig}, and
  {Weiguny}}}]{2001EPJB...20..349L}
\bibinfo{author}{\bibfnamefont{J.~C.} \bibnamefont{{Lemm}}},
  \bibinfo{author}{\bibfnamefont{J.}~\bibnamefont{{Uhlig}}}, \bibnamefont{and}
  \bibinfo{author}{\bibfnamefont{A.}~\bibnamefont{{Weiguny}}},
  \bibinfo{journal}{European Physical Journal B} \textbf{\bibinfo{volume}{20}},
  \bibinfo{pages}{349} (\bibinfo{year}{2001}), \eprint{arXiv:quant-ph/0005122}.

\bibitem[{\citenamefont{{Lemm} et~al.}(2005)\citenamefont{{Lemm}, {Uhlig}, and
  {Weiguny}}}]{2005EPJB...46...41L}
\bibinfo{author}{\bibfnamefont{J.~C.} \bibnamefont{{Lemm}}},
  \bibinfo{author}{\bibfnamefont{J.}~\bibnamefont{{Uhlig}}}, \bibnamefont{and}
  \bibinfo{author}{\bibfnamefont{A.}~\bibnamefont{{Weiguny}}},
  \bibinfo{journal}{European Physical Journal B} \textbf{\bibinfo{volume}{46}},
  \bibinfo{pages}{41} (\bibinfo{year}{2005}).

\bibitem[{\citenamefont{{Lemm}}(1998)}]{1998cond.mat..8039L}
\bibinfo{author}{\bibfnamefont{J.~C.} \bibnamefont{{Lemm}}},
  \bibinfo{journal}{ArXiv Condensed Matter e-prints}  (\bibinfo{year}{1998}),
  \eprint{cond-mat/9808039}.

\bibitem[{\citenamefont{{Binney} et~al.}(1992)\citenamefont{{Binney},
  {Dowrick}, {Fisher}, and {Newman}}}]{Binney1992}
\bibinfo{author}{\bibfnamefont{J.}~\bibnamefont{{Binney}}},
  \bibinfo{author}{\bibfnamefont{N.}~\bibnamefont{{Dowrick}}},
  \bibinfo{author}{\bibfnamefont{A.}~\bibnamefont{{Fisher}}}, \bibnamefont{and}
  \bibinfo{author}{\bibfnamefont{M.}~\bibnamefont{{Newman}}},
  \emph{\bibinfo{title}{{The theory of critical phenomena}}}
  (\bibinfo{publisher}{{Oxford University Press, Oxford, UK:
  ISBN0-19-851394-1}}, \bibinfo{year}{1992}).

\bibitem[{\citenamefont{{Peskin} and {Schroeder}}(1995)}]{PeskinSchroeder}
\bibinfo{author}{\bibfnamefont{M.~E.} \bibnamefont{{Peskin}}} \bibnamefont{and}
  \bibinfo{author}{\bibfnamefont{D.~V.} \bibnamefont{{Schroeder}}},
  \emph{\bibinfo{title}{{An Introduction to Quantum Field Theory}}}
  (\bibinfo{publisher}{Westview Press~Boulder, Colorado: 1995, ISBN-13
  978-0-201-50397-5.}, \bibinfo{year}{1995}).

\bibitem[{\citenamefont{{Zee}}(2003)}]{2003qftn.book.....Z}
\bibinfo{author}{\bibfnamefont{A.}~\bibnamefont{{Zee}}},
  \emph{\bibinfo{title}{{Quantum field theory in a nutshell}}}
  (\bibinfo{publisher}{Quantum field theory in a nutshell, by
  A.~Zee.~Princeton, NJ: Princeton University Press, 2003, ISBN 0691010196.},
  \bibinfo{year}{2003}).

\bibitem[{\citenamefont{{Matarrese} et~al.}(1986)\citenamefont{{Matarrese},
  {Lucchin}, and {Bonometto}}}]{1986ApJ...310L..21M}
\bibinfo{author}{\bibfnamefont{S.}~\bibnamefont{{Matarrese}}},
  \bibinfo{author}{\bibfnamefont{F.}~\bibnamefont{{Lucchin}}},
  \bibnamefont{and} \bibinfo{author}{\bibfnamefont{S.~A.}
  \bibnamefont{{Bonometto}}}, \bibinfo{journal}{\apjl}
  \textbf{\bibinfo{volume}{310}}, \bibinfo{pages}{L21} (\bibinfo{year}{1986}).

\bibitem[{\citenamefont{{Zel'dovich}}(1970)}]{1970A&A.....5...84Z}
\bibinfo{author}{\bibfnamefont{Y.~B.} \bibnamefont{{Zel'dovich}}},
  \bibinfo{journal}{\aap} \textbf{\bibinfo{volume}{5}}, \bibinfo{pages}{84}
  (\bibinfo{year}{1970}).

\bibitem[{\citenamefont{{Bardeen} et~al.}(1986)\citenamefont{{Bardeen}, {Bond},
  {Kaiser}, and {Szalay}}}]{1986ApJ...304...15B}
\bibinfo{author}{\bibfnamefont{J.~M.} \bibnamefont{{Bardeen}}},
  \bibinfo{author}{\bibfnamefont{J.~R.} \bibnamefont{{Bond}}},
  \bibinfo{author}{\bibfnamefont{N.}~\bibnamefont{{Kaiser}}}, \bibnamefont{and}
  \bibinfo{author}{\bibfnamefont{A.~S.} \bibnamefont{{Szalay}}},
  \bibinfo{journal}{\apj} \textbf{\bibinfo{volume}{304}}, \bibinfo{pages}{15}
  (\bibinfo{year}{1986}).

\bibitem[{\citenamefont{{Peebles}}(1980)}]{1980lssu.book.....P}
\bibinfo{author}{\bibfnamefont{P.~J.~E.} \bibnamefont{{Peebles}}},
  \emph{\bibinfo{title}{{The large-scale structure of the universe}}}
  (\bibinfo{publisher}{Research supported by the National Science
  Foundation.~Princeton, N.J., Princeton University Press, 1980.~435 p.},
  \bibinfo{year}{1980}).

\bibitem[{\citenamefont{{Kaiser}}(1987)}]{1987MNRAS.227....1K}
\bibinfo{author}{\bibfnamefont{N.}~\bibnamefont{{Kaiser}}},
  \bibinfo{journal}{\mnras} \textbf{\bibinfo{volume}{227}}, \bibinfo{pages}{1}
  (\bibinfo{year}{1987}).

\bibitem[{\citenamefont{{Peebles}}(1990)}]{1990ApJ...362....1P}
\bibinfo{author}{\bibfnamefont{P.~J.~E.} \bibnamefont{{Peebles}}},
  \bibinfo{journal}{\apj} \textbf{\bibinfo{volume}{362}}, \bibinfo{pages}{1}
  (\bibinfo{year}{1990}).

\bibitem[{\citenamefont{{Bernardeau}}(1992)}]{1992ApJ...390L..61B}
\bibinfo{author}{\bibfnamefont{F.}~\bibnamefont{{Bernardeau}}},
  \bibinfo{journal}{\apjl} \textbf{\bibinfo{volume}{390}}, \bibinfo{pages}{L61}
  (\bibinfo{year}{1992}).

\bibitem[{\citenamefont{{Zaroubi} and {Hoffman}}(1996)}]{1996ApJ...462...25Z}
\bibinfo{author}{\bibfnamefont{S.}~\bibnamefont{{Zaroubi}}} \bibnamefont{and}
  \bibinfo{author}{\bibfnamefont{Y.}~\bibnamefont{{Hoffman}}},
  \bibinfo{journal}{\apj} \textbf{\bibinfo{volume}{462}}, \bibinfo{pages}{25}
  (\bibinfo{year}{1996}).

\bibitem[{\citenamefont{{Hamilton}}(1998)}]{hamilton-1998}
\bibinfo{author}{\bibfnamefont{A.~J.~S.} \bibnamefont{{Hamilton}}}, in
  \emph{\bibinfo{booktitle}{The Evolving Universe}}, edited by
  \bibinfo{editor}{\bibfnamefont{D.}~\bibnamefont{{Hamilton}}}
  (\bibinfo{publisher}{Kluwer Academic Publishers},
  \bibinfo{address}{Dordtrecht}, \bibinfo{year}{1998}), vol.
  \bibinfo{volume}{231} of \emph{\bibinfo{series}{Astrophysics and Space
  Science Library}}, p. \bibinfo{pages}{185}.

\bibitem[{\citenamefont{{Bernardeau} et~al.}(1999)\citenamefont{{Bernardeau},
  {Chodorowski}, {{\L}okas}, {Stompor}, and {Kudlicki}}}]{1999MNRAS.309..543B}
\bibinfo{author}{\bibfnamefont{F.}~\bibnamefont{{Bernardeau}}},
  \bibinfo{author}{\bibfnamefont{M.~J.} \bibnamefont{{Chodorowski}}},
  \bibinfo{author}{\bibfnamefont{E.~L.} \bibnamefont{{{\L}okas}}},
  \bibinfo{author}{\bibfnamefont{R.}~\bibnamefont{{Stompor}}},
  \bibnamefont{and}
  \bibinfo{author}{\bibfnamefont{A.}~\bibnamefont{{Kudlicki}}},
  \bibinfo{journal}{\mnras} \textbf{\bibinfo{volume}{309}},
  \bibinfo{pages}{543} (\bibinfo{year}{1999}), \eprint{astro-ph/9901057}.

\bibitem[{\citenamefont{{Branchini} et~al.}(1999)\citenamefont{{Branchini},
  {Teodoro}, {Frenk}, {Schmoldt}, {Efstathiou}, {White}, {Saunders},
  {Sutherland}, {Rowan-Robinson}, {Keeble} et~al.}}]{1999MNRAS.308....1B}
\bibinfo{author}{\bibfnamefont{E.}~\bibnamefont{{Branchini}}},
  \bibinfo{author}{\bibfnamefont{L.}~\bibnamefont{{Teodoro}}},
  \bibinfo{author}{\bibfnamefont{C.~S.} \bibnamefont{{Frenk}}},
  \bibinfo{author}{\bibfnamefont{I.}~\bibnamefont{{Schmoldt}}},
  \bibinfo{author}{\bibfnamefont{G.}~\bibnamefont{{Efstathiou}}},
  \bibinfo{author}{\bibfnamefont{S.~D.~M.} \bibnamefont{{White}}},
  \bibinfo{author}{\bibfnamefont{W.}~\bibnamefont{{Saunders}}},
  \bibinfo{author}{\bibfnamefont{W.}~\bibnamefont{{Sutherland}}},
  \bibinfo{author}{\bibfnamefont{M.}~\bibnamefont{{Rowan-Robinson}}},
  \bibinfo{author}{\bibfnamefont{O.}~\bibnamefont{{Keeble}}},
  \bibnamefont{et~al.}, \bibinfo{journal}{\mnras}
  \textbf{\bibinfo{volume}{308}}, \bibinfo{pages}{1} (\bibinfo{year}{1999}),
  \eprint{astro-ph/9901366}.

\bibitem[{\citenamefont{{Dekel} and {Lahav}}(1999)}]{1999ApJ...520...24D}
\bibinfo{author}{\bibfnamefont{A.}~\bibnamefont{{Dekel}}} \bibnamefont{and}
  \bibinfo{author}{\bibfnamefont{O.}~\bibnamefont{{Lahav}}},
  \bibinfo{journal}{\apj} \textbf{\bibinfo{volume}{520}}, \bibinfo{pages}{24}
  (\bibinfo{year}{1999}), \eprint{astro-ph/9806193}.

\bibitem[{\citenamefont{{Zaroubi}}(2002{\natexlab{a}})}]{2002astro.ph..6052Z}
\bibinfo{author}{\bibfnamefont{S.}~\bibnamefont{{Zaroubi}}},
  \bibinfo{journal}{ArXiv Astrophysics e-prints}
  (\bibinfo{year}{2002}{\natexlab{a}}), \eprint{astro-ph/0206052}.

\bibitem[{\citenamefont{{Smith} et~al.}(2003)\citenamefont{{Smith}, {Peacock},
  {Jenkins}, {White}, {Frenk}, {Pearce}, {Thomas}, {Efstathiou}, and
  {Couchman}}}]{2003MNRAS.341.1311S}
\bibinfo{author}{\bibfnamefont{R.~E.} \bibnamefont{{Smith}}},
  \bibinfo{author}{\bibfnamefont{J.~A.} \bibnamefont{{Peacock}}},
  \bibinfo{author}{\bibfnamefont{A.}~\bibnamefont{{Jenkins}}},
  \bibinfo{author}{\bibfnamefont{S.~D.~M.} \bibnamefont{{White}}},
  \bibinfo{author}{\bibfnamefont{C.~S.} \bibnamefont{{Frenk}}},
  \bibinfo{author}{\bibfnamefont{F.~R.} \bibnamefont{{Pearce}}},
  \bibinfo{author}{\bibfnamefont{P.~A.} \bibnamefont{{Thomas}}},
  \bibinfo{author}{\bibfnamefont{G.}~\bibnamefont{{Efstathiou}}},
  \bibnamefont{and} \bibinfo{author}{\bibfnamefont{H.~M.~P.}
  \bibnamefont{{Couchman}}}, \bibinfo{journal}{\mnras}
  \textbf{\bibinfo{volume}{341}}, \bibinfo{pages}{1311} (\bibinfo{year}{2003}),
  \eprint{arXiv:astro-ph/0207664}.

\bibitem[{\citenamefont{{Scoccimarro}}(2004)}]{2004PhRvD..70h3007S}
\bibinfo{author}{\bibfnamefont{R.}~\bibnamefont{{Scoccimarro}}},
  \bibinfo{journal}{\prd} \textbf{\bibinfo{volume}{70}},
  \bibinfo{pages}{083007} (\bibinfo{year}{2004}), \eprint{astro-ph/0407214}.

\bibitem[{\citenamefont{{Springel} et~al.}(2005)\citenamefont{{Springel},
  {White}, {Jenkins}, {Frenk}, {Yoshida}, {Gao}, {Navarro}, {Thacker},
  {Croton}, {Helly} et~al.}}]{2005Natur.435..629S}
\bibinfo{author}{\bibfnamefont{V.}~\bibnamefont{{Springel}}},
  \bibinfo{author}{\bibfnamefont{S.~D.~M.} \bibnamefont{{White}}},
  \bibinfo{author}{\bibfnamefont{A.}~\bibnamefont{{Jenkins}}},
  \bibinfo{author}{\bibfnamefont{C.~S.} \bibnamefont{{Frenk}}},
  \bibinfo{author}{\bibfnamefont{N.}~\bibnamefont{{Yoshida}}},
  \bibinfo{author}{\bibfnamefont{L.}~\bibnamefont{{Gao}}},
  \bibinfo{author}{\bibfnamefont{J.}~\bibnamefont{{Navarro}}},
  \bibinfo{author}{\bibfnamefont{R.}~\bibnamefont{{Thacker}}},
  \bibinfo{author}{\bibfnamefont{D.}~\bibnamefont{{Croton}}},
  \bibinfo{author}{\bibfnamefont{J.}~\bibnamefont{{Helly}}},
  \bibnamefont{et~al.}, \bibinfo{journal}{\nat} \textbf{\bibinfo{volume}{435}},
  \bibinfo{pages}{629} (\bibinfo{year}{2005}), \eprint{arXiv:astro-ph/0504097}.

\bibitem[{\citenamefont{{Valageas}}(2004)}]{2004A&A...421...23V}
\bibinfo{author}{\bibfnamefont{P.}~\bibnamefont{{Valageas}}},
  \bibinfo{journal}{\aap} \textbf{\bibinfo{volume}{421}}, \bibinfo{pages}{23}
  (\bibinfo{year}{2004}), \eprint{arXiv:astro-ph/0307008}.

\bibitem[{\citenamefont{{Valageas}}(2007)}]{2007A&A...476...31V}
\bibinfo{author}{\bibfnamefont{P.}~\bibnamefont{{Valageas}}},
  \bibinfo{journal}{\aap} \textbf{\bibinfo{volume}{476}}, \bibinfo{pages}{31}
  (\bibinfo{year}{2007}), \eprint{arXiv:0706.2593}.

\bibitem[{\citenamefont{{Valageas}}(2008)}]{2008A&A...484...79V}
\bibinfo{author}{\bibfnamefont{P.}~\bibnamefont{{Valageas}}},
  \bibinfo{journal}{\aap} \textbf{\bibinfo{volume}{484}}, \bibinfo{pages}{79}
  (\bibinfo{year}{2008}), \eprint{arXiv:0711.3407}.

\bibitem[{\citenamefont{{Crocce} and
  {Scoccimarro}}(2006{\natexlab{a}})}]{2006PhRvD..73f3519C}
\bibinfo{author}{\bibfnamefont{M.}~\bibnamefont{{Crocce}}} \bibnamefont{and}
  \bibinfo{author}{\bibfnamefont{R.}~\bibnamefont{{Scoccimarro}}},
  \bibinfo{journal}{Physical Review D} \textbf{\bibinfo{volume}{73}},
  \bibinfo{pages}{063519} (\bibinfo{year}{2006}{\natexlab{a}}),
  \eprint{arXiv:astro-ph/0509418}.

\bibitem[{\citenamefont{{Crocce} and
  {Scoccimarro}}(2006{\natexlab{b}})}]{2006PhRvD..73f3520C}
\bibinfo{author}{\bibfnamefont{M.}~\bibnamefont{{Crocce}}} \bibnamefont{and}
  \bibinfo{author}{\bibfnamefont{R.}~\bibnamefont{{Scoccimarro}}},
  \bibinfo{journal}{Physical Review D} \textbf{\bibinfo{volume}{73}},
  \bibinfo{pages}{063520} (\bibinfo{year}{2006}{\natexlab{b}}),
  \eprint{arXiv:astro-ph/0509419}.

\bibitem[{\citenamefont{{McDonald}}(2006{\natexlab{a}})}]{2006PhRvD..74j3512M}
\bibinfo{author}{\bibfnamefont{P.}~\bibnamefont{{McDonald}}},
  \bibinfo{journal}{\prd} \textbf{\bibinfo{volume}{74}},
  \bibinfo{pages}{103512} (\bibinfo{year}{2006}{\natexlab{a}}),
  \eprint{arXiv:astro-ph/0609413}.

\bibitem[{\citenamefont{{McDonald}}(2006{\natexlab{b}})}]{2006PhRvD..74l9901M}
\bibinfo{author}{\bibfnamefont{P.}~\bibnamefont{{McDonald}}},
  \bibinfo{journal}{\prd} \textbf{\bibinfo{volume}{74}},
  \bibinfo{pages}{129901(E)} (\bibinfo{year}{2006}{\natexlab{b}}).

\bibitem[{\citenamefont{{McDonald}}(2007)}]{2007PhRvD..75d3514M}
\bibinfo{author}{\bibfnamefont{P.}~\bibnamefont{{McDonald}}},
  \bibinfo{journal}{\prd} \textbf{\bibinfo{volume}{75}},
  \bibinfo{pages}{043514} (\bibinfo{year}{2007}),
  \eprint{arXiv:astro-ph/0606028}.

\bibitem[{\citenamefont{{Jeong} and {Komatsu}}(2006)}]{2006ApJ...651..619J}
\bibinfo{author}{\bibfnamefont{D.}~\bibnamefont{{Jeong}}} \bibnamefont{and}
  \bibinfo{author}{\bibfnamefont{E.}~\bibnamefont{{Komatsu}}},
  \bibinfo{journal}{\apj} \textbf{\bibinfo{volume}{651}}, \bibinfo{pages}{619}
  (\bibinfo{year}{2006}), \eprint{arXiv:astro-ph/0604075}.

\bibitem[{\citenamefont{{Jeong} and {Komatsu}}(2008)}]{2008arXiv0805.2632J}
\bibinfo{author}{\bibfnamefont{D.}~\bibnamefont{{Jeong}}} \bibnamefont{and}
  \bibinfo{author}{\bibfnamefont{E.}~\bibnamefont{{Komatsu}}},
  \bibinfo{journal}{ArXiv e-prints} \textbf{\bibinfo{volume}{0805.2632}}
  (\bibinfo{year}{2008}), \eprint{0805.2632}.

\bibitem[{\citenamefont{{Matarrese} and
  {Pietroni}}(2007)}]{2007JCAP...06..026M}
\bibinfo{author}{\bibfnamefont{S.}~\bibnamefont{{Matarrese}}} \bibnamefont{and}
  \bibinfo{author}{\bibfnamefont{M.}~\bibnamefont{{Pietroni}}},
  \bibinfo{journal}{Journal of Cosmology and Astro-Particle Physics}
  \textbf{\bibinfo{volume}{6}}, \bibinfo{pages}{26} (\bibinfo{year}{2007}),
  \eprint{arXiv:astro-ph/0703563}.

\bibitem[{\citenamefont{{Gaite} and
  {Dom{\'{\i}}nguez}}(2007)}]{2007JPhA...40.6849G}
\bibinfo{author}{\bibfnamefont{J.}~\bibnamefont{{Gaite}}} \bibnamefont{and}
  \bibinfo{author}{\bibfnamefont{A.}~\bibnamefont{{Dom{\'{\i}}nguez}}},
  \bibinfo{journal}{Journal of Physics A Mathematical General}
  \textbf{\bibinfo{volume}{40}}, \bibinfo{pages}{6849} (\bibinfo{year}{2007}),
  \eprint{arXiv:astro-ph/0610886}.

\bibitem[{\citenamefont{{Matarrese} and
  {Pietroni}}(2008)}]{2008MPLA...23...25M}
\bibinfo{author}{\bibfnamefont{S.}~\bibnamefont{{Matarrese}}} \bibnamefont{and}
  \bibinfo{author}{\bibfnamefont{M.}~\bibnamefont{{Pietroni}}},
  \bibinfo{journal}{Modern Physics Letters A} \textbf{\bibinfo{volume}{23}},
  \bibinfo{pages}{25} (\bibinfo{year}{2008}), \eprint{arXiv:astro-ph/0702653}.

\bibitem[{\citenamefont{{Matsubara}}(2008)}]{2008PhRvD..77f3530M}
\bibinfo{author}{\bibfnamefont{T.}~\bibnamefont{{Matsubara}}},
  \bibinfo{journal}{\prd} \textbf{\bibinfo{volume}{77}},
  \bibinfo{pages}{063530} (\bibinfo{year}{2008}), \eprint{arXiv:0711.2521}.

\bibitem[{\citenamefont{{Pietroni}}(2008)}]{2008arXiv0806.0971P}
\bibinfo{author}{\bibfnamefont{M.}~\bibnamefont{{Pietroni}}},
  \bibinfo{journal}{ArXiv e-prints} \textbf{\bibinfo{volume}{0806.0971}}
  (\bibinfo{year}{2008}), \eprint{0806.0971}.

\bibitem[{\citenamefont{{Bertschinger} and
  {Dekel}}(1989)}]{1989ApJ...336L...5B}
\bibinfo{author}{\bibfnamefont{E.}~\bibnamefont{{Bertschinger}}}
  \bibnamefont{and} \bibinfo{author}{\bibfnamefont{A.}~\bibnamefont{{Dekel}}},
  \bibinfo{journal}{\apjl} \textbf{\bibinfo{volume}{336}}, \bibinfo{pages}{L5}
  (\bibinfo{year}{1989}).

\bibitem[{\citenamefont{{Bertschinger} and
  {Dekel}}(1991)}]{1991ASPC...15...67B}
\bibinfo{author}{\bibfnamefont{E.}~\bibnamefont{{Bertschinger}}}
  \bibnamefont{and} \bibinfo{author}{\bibfnamefont{A.}~\bibnamefont{{Dekel}}},
  in \emph{\bibinfo{booktitle}{ASP Conf. Ser. 15: Large-Scale Structures and
  Peculiar Motions in the Universe}}, edited by
  \bibinfo{editor}{\bibfnamefont{D.~W.} \bibnamefont{{Latham}}}
  \bibnamefont{and} \bibinfo{editor}{\bibfnamefont{L.~A.~N.} \bibnamefont{{da
  Costa}}} (\bibinfo{year}{1991}), p.~\bibinfo{pages}{67}.

\bibitem[{\citenamefont{{Peebles}}(1989)}]{1989ApJ...344L..53P}
\bibinfo{author}{\bibfnamefont{P.~J.~E.} \bibnamefont{{Peebles}}},
  \bibinfo{journal}{\apjl} \textbf{\bibinfo{volume}{344}}, \bibinfo{pages}{L53}
  (\bibinfo{year}{1989}).

\bibitem[{\citenamefont{{Dekel} et~al.}(1990)\citenamefont{{Dekel},
  {Bertschinger}, and {Faber}}}]{1990ApJ...364..349D}
\bibinfo{author}{\bibfnamefont{A.}~\bibnamefont{{Dekel}}},
  \bibinfo{author}{\bibfnamefont{E.}~\bibnamefont{{Bertschinger}}},
  \bibnamefont{and} \bibinfo{author}{\bibfnamefont{S.~M.}
  \bibnamefont{{Faber}}}, \bibinfo{journal}{\apj}
  \textbf{\bibinfo{volume}{364}}, \bibinfo{pages}{349} (\bibinfo{year}{1990}).

\bibitem[{\citenamefont{{Kaiser} and {Stebbins}}(1991)}]{1991ASPC...15..111K}
\bibinfo{author}{\bibfnamefont{N.}~\bibnamefont{{Kaiser}}} \bibnamefont{and}
  \bibinfo{author}{\bibfnamefont{A.}~\bibnamefont{{Stebbins}}}, in
  \emph{\bibinfo{booktitle}{ASP Conf. Ser. 15: Large-Scale Structures and
  Peculiar Motions in the Universe}}, edited by
  \bibinfo{editor}{\bibfnamefont{D.~W.} \bibnamefont{{Latham}}}
  \bibnamefont{and} \bibinfo{editor}{\bibfnamefont{L.~A.~N.} \bibnamefont{{da
  Costa}}} (\bibinfo{year}{1991}), p. \bibinfo{pages}{111}.

\bibitem[{\citenamefont{{Hoffman} and {Ribak}}(1991)}]{1991ApJ...380L...5H}
\bibinfo{author}{\bibfnamefont{Y.}~\bibnamefont{{Hoffman}}} \bibnamefont{and}
  \bibinfo{author}{\bibfnamefont{E.}~\bibnamefont{{Ribak}}},
  \bibinfo{journal}{\apjl} \textbf{\bibinfo{volume}{380}}, \bibinfo{pages}{L5}
  (\bibinfo{year}{1991}).

\bibitem[{\citenamefont{{Weinberg}}(1992)}]{1992MNRAS.254..315W}
\bibinfo{author}{\bibfnamefont{D.~H.} \bibnamefont{{Weinberg}}},
  \bibinfo{journal}{\mnras} \textbf{\bibinfo{volume}{254}},
  \bibinfo{pages}{315} (\bibinfo{year}{1992}).

\bibitem[{\citenamefont{{Nusser} and {Dekel}}(1992)}]{1992ApJ...391..443N}
\bibinfo{author}{\bibfnamefont{A.}~\bibnamefont{{Nusser}}} \bibnamefont{and}
  \bibinfo{author}{\bibfnamefont{A.}~\bibnamefont{{Dekel}}},
  \bibinfo{journal}{\apj} \textbf{\bibinfo{volume}{391}}, \bibinfo{pages}{443}
  (\bibinfo{year}{1992}).

\bibitem[{\citenamefont{{Rybicki} and {Press}}(1992)}]{1992ApJ...398..169R}
\bibinfo{author}{\bibfnamefont{G.~B.} \bibnamefont{{Rybicki}}}
  \bibnamefont{and} \bibinfo{author}{\bibfnamefont{W.~H.}
  \bibnamefont{{Press}}}, \bibinfo{journal}{\apj}
  \textbf{\bibinfo{volume}{398}}, \bibinfo{pages}{169} (\bibinfo{year}{1992}).

\bibitem[{\citenamefont{{Gramann}}(1993)}]{1993ApJ...405..449G}
\bibinfo{author}{\bibfnamefont{M.}~\bibnamefont{{Gramann}}},
  \bibinfo{journal}{\apj} \textbf{\bibinfo{volume}{405}}, \bibinfo{pages}{449}
  (\bibinfo{year}{1993}).

\bibitem[{\citenamefont{{Ganon} and {Hoffman}}(1993)}]{1993ApJ...415L...5G}
\bibinfo{author}{\bibfnamefont{G.}~\bibnamefont{{Ganon}}} \bibnamefont{and}
  \bibinfo{author}{\bibfnamefont{Y.}~\bibnamefont{{Hoffman}}},
  \bibinfo{journal}{\apjl} \textbf{\bibinfo{volume}{415}}, \bibinfo{pages}{L5}
  (\bibinfo{year}{1993}).

\bibitem[{\citenamefont{{Bernardeau}}(1994)}]{1994A&A...291..697B}
\bibinfo{author}{\bibfnamefont{F.}~\bibnamefont{{Bernardeau}}},
  \bibinfo{journal}{\aap} \textbf{\bibinfo{volume}{291}}, \bibinfo{pages}{697}
  (\bibinfo{year}{1994}), \eprint{astro-ph/9403020}.

\bibitem[{\citenamefont{{Nusser} and {Davis}}(1994)}]{1994ApJ...421L...1N}
\bibinfo{author}{\bibfnamefont{A.}~\bibnamefont{{Nusser}}} \bibnamefont{and}
  \bibinfo{author}{\bibfnamefont{M.}~\bibnamefont{{Davis}}},
  \bibinfo{journal}{\apjl} \textbf{\bibinfo{volume}{421}}, \bibinfo{pages}{L1}
  (\bibinfo{year}{1994}), \eprint{astro-ph/9309009}.

\bibitem[{\citenamefont{{Lahav}}(1994)}]{1994ASPC...67..171L}
\bibinfo{author}{\bibfnamefont{O.}~\bibnamefont{{Lahav}}}, in
  \emph{\bibinfo{booktitle}{ASP Conf. Ser. 67: Unveiling Large-Scale Structures
  Behind the Milky Way}}, edited by
  \bibinfo{editor}{\bibfnamefont{C.}~\bibnamefont{{Balkowski}}}
  \bibnamefont{and} \bibinfo{editor}{\bibfnamefont{R.~C.}
  \bibnamefont{{Kraan-Korteweg}}} (\bibinfo{year}{1994}), p.
  \bibinfo{pages}{171}.

\bibitem[{\citenamefont{{Lahav} et~al.}(1994)\citenamefont{{Lahav}, {Fisher},
  {Hoffman}, {Scharf}, and {Zaroubi}}}]{1994ApJ...423L..93L}
\bibinfo{author}{\bibfnamefont{O.}~\bibnamefont{{Lahav}}},
  \bibinfo{author}{\bibfnamefont{K.~B.} \bibnamefont{{Fisher}}},
  \bibinfo{author}{\bibfnamefont{Y.}~\bibnamefont{{Hoffman}}},
  \bibinfo{author}{\bibfnamefont{C.~A.} \bibnamefont{{Scharf}}},
  \bibnamefont{and}
  \bibinfo{author}{\bibfnamefont{S.}~\bibnamefont{{Zaroubi}}},
  \bibinfo{journal}{\apjl} \textbf{\bibinfo{volume}{423}}, \bibinfo{pages}{L93}
  (\bibinfo{year}{1994}), \eprint{astro-ph/9311059}.

\bibitem[{\citenamefont{{Fisher} et~al.}(1995)\citenamefont{{Fisher}, {Lahav},
  {Hoffman}, {Lynden-Bell}, and {Zaroubi}}}]{1995MNRAS.272..885F}
\bibinfo{author}{\bibfnamefont{K.~B.} \bibnamefont{{Fisher}}},
  \bibinfo{author}{\bibfnamefont{O.}~\bibnamefont{{Lahav}}},
  \bibinfo{author}{\bibfnamefont{Y.}~\bibnamefont{{Hoffman}}},
  \bibinfo{author}{\bibfnamefont{D.}~\bibnamefont{{Lynden-Bell}}},
  \bibnamefont{and}
  \bibinfo{author}{\bibfnamefont{S.}~\bibnamefont{{Zaroubi}}},
  \bibinfo{journal}{\mnras} \textbf{\bibinfo{volume}{272}},
  \bibinfo{pages}{885} (\bibinfo{year}{1995}), \eprint{astro-ph/9406009}.

\bibitem[{\citenamefont{{Sheth}}(1995)}]{1995MNRAS.277..933S}
\bibinfo{author}{\bibfnamefont{R.~K.} \bibnamefont{{Sheth}}},
  \bibinfo{journal}{\mnras} \textbf{\bibinfo{volume}{277}},
  \bibinfo{pages}{933} (\bibinfo{year}{1995}), \eprint{astro-ph/9511096}.

\bibitem[{\citenamefont{{Zaroubi} et~al.}(1995)\citenamefont{{Zaroubi},
  {Hoffman}, {Fisher}, and {Lahav}}}]{1995ApJ...449..446Z}
\bibinfo{author}{\bibfnamefont{S.}~\bibnamefont{{Zaroubi}}},
  \bibinfo{author}{\bibfnamefont{Y.}~\bibnamefont{{Hoffman}}},
  \bibinfo{author}{\bibfnamefont{K.~B.} \bibnamefont{{Fisher}}},
  \bibnamefont{and} \bibinfo{author}{\bibfnamefont{O.}~\bibnamefont{{Lahav}}},
  \bibinfo{journal}{\apj} \textbf{\bibinfo{volume}{449}}, \bibinfo{pages}{446}
  (\bibinfo{year}{1995}), \eprint{astro-ph/9410080}.

\bibitem[{\citenamefont{{Tegmark} and {Bromley}}(1995)}]{1995ApJ...453..533T}
\bibinfo{author}{\bibfnamefont{M.}~\bibnamefont{{Tegmark}}} \bibnamefont{and}
  \bibinfo{author}{\bibfnamefont{B.~C.} \bibnamefont{{Bromley}}},
  \bibinfo{journal}{\apj} \textbf{\bibinfo{volume}{453}}, \bibinfo{pages}{533}
  (\bibinfo{year}{1995}), \eprint{astro-ph/9409038}.

\bibitem[{\citenamefont{{Croft} and {Gaztanaga}}(1997)}]{1997MNRAS.285..793C}
\bibinfo{author}{\bibfnamefont{R.~A.~C.} \bibnamefont{{Croft}}}
  \bibnamefont{and}
  \bibinfo{author}{\bibfnamefont{E.}~\bibnamefont{{Gaztanaga}}},
  \bibinfo{journal}{\mnras} \textbf{\bibinfo{volume}{285}},
  \bibinfo{pages}{793} (\bibinfo{year}{1997}), \eprint{astro-ph/9602100}.

\bibitem[{\citenamefont{{Narayanan} and
  {Weinberg}}(1998)}]{1998ApJ...508..440N}
\bibinfo{author}{\bibfnamefont{V.~K.} \bibnamefont{{Narayanan}}}
  \bibnamefont{and} \bibinfo{author}{\bibfnamefont{D.~H.}
  \bibnamefont{{Weinberg}}}, \bibinfo{journal}{\apj}
  \textbf{\bibinfo{volume}{508}}, \bibinfo{pages}{440} (\bibinfo{year}{1998}),
  \eprint{astro-ph/9806238}.

\bibitem[{\citenamefont{{Pen}}(1998)}]{1998ApJ...504..601P}
\bibinfo{author}{\bibfnamefont{U.-L.} \bibnamefont{{Pen}}},
  \bibinfo{journal}{\apj} \textbf{\bibinfo{volume}{504}}, \bibinfo{pages}{601}
  (\bibinfo{year}{1998}), \eprint{astro-ph/9711180}.

\bibitem[{\citenamefont{{Seljak}}(1998{\natexlab{a}})}]{1998ApJ...503..492S}
\bibinfo{author}{\bibfnamefont{U.}~\bibnamefont{{Seljak}}},
  \bibinfo{journal}{\apj} \textbf{\bibinfo{volume}{503}}, \bibinfo{pages}{492}
  (\bibinfo{year}{1998}{\natexlab{a}}), \eprint{astro-ph/9710269}.

\bibitem[{\citenamefont{{Seljak}}(1998{\natexlab{b}})}]{1998ApJ...506...64S}
\bibinfo{author}{\bibfnamefont{U.}~\bibnamefont{{Seljak}}},
  \bibinfo{journal}{\apj} \textbf{\bibinfo{volume}{506}}, \bibinfo{pages}{64}
  (\bibinfo{year}{1998}{\natexlab{b}}), \eprint{astro-ph/9711124}.

\bibitem[{\citenamefont{{Bistolas} and {Hoffman}}(1998)}]{1998ApJ...492..439B}
\bibinfo{author}{\bibfnamefont{V.}~\bibnamefont{{Bistolas}}} \bibnamefont{and}
  \bibinfo{author}{\bibfnamefont{Y.}~\bibnamefont{{Hoffman}}},
  \bibinfo{journal}{\apj} \textbf{\bibinfo{volume}{492}}, \bibinfo{pages}{439}
  (\bibinfo{year}{1998}), \eprint{astro-ph/9707243}.

\bibitem[{\citenamefont{{Taylor} and {Valentine}}(1999)}]{1999MNRAS.306..491T}
\bibinfo{author}{\bibfnamefont{A.}~\bibnamefont{{Taylor}}} \bibnamefont{and}
  \bibinfo{author}{\bibfnamefont{H.}~\bibnamefont{{Valentine}}},
  \bibinfo{journal}{\mnras} \textbf{\bibinfo{volume}{306}},
  \bibinfo{pages}{491} (\bibinfo{year}{1999}), \eprint{astro-ph/9901171}.

\bibitem[{\citenamefont{{Narayanan} and {Croft}}(1999)}]{1999ApJ...515..471N}
\bibinfo{author}{\bibfnamefont{V.~K.} \bibnamefont{{Narayanan}}}
  \bibnamefont{and} \bibinfo{author}{\bibfnamefont{R.~A.~C.}
  \bibnamefont{{Croft}}}, \bibinfo{journal}{\apj}
  \textbf{\bibinfo{volume}{515}}, \bibinfo{pages}{471} (\bibinfo{year}{1999}),
  \eprint{astro-ph/9806255}.

\bibitem[{\citenamefont{{Zaroubi} et~al.}(1999)\citenamefont{{Zaroubi},
  {Hoffman}, and {Dekel}}}]{1999ApJ...520..413Z}
\bibinfo{author}{\bibfnamefont{S.}~\bibnamefont{{Zaroubi}}},
  \bibinfo{author}{\bibfnamefont{Y.}~\bibnamefont{{Hoffman}}},
  \bibnamefont{and} \bibinfo{author}{\bibfnamefont{A.}~\bibnamefont{{Dekel}}},
  \bibinfo{journal}{\apj} \textbf{\bibinfo{volume}{520}}, \bibinfo{pages}{413}
  (\bibinfo{year}{1999}), \eprint{astro-ph/9810279}.

\bibitem[{\citenamefont{{Goldberg} and
  {Spergel}}(2000{\natexlab{a}})}]{2000ASPC..201..282G}
\bibinfo{author}{\bibfnamefont{D.~M.} \bibnamefont{{Goldberg}}}
  \bibnamefont{and} \bibinfo{author}{\bibfnamefont{D.~N.}
  \bibnamefont{{Spergel}}}, in \emph{\bibinfo{booktitle}{ASP Conf. Ser. 201:
  Cosmic Flows Workshop}}, edited by
  \bibinfo{editor}{\bibfnamefont{S.}~\bibnamefont{{Courteau}}}
  \bibnamefont{and} \bibinfo{editor}{\bibfnamefont{J.}~\bibnamefont{{Willick}}}
  (\bibinfo{year}{2000}{\natexlab{a}}), p. \bibinfo{pages}{282}.

\bibitem[{\citenamefont{{Goldberg} and
  {Spergel}}(2000{\natexlab{b}})}]{2000ApJ...544...21G}
\bibinfo{author}{\bibfnamefont{D.~M.} \bibnamefont{{Goldberg}}}
  \bibnamefont{and} \bibinfo{author}{\bibfnamefont{D.~N.}
  \bibnamefont{{Spergel}}}, \bibinfo{journal}{\apj}
  \textbf{\bibinfo{volume}{544}}, \bibinfo{pages}{21}
  (\bibinfo{year}{2000}{\natexlab{b}}), \eprint{astro-ph/9912408}.

\bibitem[{\citenamefont{{Kudlicki} et~al.}(2000)\citenamefont{{Kudlicki},
  {Chodorowski}, {Plewa}, and {R{\'o}{\.z}yczka}}}]{2000MNRAS.316..464K}
\bibinfo{author}{\bibfnamefont{A.}~\bibnamefont{{Kudlicki}}},
  \bibinfo{author}{\bibfnamefont{M.}~\bibnamefont{{Chodorowski}}},
  \bibinfo{author}{\bibfnamefont{T.}~\bibnamefont{{Plewa}}}, \bibnamefont{and}
  \bibinfo{author}{\bibfnamefont{M.}~\bibnamefont{{R{\'o}{\.z}yczka}}},
  \bibinfo{journal}{\mnras} \textbf{\bibinfo{volume}{316}},
  \bibinfo{pages}{464} (\bibinfo{year}{2000}), \eprint{astro-ph/9910018}.

\bibitem[{\citenamefont{{Basilakos} and {Plionis}}(2001)}]{2001ApJ...550..522B}
\bibinfo{author}{\bibfnamefont{S.}~\bibnamefont{{Basilakos}}} \bibnamefont{and}
  \bibinfo{author}{\bibfnamefont{M.}~\bibnamefont{{Plionis}}},
  \bibinfo{journal}{\apj} \textbf{\bibinfo{volume}{550}}, \bibinfo{pages}{522}
  (\bibinfo{year}{2001}), \eprint{astro-ph/0011265}.

\bibitem[{\citenamefont{{Goldberg}}(2001{\natexlab{a}})}]{2001ApJ...552..413G}
\bibinfo{author}{\bibfnamefont{D.~M.} \bibnamefont{{Goldberg}}},
  \bibinfo{journal}{\apj} \textbf{\bibinfo{volume}{552}}, \bibinfo{pages}{413}
  (\bibinfo{year}{2001}{\natexlab{a}}), \eprint{astro-ph/0008266}.

\bibitem[{\citenamefont{{Frisch} et~al.}(2002)\citenamefont{{Frisch},
  {Matarrese}, {Mohayaee}, and {Sobolevski}}}]{2002Natur.417..260F}
\bibinfo{author}{\bibfnamefont{U.}~\bibnamefont{{Frisch}}},
  \bibinfo{author}{\bibfnamefont{S.}~\bibnamefont{{Matarrese}}},
  \bibinfo{author}{\bibfnamefont{R.}~\bibnamefont{{Mohayaee}}},
  \bibnamefont{and}
  \bibinfo{author}{\bibfnamefont{A.}~\bibnamefont{{Sobolevski}}},
  \bibinfo{journal}{\nat} \textbf{\bibinfo{volume}{417}}, \bibinfo{pages}{260}
  (\bibinfo{year}{2002}), \eprint{arXiv:astro-ph/0109483}.

\bibitem[{\citenamefont{{Zaroubi}}(2002{\natexlab{b}})}]{2002MNRAS.331..901Z}
\bibinfo{author}{\bibfnamefont{S.}~\bibnamefont{{Zaroubi}}},
  \bibinfo{journal}{\mnras} \textbf{\bibinfo{volume}{331}},
  \bibinfo{pages}{901} (\bibinfo{year}{2002}{\natexlab{b}}),
  \eprint{astro-ph/0010561}.

\bibitem[{\citenamefont{{Brenier} et~al.}(2003)\citenamefont{{Brenier},
  {Frisch}, {H{\'e}non}, {Loeper}, {Matarrese}, {Mohayaee}, and {Sobolevski{\u
  i}}}}]{2003MNRAS.346..501B}
\bibinfo{author}{\bibfnamefont{Y.}~\bibnamefont{{Brenier}}},
  \bibinfo{author}{\bibfnamefont{U.}~\bibnamefont{{Frisch}}},
  \bibinfo{author}{\bibfnamefont{M.}~\bibnamefont{{H{\'e}non}}},
  \bibinfo{author}{\bibfnamefont{G.}~\bibnamefont{{Loeper}}},
  \bibinfo{author}{\bibfnamefont{S.}~\bibnamefont{{Matarrese}}},
  \bibinfo{author}{\bibfnamefont{R.}~\bibnamefont{{Mohayaee}}},
  \bibnamefont{and}
  \bibinfo{author}{\bibfnamefont{A.}~\bibnamefont{{Sobolevski{\u i}}}},
  \bibinfo{journal}{\mnras} \textbf{\bibinfo{volume}{346}},
  \bibinfo{pages}{501} (\bibinfo{year}{2003}), \eprint{astro-ph/0304214}.

\bibitem[{\citenamefont{{Mohayaee} et~al.}(2003)\citenamefont{{Mohayaee},
  {Frisch}, {Matarrese}, and {Sobolevskii}}}]{2003A&A...406..393M}
\bibinfo{author}{\bibfnamefont{R.}~\bibnamefont{{Mohayaee}}},
  \bibinfo{author}{\bibfnamefont{U.}~\bibnamefont{{Frisch}}},
  \bibinfo{author}{\bibfnamefont{S.}~\bibnamefont{{Matarrese}}},
  \bibnamefont{and}
  \bibinfo{author}{\bibfnamefont{A.}~\bibnamefont{{Sobolevskii}}},
  \bibinfo{journal}{\aap} \textbf{\bibinfo{volume}{406}}, \bibinfo{pages}{393}
  (\bibinfo{year}{2003}), \eprint{arXiv:astro-ph/0301641}.

\bibitem[{\citenamefont{{Mohayaee} et~al.}(2004)\citenamefont{{Mohayaee},
  {Tully}, and {Frisch}}}]{2004astro.ph.10063M}
\bibinfo{author}{\bibfnamefont{R.}~\bibnamefont{{Mohayaee}}},
  \bibinfo{author}{\bibfnamefont{B.}~\bibnamefont{{Tully}}}, \bibnamefont{and}
  \bibinfo{author}{\bibfnamefont{U.}~\bibnamefont{{Frisch}}},
  \bibinfo{journal}{ArXiv Astrophysics e-prints}  (\bibinfo{year}{2004}),
  \eprint{astro-ph/0410063}.

\bibitem[{\citenamefont{{Botzler} et~al.}(2004)\citenamefont{{Botzler},
  {Snigula}, {Bender}, and {Hopp}}}]{2004MNRAS.349..425B}
\bibinfo{author}{\bibfnamefont{C.~S.} \bibnamefont{{Botzler}}},
  \bibinfo{author}{\bibfnamefont{J.}~\bibnamefont{{Snigula}}},
  \bibinfo{author}{\bibfnamefont{R.}~\bibnamefont{{Bender}}}, \bibnamefont{and}
  \bibinfo{author}{\bibfnamefont{U.}~\bibnamefont{{Hopp}}},
  \bibinfo{journal}{\mnras} \textbf{\bibinfo{volume}{349}},
  \bibinfo{pages}{425} (\bibinfo{year}{2004}), \eprint{arXiv:astro-ph/0312018}.

\bibitem[{\citenamefont{{Mohayaee} and {Tully}}(2005)}]{2005ApJ...635L.113M}
\bibinfo{author}{\bibfnamefont{R.}~\bibnamefont{{Mohayaee}}} \bibnamefont{and}
  \bibinfo{author}{\bibfnamefont{R.~B.} \bibnamefont{{Tully}}},
  \bibinfo{journal}{\apjl} \textbf{\bibinfo{volume}{635}},
  \bibinfo{pages}{L113} (\bibinfo{year}{2005}), \eprint{astro-ph/0509313}.

\bibitem[{\citenamefont{{Mohayaee} et~al.}(2006)\citenamefont{{Mohayaee},
  {Mathis}, {Colombi}, and {Silk}}}]{2006MNRAS.365..939M}
\bibinfo{author}{\bibfnamefont{R.}~\bibnamefont{{Mohayaee}}},
  \bibinfo{author}{\bibfnamefont{H.}~\bibnamefont{{Mathis}}},
  \bibinfo{author}{\bibfnamefont{S.}~\bibnamefont{{Colombi}}},
  \bibnamefont{and} \bibinfo{author}{\bibfnamefont{J.}~\bibnamefont{{Silk}}},
  \bibinfo{journal}{\mnras} \textbf{\bibinfo{volume}{365}},
  \bibinfo{pages}{939} (\bibinfo{year}{2006}), \eprint{astro-ph/0501217}.

\bibitem[{\citenamefont{{Icke} and {van de
  Weygaert}}(1991)}]{1991QJRAS..32...85I}
\bibinfo{author}{\bibfnamefont{V.}~\bibnamefont{{Icke}}} \bibnamefont{and}
  \bibinfo{author}{\bibfnamefont{R.}~\bibnamefont{{van de Weygaert}}},
  \bibinfo{journal}{\qjras} \textbf{\bibinfo{volume}{32}}, \bibinfo{pages}{85}
  (\bibinfo{year}{1991}).

\bibitem[{\citenamefont{{Ikeuchi} and {Turner}}(1991)}]{1991MNRAS.250..519I}
\bibinfo{author}{\bibfnamefont{S.}~\bibnamefont{{Ikeuchi}}} \bibnamefont{and}
  \bibinfo{author}{\bibfnamefont{E.~L.} \bibnamefont{{Turner}}},
  \bibinfo{journal}{\mnras} \textbf{\bibinfo{volume}{250}},
  \bibinfo{pages}{519} (\bibinfo{year}{1991}).

\bibitem[{\citenamefont{{Bernardeau} and {van de
  Weygaert}}(1996)}]{1996MNRAS.279..693B}
\bibinfo{author}{\bibfnamefont{F.}~\bibnamefont{{Bernardeau}}}
  \bibnamefont{and} \bibinfo{author}{\bibfnamefont{R.}~\bibnamefont{{van de
  Weygaert}}}, \bibinfo{journal}{\mnras} \textbf{\bibinfo{volume}{279}},
  \bibinfo{pages}{693} (\bibinfo{year}{1996}).

\bibitem[{\citenamefont{{Schaap} and {van de
  Weygaert}}(2000)}]{2000A&A...363L..29S}
\bibinfo{author}{\bibfnamefont{W.~E.} \bibnamefont{{Schaap}}} \bibnamefont{and}
  \bibinfo{author}{\bibfnamefont{R.}~\bibnamefont{{van de Weygaert}}},
  \bibinfo{journal}{\aap} \textbf{\bibinfo{volume}{363}}, \bibinfo{pages}{L29}
  (\bibinfo{year}{2000}), \eprint{astro-ph/0011007}.

\bibitem[{\citenamefont{{van de Weygaert} and
  {Schaap}}(2001)}]{2001misk.conf..268V}
\bibinfo{author}{\bibfnamefont{R.}~\bibnamefont{{van de Weygaert}}}
  \bibnamefont{and} \bibinfo{author}{\bibfnamefont{W.}~\bibnamefont{{Schaap}}},
  in \emph{\bibinfo{booktitle}{Mining the Sky}}, edited by
  \bibinfo{editor}{\bibfnamefont{A.~J.} \bibnamefont{{Banday}}},
  \bibinfo{editor}{\bibfnamefont{S.}~\bibnamefont{{Zaroubi}}},
  \bibnamefont{and}
  \bibinfo{editor}{\bibfnamefont{M.}~\bibnamefont{{Bartelmann}}}
  (\bibinfo{year}{2001}), p. \bibinfo{pages}{268}.

\bibitem[{\citenamefont{{Ramella} et~al.}(2001)\citenamefont{{Ramella},
  {Boschin}, {Fadda}, and {Nonino}}}]{2001A&A...368..776R}
\bibinfo{author}{\bibfnamefont{M.}~\bibnamefont{{Ramella}}},
  \bibinfo{author}{\bibfnamefont{W.}~\bibnamefont{{Boschin}}},
  \bibinfo{author}{\bibfnamefont{D.}~\bibnamefont{{Fadda}}}, \bibnamefont{and}
  \bibinfo{author}{\bibfnamefont{M.}~\bibnamefont{{Nonino}}},
  \bibinfo{journal}{\aap} \textbf{\bibinfo{volume}{368}}, \bibinfo{pages}{776}
  (\bibinfo{year}{2001}), \eprint{arXiv:astro-ph/0101411}.

\bibitem[{\citenamefont{{Zaninetti}}(2006)}]{2006ChJAA...6..387Z}
\bibinfo{author}{\bibfnamefont{L.}~\bibnamefont{{Zaninetti}}},
  \bibinfo{journal}{Chinese Journal of Astronomy and Astrophysics}
  \textbf{\bibinfo{volume}{6}}, \bibinfo{pages}{387} (\bibinfo{year}{2006}),
  \eprint{arXiv:astro-ph/0602431}.

\bibitem[{\citenamefont{{Bertschinger}
  et~al.}(1990)\citenamefont{{Bertschinger}, {Dekel}, {Faber}, {Dressler}, and
  {Burstein}}}]{1990ApJ...364..370B}
\bibinfo{author}{\bibfnamefont{E.}~\bibnamefont{{Bertschinger}}},
  \bibinfo{author}{\bibfnamefont{A.}~\bibnamefont{{Dekel}}},
  \bibinfo{author}{\bibfnamefont{S.~M.} \bibnamefont{{Faber}}},
  \bibinfo{author}{\bibfnamefont{A.}~\bibnamefont{{Dressler}}},
  \bibnamefont{and}
  \bibinfo{author}{\bibfnamefont{D.}~\bibnamefont{{Burstein}}},
  \bibinfo{journal}{\apj} \textbf{\bibinfo{volume}{364}}, \bibinfo{pages}{370}
  (\bibinfo{year}{1990}).

\bibitem[{\citenamefont{{Yahil} et~al.}(1991)\citenamefont{{Yahil}, {Strauss},
  {Davis}, and {Huchra}}}]{1991ApJ...372..380Y}
\bibinfo{author}{\bibfnamefont{A.}~\bibnamefont{{Yahil}}},
  \bibinfo{author}{\bibfnamefont{M.~A.} \bibnamefont{{Strauss}}},
  \bibinfo{author}{\bibfnamefont{M.}~\bibnamefont{{Davis}}}, \bibnamefont{and}
  \bibinfo{author}{\bibfnamefont{J.~P.} \bibnamefont{{Huchra}}},
  \bibinfo{journal}{\apj} \textbf{\bibinfo{volume}{372}}, \bibinfo{pages}{380}
  (\bibinfo{year}{1991}).

\bibitem[{\citenamefont{{Fisher} et~al.}(1994)\citenamefont{{Fisher}, {Scharf},
  and {Lahav}}}]{WienerFSL}
\bibinfo{author}{\bibfnamefont{K.~B.} \bibnamefont{{Fisher}}},
  \bibinfo{author}{\bibfnamefont{C.~A.} \bibnamefont{{Scharf}}},
  \bibnamefont{and} \bibinfo{author}{\bibfnamefont{O.}~\bibnamefont{{Lahav}}},
  \bibinfo{journal}{\mnras} \textbf{\bibinfo{volume}{266}},
  \bibinfo{pages}{219} (\bibinfo{year}{1994}), \eprint{astro-ph/9309027}.

\bibitem[{\citenamefont{{Shaya} et~al.}(1995)\citenamefont{{Shaya}, {Peebles},
  and {Tully}}}]{1995ApJ...454...15S}
\bibinfo{author}{\bibfnamefont{E.~J.} \bibnamefont{{Shaya}}},
  \bibinfo{author}{\bibfnamefont{P.~J.~E.} \bibnamefont{{Peebles}}},
  \bibnamefont{and} \bibinfo{author}{\bibfnamefont{R.~B.}
  \bibnamefont{{Tully}}}, \bibinfo{journal}{\apj}
  \textbf{\bibinfo{volume}{454}}, \bibinfo{pages}{15} (\bibinfo{year}{1995}),
  \eprint{astro-ph/9506144}.

\bibitem[{\citenamefont{{Branchini} et~al.}(1996)\citenamefont{{Branchini},
  {Plionis}, and {Sciama}}}]{1996ApJ...461L..17B}
\bibinfo{author}{\bibfnamefont{E.}~\bibnamefont{{Branchini}}},
  \bibinfo{author}{\bibfnamefont{M.}~\bibnamefont{{Plionis}}},
  \bibnamefont{and} \bibinfo{author}{\bibfnamefont{D.~W.}
  \bibnamefont{{Sciama}}}, \bibinfo{journal}{\apjl}
  \textbf{\bibinfo{volume}{461}}, \bibinfo{pages}{L17} (\bibinfo{year}{1996}),
  \eprint{astro-ph/9512055}.

\bibitem[{\citenamefont{{Webster} et~al.}(1997)\citenamefont{{Webster},
  {Lahav}, and {Fisher}}}]{1997MNRAS.287..425W}
\bibinfo{author}{\bibfnamefont{M.}~\bibnamefont{{Webster}}},
  \bibinfo{author}{\bibfnamefont{O.}~\bibnamefont{{Lahav}}}, \bibnamefont{and}
  \bibinfo{author}{\bibfnamefont{K.}~\bibnamefont{{Fisher}}},
  \bibinfo{journal}{\mnras} \textbf{\bibinfo{volume}{287}},
  \bibinfo{pages}{425} (\bibinfo{year}{1997}), \eprint{astro-ph/9608021}.

\bibitem[{\citenamefont{{Yess} et~al.}(1997)\citenamefont{{Yess}, {Shandarin},
  and {Fisher}}}]{1997ApJ...474..553Y}
\bibinfo{author}{\bibfnamefont{C.}~\bibnamefont{{Yess}}},
  \bibinfo{author}{\bibfnamefont{S.~F.} \bibnamefont{{Shandarin}}},
  \bibnamefont{and} \bibinfo{author}{\bibfnamefont{K.~B.}
  \bibnamefont{{Fisher}}}, \bibinfo{journal}{\apj}
  \textbf{\bibinfo{volume}{474}}, \bibinfo{pages}{553} (\bibinfo{year}{1997}),
  \eprint{astro-ph/9605041}.

\bibitem[{\citenamefont{{Schmoldt} et~al.}(1999)\citenamefont{{Schmoldt},
  {Saar}, {Saha}, {Branchini}, {Efstathiou}, {Frenk}, {Keeble}, {Maddox},
  {McMahon}, {Oliver} et~al.}}]{1999AJ....118.1146S}
\bibinfo{author}{\bibfnamefont{I.~M.} \bibnamefont{{Schmoldt}}},
  \bibinfo{author}{\bibfnamefont{V.}~\bibnamefont{{Saar}}},
  \bibinfo{author}{\bibfnamefont{P.}~\bibnamefont{{Saha}}},
  \bibinfo{author}{\bibfnamefont{E.}~\bibnamefont{{Branchini}}},
  \bibinfo{author}{\bibfnamefont{G.~P.} \bibnamefont{{Efstathiou}}},
  \bibinfo{author}{\bibfnamefont{C.~S.} \bibnamefont{{Frenk}}},
  \bibinfo{author}{\bibfnamefont{O.}~\bibnamefont{{Keeble}}},
  \bibinfo{author}{\bibfnamefont{S.}~\bibnamefont{{Maddox}}},
  \bibinfo{author}{\bibfnamefont{R.}~\bibnamefont{{McMahon}}},
  \bibinfo{author}{\bibfnamefont{S.}~\bibnamefont{{Oliver}}},
  \bibnamefont{et~al.}, \bibinfo{journal}{\apj} \textbf{\bibinfo{volume}{118}},
  \bibinfo{pages}{1146} (\bibinfo{year}{1999}), \eprint{astro-ph/9906035}.

\bibitem[{\citenamefont{{Nusser} and {Haehnelt}}(1999)}]{1999MNRAS.303..179N}
\bibinfo{author}{\bibfnamefont{A.}~\bibnamefont{{Nusser}}} \bibnamefont{and}
  \bibinfo{author}{\bibfnamefont{M.}~\bibnamefont{{Haehnelt}}},
  \bibinfo{journal}{\mnras} \textbf{\bibinfo{volume}{303}},
  \bibinfo{pages}{179} (\bibinfo{year}{1999}), \eprint{astro-ph/9806109}.

\bibitem[{\citenamefont{Tegmark and Bromley}(1999)}]{tegmark-1999-518}
\bibinfo{author}{\bibfnamefont{M.}~\bibnamefont{Tegmark}} \bibnamefont{and}
  \bibinfo{author}{\bibfnamefont{B.~C.} \bibnamefont{Bromley}},
  \bibinfo{journal}{The Astrophysical Journal} \textbf{\bibinfo{volume}{518}},
  \bibinfo{pages}{L69} (\bibinfo{year}{1999}),
  \urlprefix\url{http://www.citebase.org/abstract?id=oai:arXiv.org:astro-ph/98%
09324}.

\bibitem[{\citenamefont{{Hoffman} and {Zaroubi}}(2000)}]{2000ApJ...535L...5H}
\bibinfo{author}{\bibfnamefont{Y.}~\bibnamefont{{Hoffman}}} \bibnamefont{and}
  \bibinfo{author}{\bibfnamefont{S.}~\bibnamefont{{Zaroubi}}},
  \bibinfo{journal}{\apjl} \textbf{\bibinfo{volume}{535}}, \bibinfo{pages}{L5}
  (\bibinfo{year}{2000}), \eprint{astro-ph/0003306}.

\bibitem[{\citenamefont{{Goldberg}}(2001{\natexlab{b}})}]{2001ApJ...550...87G}
\bibinfo{author}{\bibfnamefont{D.~M.} \bibnamefont{{Goldberg}}},
  \bibinfo{journal}{\apj} \textbf{\bibinfo{volume}{550}}, \bibinfo{pages}{87}
  (\bibinfo{year}{2001}{\natexlab{b}}), \eprint{astro-ph/0009046}.

\bibitem[{\citenamefont{{Mathis} et~al.}(2002)\citenamefont{{Mathis}, {Lemson},
  {Springel}, {Kauffmann}, {White}, {Eldar}, and
  {Dekel}}}]{2002MNRAS.333..739M}
\bibinfo{author}{\bibfnamefont{H.}~\bibnamefont{{Mathis}}},
  \bibinfo{author}{\bibfnamefont{G.}~\bibnamefont{{Lemson}}},
  \bibinfo{author}{\bibfnamefont{V.}~\bibnamefont{{Springel}}},
  \bibinfo{author}{\bibfnamefont{G.}~\bibnamefont{{Kauffmann}}},
  \bibinfo{author}{\bibfnamefont{S.~D.~M.} \bibnamefont{{White}}},
  \bibinfo{author}{\bibfnamefont{A.}~\bibnamefont{{Eldar}}}, \bibnamefont{and}
  \bibinfo{author}{\bibfnamefont{A.}~\bibnamefont{{Dekel}}},
  \bibinfo{journal}{\mnras} \textbf{\bibinfo{volume}{333}},
  \bibinfo{pages}{739} (\bibinfo{year}{2002}), \eprint{astro-ph/0111099}.

\bibitem[{\citenamefont{{Erdo{\u g}du} et~al.}(2004)\citenamefont{{Erdo{\u
  g}du}, {Lahav}, {Zaroubi}, and {et al.}}}]{2004MNRAS.352..939E}
\bibinfo{author}{\bibfnamefont{P.}~\bibnamefont{{Erdo{\u g}du}}},
  \bibinfo{author}{\bibfnamefont{O.}~\bibnamefont{{Lahav}}},
  \bibinfo{author}{\bibfnamefont{S.}~\bibnamefont{{Zaroubi}}},
  \bibnamefont{and} \bibinfo{author}{\bibnamefont{{et al.}}},
  \bibinfo{journal}{\mnras} \textbf{\bibinfo{volume}{352}},
  \bibinfo{pages}{939} (\bibinfo{year}{2004}), \eprint{astro-ph/0312546}.

\bibitem[{\citenamefont{{Vogeley} et~al.}(2004)\citenamefont{{Vogeley},
  {Hoyle}, {Rojas}, and {Goldberg}}}]{2004ogci.conf....5V}
\bibinfo{author}{\bibfnamefont{M.~S.} \bibnamefont{{Vogeley}}},
  \bibinfo{author}{\bibfnamefont{F.}~\bibnamefont{{Hoyle}}},
  \bibinfo{author}{\bibfnamefont{R.~R.} \bibnamefont{{Rojas}}},
  \bibnamefont{and} \bibinfo{author}{\bibfnamefont{D.~M.}
  \bibnamefont{{Goldberg}}}, in \emph{\bibinfo{booktitle}{IAU Colloq. 195:
  Outskirts of Galaxy Clusters: Intense Life in the Suburbs}}, edited by
  \bibinfo{editor}{\bibfnamefont{A.}~\bibnamefont{{Diaferio}}}
  (\bibinfo{year}{2004}), pp. \bibinfo{pages}{5--11}.

\bibitem[{\citenamefont{{Huchra} et~al.}(2005)\citenamefont{{Huchra},
  {Jarrett}, {Skrutskie}, {Cutri}, {Schneider}, {Macri}, {Steining}, {Mader},
  {Martimbeau}, and {George}}}]{2005ASPC..329..135H}
\bibinfo{author}{\bibfnamefont{J.}~\bibnamefont{{Huchra}}},
  \bibinfo{author}{\bibfnamefont{T.}~\bibnamefont{{Jarrett}}},
  \bibinfo{author}{\bibfnamefont{M.}~\bibnamefont{{Skrutskie}}},
  \bibinfo{author}{\bibfnamefont{R.}~\bibnamefont{{Cutri}}},
  \bibinfo{author}{\bibfnamefont{S.}~\bibnamefont{{Schneider}}},
  \bibinfo{author}{\bibfnamefont{L.}~\bibnamefont{{Macri}}},
  \bibinfo{author}{\bibfnamefont{R.}~\bibnamefont{{Steining}}},
  \bibinfo{author}{\bibfnamefont{J.}~\bibnamefont{{Mader}}},
  \bibinfo{author}{\bibfnamefont{N.}~\bibnamefont{{Martimbeau}}},
  \bibnamefont{and} \bibinfo{author}{\bibfnamefont{T.}~\bibnamefont{{George}}},
  in \emph{\bibinfo{booktitle}{ASP Conf. Ser. 329: Nearby Large-Scale
  Structures and the Zone of Avoidance}}, edited by
  \bibinfo{editor}{\bibfnamefont{A.~P.} \bibnamefont{{Fairall}}}
  \bibnamefont{and} \bibinfo{editor}{\bibfnamefont{P.~A.}
  \bibnamefont{{Woudt}}} (\bibinfo{year}{2005}), p. \bibinfo{pages}{135}.

\bibitem[{\citenamefont{{Percival}}(2005)}]{2005MNRAS.356.1168P}
\bibinfo{author}{\bibfnamefont{W.~J.} \bibnamefont{{Percival}}},
  \bibinfo{journal}{\mnras} \textbf{\bibinfo{volume}{356}},
  \bibinfo{pages}{1168} (\bibinfo{year}{2005}), \eprint{astro-ph/0410631}.

\bibitem[{\citenamefont{{Erdo{\u g}du} et~al.}(2006)\citenamefont{{Erdo{\u
  g}du}, {Lahav}, {Huchra}, and {et al.}}}]{2006MNRAS.373...45E}
\bibinfo{author}{\bibfnamefont{P.}~\bibnamefont{{Erdo{\u g}du}}},
  \bibinfo{author}{\bibfnamefont{O.}~\bibnamefont{{Lahav}}},
  \bibinfo{author}{\bibfnamefont{J.}~\bibnamefont{{Huchra}}}, \bibnamefont{and}
  \bibinfo{author}{\bibnamefont{{et al.}}}, \bibinfo{journal}{\mnras}
  \textbf{\bibinfo{volume}{373}}, \bibinfo{pages}{45} (\bibinfo{year}{2006}),
  \eprint{astro-ph/0610005}.

\bibitem[{\citenamefont{{Hoffman}}(1994)}]{1994ASPC...67..185H}
\bibinfo{author}{\bibfnamefont{Y.}~\bibnamefont{{Hoffman}}}, in
  \emph{\bibinfo{booktitle}{ASP Conf. Ser. 67: Unveiling Large-Scale Structures
  Behind the Milky Way}}, edited by
  \bibinfo{editor}{\bibfnamefont{C.}~\bibnamefont{{Balkowski}}}
  \bibnamefont{and} \bibinfo{editor}{\bibfnamefont{R.~C.}
  \bibnamefont{{Kraan-Korteweg}}} (\bibinfo{year}{1994}), p.
  \bibinfo{pages}{185}.

\bibitem[{\citenamefont{{Zaroubi}}(2000)}]{2000ASPC..218..173Z}
\bibinfo{author}{\bibfnamefont{S.}~\bibnamefont{{Zaroubi}}}, in
  \emph{\bibinfo{booktitle}{ASP Conf. Ser. 218: Mapping the Hidden Universe:
  The Universe behind the Mily Way - The Universe in HI}}, edited by
  \bibinfo{editor}{\bibfnamefont{R.~C.} \bibnamefont{{Kraan-Korteweg}}},
  \bibinfo{editor}{\bibfnamefont{P.~A.} \bibnamefont{{Henning}}},
  \bibnamefont{and}
  \bibinfo{editor}{\bibfnamefont{H.}~\bibnamefont{{Andernach}}}
  (\bibinfo{year}{2000}), p. \bibinfo{pages}{173}.

\bibitem[{\citenamefont{{Kraan-Korteweg} and
  {Lahav}}(2000)}]{2000A&ARv..10..211K}
\bibinfo{author}{\bibfnamefont{R.~C.} \bibnamefont{{Kraan-Korteweg}}}
  \bibnamefont{and} \bibinfo{author}{\bibfnamefont{O.}~\bibnamefont{{Lahav}}},
  \bibinfo{journal}{\aapr} \textbf{\bibinfo{volume}{10}}, \bibinfo{pages}{211}
  (\bibinfo{year}{2000}), \eprint{astro-ph/0005501}.

\bibitem[{\citenamefont{{Peacock} and {Dodds}}(1994)}]{1994MNRAS.267.1020P}
\bibinfo{author}{\bibfnamefont{J.~A.} \bibnamefont{{Peacock}}}
  \bibnamefont{and} \bibinfo{author}{\bibfnamefont{S.~J.}
  \bibnamefont{{Dodds}}}, \bibinfo{journal}{\mnras}
  \textbf{\bibinfo{volume}{267}}, \bibinfo{pages}{1020} (\bibinfo{year}{1994}),
  \eprint{astro-ph/9311057}.

\bibitem[{\citenamefont{{Vogeley} and {Szalay}}(1996)}]{1996ApJ...465...34V}
\bibinfo{author}{\bibfnamefont{M.~S.} \bibnamefont{{Vogeley}}}
  \bibnamefont{and} \bibinfo{author}{\bibfnamefont{A.~S.}
  \bibnamefont{{Szalay}}}, \bibinfo{journal}{\apj}
  \textbf{\bibinfo{volume}{465}}, \bibinfo{pages}{34} (\bibinfo{year}{1996}),
  \eprint{astro-ph/9601185}.

\bibitem[{\citenamefont{{Zaroubi} et~al.}(1997)\citenamefont{{Zaroubi},
  {Zehavi}, {Dekel}, {Hoffman}, and {Kolatt}}}]{1997ApJ...486...21Z}
\bibinfo{author}{\bibfnamefont{S.}~\bibnamefont{{Zaroubi}}},
  \bibinfo{author}{\bibfnamefont{I.}~\bibnamefont{{Zehavi}}},
  \bibinfo{author}{\bibfnamefont{A.}~\bibnamefont{{Dekel}}},
  \bibinfo{author}{\bibfnamefont{Y.}~\bibnamefont{{Hoffman}}},
  \bibnamefont{and} \bibinfo{author}{\bibfnamefont{T.}~\bibnamefont{{Kolatt}}},
  \bibinfo{journal}{\apj} \textbf{\bibinfo{volume}{486}}, \bibinfo{pages}{21}
  (\bibinfo{year}{1997}), \eprint{astro-ph/9610226}.

\bibitem[{\citenamefont{{Tegmark}}(1997{\natexlab{a}})}]{1997PhRvL..79.3806T}
\bibinfo{author}{\bibfnamefont{M.}~\bibnamefont{{Tegmark}}},
  \bibinfo{journal}{Physical Review Letters} \textbf{\bibinfo{volume}{79}},
  \bibinfo{pages}{3806} (\bibinfo{year}{1997}{\natexlab{a}}),
  \eprint{astro-ph/9706198}.

\bibitem[{\citenamefont{{Eisenstein} and {Hu}}(1999)}]{1999ApJ...511....5E}
\bibinfo{author}{\bibfnamefont{D.~J.} \bibnamefont{{Eisenstein}}}
  \bibnamefont{and} \bibinfo{author}{\bibfnamefont{W.}~\bibnamefont{{Hu}}},
  \bibinfo{journal}{\apj} \textbf{\bibinfo{volume}{511}}, \bibinfo{pages}{5}
  (\bibinfo{year}{1999}), \eprint{astro-ph/9710252}.

\bibitem[{\citenamefont{{Efstathiou} et~al.}(1992)\citenamefont{{Efstathiou},
  {Bond}, and {White}}}]{1992MNRAS.258P...1E}
\bibinfo{author}{\bibfnamefont{G.}~\bibnamefont{{Efstathiou}}},
  \bibinfo{author}{\bibfnamefont{J.~R.} \bibnamefont{{Bond}}},
  \bibnamefont{and} \bibinfo{author}{\bibfnamefont{S.~D.~M.}
  \bibnamefont{{White}}}, \bibinfo{journal}{\mnras}
  \textbf{\bibinfo{volume}{258}}, \bibinfo{pages}{1P} (\bibinfo{year}{1992}).

\bibitem[{\citenamefont{{Bunn} et~al.}(1995)\citenamefont{{Bunn}, {Scott}, and
  {White}}}]{1995ApJ...441L...9B}
\bibinfo{author}{\bibfnamefont{E.~F.} \bibnamefont{{Bunn}}},
  \bibinfo{author}{\bibfnamefont{D.}~\bibnamefont{{Scott}}}, \bibnamefont{and}
  \bibinfo{author}{\bibfnamefont{M.}~\bibnamefont{{White}}},
  \bibinfo{journal}{\apjl} \textbf{\bibinfo{volume}{441}}, \bibinfo{pages}{L9}
  (\bibinfo{year}{1995}), \eprint{astro-ph/9409003}.

\bibitem[{\citenamefont{{Janssen} and {Gulkis}}(1992)}]{1992issa.proc..391J}
\bibinfo{author}{\bibfnamefont{M.~A.} \bibnamefont{{Janssen}}}
  \bibnamefont{and} \bibinfo{author}{\bibfnamefont{S.}~\bibnamefont{{Gulkis}}},
  in \emph{\bibinfo{booktitle}{NATO ASIC Proc. 359: The Infrared and
  Submillimetre Sky after COBE}}, edited by
  \bibinfo{editor}{\bibfnamefont{M.}~\bibnamefont{{Signore}}} \bibnamefont{and}
  \bibinfo{editor}{\bibfnamefont{C.}~\bibnamefont{{Dupraz}}}
  (\bibinfo{publisher}{Kluwer Academic Publishers},
  \bibinfo{address}{Dordtrecht}, \bibinfo{year}{1992}), pp.
  \bibinfo{pages}{391--408}.

\bibitem[{\citenamefont{{Bunn} et~al.}(1994)\citenamefont{{Bunn}, {Fisher},
  {Hoffman}, {Lahav}, {Silk}, and {Zaroubi}}}]{1994ApJ...432L..75B}
\bibinfo{author}{\bibfnamefont{E.~F.} \bibnamefont{{Bunn}}},
  \bibinfo{author}{\bibfnamefont{K.~B.} \bibnamefont{{Fisher}}},
  \bibinfo{author}{\bibfnamefont{Y.}~\bibnamefont{{Hoffman}}},
  \bibinfo{author}{\bibfnamefont{O.}~\bibnamefont{{Lahav}}},
  \bibinfo{author}{\bibfnamefont{J.}~\bibnamefont{{Silk}}}, \bibnamefont{and}
  \bibinfo{author}{\bibfnamefont{S.}~\bibnamefont{{Zaroubi}}},
  \bibinfo{journal}{\apjl} \textbf{\bibinfo{volume}{432}}, \bibinfo{pages}{L75}
  (\bibinfo{year}{1994}), \eprint{astro-ph/9404007}.

\bibitem[{\citenamefont{{Maisinger} et~al.}(1997)\citenamefont{{Maisinger},
  {Hobson}, and {Lasenby}}}]{1997MNRAS.290..313M}
\bibinfo{author}{\bibfnamefont{K.}~\bibnamefont{{Maisinger}}},
  \bibinfo{author}{\bibfnamefont{M.~P.} \bibnamefont{{Hobson}}},
  \bibnamefont{and} \bibinfo{author}{\bibfnamefont{A.~N.}
  \bibnamefont{{Lasenby}}}, \bibinfo{journal}{\mnras}
  \textbf{\bibinfo{volume}{290}}, \bibinfo{pages}{313} (\bibinfo{year}{1997}).

\bibitem[{\citenamefont{{Tegmark}}(1997{\natexlab{b}})}]{1997PhRvD..56.4514T}
\bibinfo{author}{\bibfnamefont{M.}~\bibnamefont{{Tegmark}}},
  \bibinfo{journal}{\prd} \textbf{\bibinfo{volume}{56}}, \bibinfo{pages}{4514}
  (\bibinfo{year}{1997}{\natexlab{b}}), \eprint{astro-ph/9705188}.

\bibitem[{\citenamefont{{Tegmark}}(1997{\natexlab{c}})}]{1997ApJ...480L..87T}
\bibinfo{author}{\bibfnamefont{M.}~\bibnamefont{{Tegmark}}},
  \bibinfo{journal}{\apjl} \textbf{\bibinfo{volume}{480}}, \bibinfo{pages}{L87}
  (\bibinfo{year}{1997}{\natexlab{c}}), \eprint{astro-ph/9611130}.

\bibitem[{\citenamefont{{Dodelson}}(1997)}]{1997ApJ...482..577D}
\bibinfo{author}{\bibfnamefont{S.}~\bibnamefont{{Dodelson}}},
  \bibinfo{journal}{\apj} \textbf{\bibinfo{volume}{482}}, \bibinfo{pages}{577}
  (\bibinfo{year}{1997}), \eprint{astro-ph/9512021}.

\bibitem[{\citenamefont{{Hobson} et~al.}(1998)\citenamefont{{Hobson}, {Jones},
  {Lasenby}, and {Bouchet}}}]{1998MNRAS.300....1H}
\bibinfo{author}{\bibfnamefont{M.~P.} \bibnamefont{{Hobson}}},
  \bibinfo{author}{\bibfnamefont{A.~W.} \bibnamefont{{Jones}}},
  \bibinfo{author}{\bibfnamefont{A.~N.} \bibnamefont{{Lasenby}}},
  \bibnamefont{and} \bibinfo{author}{\bibfnamefont{F.~R.}
  \bibnamefont{{Bouchet}}}, \bibinfo{journal}{\mnras}
  \textbf{\bibinfo{volume}{300}}, \bibinfo{pages}{1} (\bibinfo{year}{1998}),
  \eprint{astro-ph/9806387}.

\bibitem[{\citenamefont{{Natoli} et~al.}(2001)\citenamefont{{Natoli}, {de
  Gasperis}, {Gheller}, and {Vittorio}}}]{natoli}
\bibinfo{author}{\bibfnamefont{P.}~\bibnamefont{{Natoli}}},
  \bibinfo{author}{\bibfnamefont{G.}~\bibnamefont{{de Gasperis}}},
  \bibinfo{author}{\bibfnamefont{C.}~\bibnamefont{{Gheller}}},
  \bibnamefont{and}
  \bibinfo{author}{\bibfnamefont{N.}~\bibnamefont{{Vittorio}}},
  \bibinfo{journal}{\aap} \textbf{\bibinfo{volume}{372}}, \bibinfo{pages}{346}
  (\bibinfo{year}{2001}), \eprint{astro-ph/0101252}.

\bibitem[{\citenamefont{{Dor{\'e}} et~al.}(2001)\citenamefont{{Dor{\'e}},
  {Teyssier}, {Bouchet}, {Vibert}, and {Prunet}}}]{2001A&A...374..358D}
\bibinfo{author}{\bibfnamefont{O.}~\bibnamefont{{Dor{\'e}}}},
  \bibinfo{author}{\bibfnamefont{R.}~\bibnamefont{{Teyssier}}},
  \bibinfo{author}{\bibfnamefont{F.~R.} \bibnamefont{{Bouchet}}},
  \bibinfo{author}{\bibfnamefont{D.}~\bibnamefont{{Vibert}}}, \bibnamefont{and}
  \bibinfo{author}{\bibfnamefont{S.}~\bibnamefont{{Prunet}}},
  \bibinfo{journal}{\aap} \textbf{\bibinfo{volume}{374}}, \bibinfo{pages}{358}
  (\bibinfo{year}{2001}), \eprint{astro-ph/0101112}.

\bibitem[{\citenamefont{{Stompor} et~al.}(2001)\citenamefont{{Stompor},
  {Balbi}, {Borrill}, {Ferreira}, {Hanany}, {Jaffe}, {Lee}, {Oh}, {Rabii},
  {Richards} et~al.}}]{maxima}
\bibinfo{author}{\bibfnamefont{R.}~\bibnamefont{{Stompor}}},
  \bibinfo{author}{\bibfnamefont{A.}~\bibnamefont{{Balbi}}},
  \bibinfo{author}{\bibfnamefont{J.~D.} \bibnamefont{{Borrill}}},
  \bibinfo{author}{\bibfnamefont{P.~G.} \bibnamefont{{Ferreira}}},
  \bibinfo{author}{\bibfnamefont{S.}~\bibnamefont{{Hanany}}},
  \bibinfo{author}{\bibfnamefont{A.~H.} \bibnamefont{{Jaffe}}},
  \bibinfo{author}{\bibfnamefont{A.~T.} \bibnamefont{{Lee}}},
  \bibinfo{author}{\bibfnamefont{S.}~\bibnamefont{{Oh}}},
  \bibinfo{author}{\bibfnamefont{B.}~\bibnamefont{{Rabii}}},
  \bibinfo{author}{\bibfnamefont{P.~L.} \bibnamefont{{Richards}}},
  \bibnamefont{et~al.}, \bibinfo{journal}{\prd} \textbf{\bibinfo{volume}{65}},
  \bibinfo{pages}{022003} (\bibinfo{year}{2001}), \eprint{astro-ph/0106451}.

\bibitem[{\citenamefont{{Wandelt} et~al.}(2004)\citenamefont{{Wandelt},
  {Larson}, and {Lakshminarayanan}}}]{2004PhRvD..70h3511W}
\bibinfo{author}{\bibfnamefont{B.~D.} \bibnamefont{{Wandelt}}},
  \bibinfo{author}{\bibfnamefont{D.~L.} \bibnamefont{{Larson}}},
  \bibnamefont{and}
  \bibinfo{author}{\bibfnamefont{A.}~\bibnamefont{{Lakshminarayanan}}},
  \bibinfo{journal}{\prd} \textbf{\bibinfo{volume}{70}},
  \bibinfo{pages}{083511} (\bibinfo{year}{2004}), \eprint{astro-ph/0310080}.

\bibitem[{\citenamefont{{Eriksen} et~al.}(2004)\citenamefont{{Eriksen},
  {O'Dwyer}, {Jewell}, {Wandelt}, {Larson}, {G{\'o}rski}, {Levin}, {Banday},
  and {Lilje}}}]{2004ApJS..155..227E}
\bibinfo{author}{\bibfnamefont{H.~K.} \bibnamefont{{Eriksen}}},
  \bibinfo{author}{\bibfnamefont{I.~J.} \bibnamefont{{O'Dwyer}}},
  \bibinfo{author}{\bibfnamefont{J.~B.} \bibnamefont{{Jewell}}},
  \bibinfo{author}{\bibfnamefont{B.~D.} \bibnamefont{{Wandelt}}},
  \bibinfo{author}{\bibfnamefont{D.~L.} \bibnamefont{{Larson}}},
  \bibinfo{author}{\bibfnamefont{K.~M.} \bibnamefont{{G{\'o}rski}}},
  \bibinfo{author}{\bibfnamefont{S.}~\bibnamefont{{Levin}}},
  \bibinfo{author}{\bibfnamefont{A.~J.} \bibnamefont{{Banday}}},
  \bibnamefont{and} \bibinfo{author}{\bibfnamefont{P.~B.}
  \bibnamefont{{Lilje}}}, \bibinfo{journal}{\apjs}
  \textbf{\bibinfo{volume}{155}}, \bibinfo{pages}{227} (\bibinfo{year}{2004}),
  \eprint{astro-ph/0407028}.

\bibitem[{\citenamefont{{Jewell} et~al.}(2004)\citenamefont{{Jewell}, {Levin},
  and {Anderson}}}]{2004ApJ...609....1J}
\bibinfo{author}{\bibfnamefont{J.}~\bibnamefont{{Jewell}}},
  \bibinfo{author}{\bibfnamefont{S.}~\bibnamefont{{Levin}}}, \bibnamefont{and}
  \bibinfo{author}{\bibfnamefont{C.~H.} \bibnamefont{{Anderson}}},
  \bibinfo{journal}{\apj} \textbf{\bibinfo{volume}{609}}, \bibinfo{pages}{1}
  (\bibinfo{year}{2004}), \eprint{astro-ph/0209560}.

\bibitem[{\citenamefont{{Yvon} and {Mayet}}(2005)}]{mirage}
\bibinfo{author}{\bibfnamefont{D.}~\bibnamefont{{Yvon}}} \bibnamefont{and}
  \bibinfo{author}{\bibfnamefont{F.}~\bibnamefont{{Mayet}}},
  \bibinfo{journal}{\aap} \textbf{\bibinfo{volume}{436}}, \bibinfo{pages}{729}
  (\bibinfo{year}{2005}), \eprint{astro-ph/0401505}.

\bibitem[{\citenamefont{{Keih{\"a}nen}
  et~al.}(2005)\citenamefont{{Keih{\"a}nen}, {Kurki-Suonio}, and
  {Poutanen}}}]{madam}
\bibinfo{author}{\bibfnamefont{E.}~\bibnamefont{{Keih{\"a}nen}}},
  \bibinfo{author}{\bibfnamefont{H.}~\bibnamefont{{Kurki-Suonio}}},
  \bibnamefont{and}
  \bibinfo{author}{\bibfnamefont{T.}~\bibnamefont{{Poutanen}}},
  \bibinfo{journal}{\mnras} \textbf{\bibinfo{volume}{360}},
  \bibinfo{pages}{390} (\bibinfo{year}{2005}), \eprint{astro-ph/0412517}.

\bibitem[{\citenamefont{{Sutton} and {Wandelt}}(2006)}]{2006ApJS..162..401S}
\bibinfo{author}{\bibfnamefont{E.~C.} \bibnamefont{{Sutton}}} \bibnamefont{and}
  \bibinfo{author}{\bibfnamefont{B.~D.} \bibnamefont{{Wandelt}}},
  \bibinfo{journal}{\apjs} \textbf{\bibinfo{volume}{162}}, \bibinfo{pages}{401}
  (\bibinfo{year}{2006}).

\bibitem[{\citenamefont{{Larson} et~al.}(2007)\citenamefont{{Larson},
  {Eriksen}, {Wandelt}, {G{\'o}rski}, {Huey}, {Jewell}, and
  {O'Dwyer}}}]{2007ApJ...656..653L}
\bibinfo{author}{\bibfnamefont{D.~L.} \bibnamefont{{Larson}}},
  \bibinfo{author}{\bibfnamefont{H.~K.} \bibnamefont{{Eriksen}}},
  \bibinfo{author}{\bibfnamefont{B.~D.} \bibnamefont{{Wandelt}}},
  \bibinfo{author}{\bibfnamefont{K.~M.} \bibnamefont{{G{\'o}rski}}},
  \bibinfo{author}{\bibfnamefont{G.}~\bibnamefont{{Huey}}},
  \bibinfo{author}{\bibfnamefont{J.~B.} \bibnamefont{{Jewell}}},
  \bibnamefont{and} \bibinfo{author}{\bibfnamefont{I.~J.}
  \bibnamefont{{O'Dwyer}}}, \bibinfo{journal}{\apj}
  \textbf{\bibinfo{volume}{656}}, \bibinfo{pages}{653} (\bibinfo{year}{2007}),
  \eprint{astro-ph/0608007}.

\bibitem[{\citenamefont{Hinshaw et~al.}(2008)}]{Hinshaw:2008kr}
\bibinfo{author}{\bibfnamefont{G.}~\bibnamefont{Hinshaw}} \bibnamefont{et~al.}
  (\bibinfo{collaboration}{WMAP}), \bibinfo{journal}{arXiv}
  \textbf{\bibinfo{volume}{0803.0732}} (\bibinfo{year}{2008}),
  \eprint{0803.0732}.

\bibitem[{\citenamefont{{Seljak} and
  {Zaldarriaga}}(1996)}]{1996ApJ...469..437S}
\bibinfo{author}{\bibfnamefont{U.}~\bibnamefont{{Seljak}}} \bibnamefont{and}
  \bibinfo{author}{\bibfnamefont{M.}~\bibnamefont{{Zaldarriaga}}},
  \bibinfo{journal}{\apj} \textbf{\bibinfo{volume}{469}}, \bibinfo{pages}{437}
  (\bibinfo{year}{1996}), \eprint{arXiv:astro-ph/9603033}.

\bibitem[{\citenamefont{Lewis et~al.}(2000)\citenamefont{Lewis, Challinor, and
  Lasenby}}]{Lewis:1999bs}
\bibinfo{author}{\bibfnamefont{A.}~\bibnamefont{Lewis}},
  \bibinfo{author}{\bibfnamefont{A.}~\bibnamefont{Challinor}},
  \bibnamefont{and} \bibinfo{author}{\bibfnamefont{A.}~\bibnamefont{Lasenby}},
  \bibinfo{journal}{Astrophys. J.} \textbf{\bibinfo{volume}{538}},
  \bibinfo{pages}{473} (\bibinfo{year}{2000}), \eprint{astro-ph/9911177}.

\bibitem[{\citenamefont{{Doran}}(2005)}]{2005JCAP...10..011D}
\bibinfo{author}{\bibfnamefont{M.}~\bibnamefont{{Doran}}},
  \bibinfo{journal}{Journal of Cosmology and Astro-Particle Physics}
  \textbf{\bibinfo{volume}{10}}, \bibinfo{pages}{11} (\bibinfo{year}{2005}),
  \eprint{arXiv:astro-ph/0302138}.

\bibitem[{\citenamefont{{Bunn} and {Sugiyama}}(1995)}]{1995ApJ...446...49B}
\bibinfo{author}{\bibfnamefont{E.~F.} \bibnamefont{{Bunn}}} \bibnamefont{and}
  \bibinfo{author}{\bibfnamefont{N.}~\bibnamefont{{Sugiyama}}},
  \bibinfo{journal}{\apj} \textbf{\bibinfo{volume}{446}}, \bibinfo{pages}{49}
  (\bibinfo{year}{1995}), \eprint{astro-ph/9407069}.

\bibitem[{\citenamefont{{Tegmark} et~al.}(1997)\citenamefont{{Tegmark},
  {Taylor}, and {Heavens}}}]{1997ApJ...480...22T}
\bibinfo{author}{\bibfnamefont{M.}~\bibnamefont{{Tegmark}}},
  \bibinfo{author}{\bibfnamefont{A.~N.} \bibnamefont{{Taylor}}},
  \bibnamefont{and} \bibinfo{author}{\bibfnamefont{A.~F.}
  \bibnamefont{{Heavens}}}, \bibinfo{journal}{\apj}
  \textbf{\bibinfo{volume}{480}}, \bibinfo{pages}{22} (\bibinfo{year}{1997}),
  \eprint{astro-ph/9603021}.

\bibitem[{\citenamefont{{Tegmark}}(1997{\natexlab{d}})}]{1997PhRvD..55.5895T}
\bibinfo{author}{\bibfnamefont{M.}~\bibnamefont{{Tegmark}}},
  \bibinfo{journal}{\prd} \textbf{\bibinfo{volume}{55}}, \bibinfo{pages}{5895}
  (\bibinfo{year}{1997}{\natexlab{d}}), \eprint{astro-ph/9611174}.

\bibitem[{\citenamefont{Nolta et~al.}(2008)}]{Nolta:2008ih}
\bibinfo{author}{\bibfnamefont{M.~R.} \bibnamefont{Nolta}} \bibnamefont{et~al.}
  (\bibinfo{collaboration}{WMAP}), \bibinfo{journal}{arXiv}
  \textbf{\bibinfo{volume}{0803.0593}} (\bibinfo{year}{2008}),
  \eprint{0803.0593}.

\bibitem[{\citenamefont{{Guth}}(1981)}]{1981PhRvD..23..347G}
\bibinfo{author}{\bibfnamefont{A.~H.} \bibnamefont{{Guth}}},
  \bibinfo{journal}{\prd} \textbf{\bibinfo{volume}{23}}, \bibinfo{pages}{347}
  (\bibinfo{year}{1981}).

\bibitem[{\citenamefont{{Linde}}(1982)}]{1982PhLB..108..389L}
\bibinfo{author}{\bibfnamefont{A.~D.} \bibnamefont{{Linde}}},
  \bibinfo{journal}{Physics Letters B} \textbf{\bibinfo{volume}{108}},
  \bibinfo{pages}{389} (\bibinfo{year}{1982}).

\bibitem[{\citenamefont{{Albrecht} and
  {Steinhardt}}(1982)}]{1982PhRvL..48.1220A}
\bibinfo{author}{\bibfnamefont{A.}~\bibnamefont{{Albrecht}}} \bibnamefont{and}
  \bibinfo{author}{\bibfnamefont{P.~J.} \bibnamefont{{Steinhardt}}},
  \bibinfo{journal}{Physical Review Letters} \textbf{\bibinfo{volume}{48}},
  \bibinfo{pages}{1220} (\bibinfo{year}{1982}).

\bibitem[{\citenamefont{{Guth} and {Pi}}(1982)}]{1982PhRvL..49.1110G}
\bibinfo{author}{\bibfnamefont{A.~H.} \bibnamefont{{Guth}}} \bibnamefont{and}
  \bibinfo{author}{\bibfnamefont{S.-Y.} \bibnamefont{{Pi}}},
  \bibinfo{journal}{Physical Review Letters} \textbf{\bibinfo{volume}{49}},
  \bibinfo{pages}{1110} (\bibinfo{year}{1982}).

\bibitem[{\citenamefont{{Starobinsky}}(1982)}]{1982PhLB..117..175S}
\bibinfo{author}{\bibfnamefont{A.~A.} \bibnamefont{{Starobinsky}}},
  \bibinfo{journal}{Physics Letters B} \textbf{\bibinfo{volume}{117}},
  \bibinfo{pages}{175} (\bibinfo{year}{1982}).

\bibitem[{\citenamefont{{Bardeen} et~al.}(1983)\citenamefont{{Bardeen},
  {Steinhardt}, and {Turner}}}]{1983PhRvD..28..679B}
\bibinfo{author}{\bibfnamefont{J.~M.} \bibnamefont{{Bardeen}}},
  \bibinfo{author}{\bibfnamefont{P.~J.} \bibnamefont{{Steinhardt}}},
  \bibnamefont{and} \bibinfo{author}{\bibfnamefont{M.~S.}
  \bibnamefont{{Turner}}}, \bibinfo{journal}{\prd}
  \textbf{\bibinfo{volume}{28}}, \bibinfo{pages}{679} (\bibinfo{year}{1983}).

\bibitem[{\citenamefont{{Hu}}(2001)}]{2001PhRvD..64h3005H}
\bibinfo{author}{\bibfnamefont{W.}~\bibnamefont{{Hu}}}, \bibinfo{journal}{\prd}
  \textbf{\bibinfo{volume}{64}}, \bibinfo{pages}{083005}
  (\bibinfo{year}{2001}), \eprint{astro-ph/0105117}.

\bibitem[{\citenamefont{{Bernardeau} and {Uzan}}(2002)}]{2002PhRvD..66j3506B}
\bibinfo{author}{\bibfnamefont{F.}~\bibnamefont{{Bernardeau}}}
  \bibnamefont{and} \bibinfo{author}{\bibfnamefont{J.-P.}
  \bibnamefont{{Uzan}}}, \bibinfo{journal}{\prd} \textbf{\bibinfo{volume}{66}},
  \bibinfo{pages}{103506} (\bibinfo{year}{2002}), \eprint{hep-ph/0207295}.

\bibitem[{\citenamefont{{Bartolo} et~al.}(2004)\citenamefont{{Bartolo},
  {Komatsu}, {Matarrese}, and {Riotto}}}]{2004PhR...402..103B}
\bibinfo{author}{\bibfnamefont{N.}~\bibnamefont{{Bartolo}}},
  \bibinfo{author}{\bibfnamefont{E.}~\bibnamefont{{Komatsu}}},
  \bibinfo{author}{\bibfnamefont{S.}~\bibnamefont{{Matarrese}}},
  \bibnamefont{and} \bibinfo{author}{\bibfnamefont{A.}~\bibnamefont{{Riotto}}},
  \bibinfo{journal}{\physrep} \textbf{\bibinfo{volume}{402}},
  \bibinfo{pages}{103} (\bibinfo{year}{2004}), \eprint{astro-ph/0406398}.

\bibitem[{\citenamefont{{Babich} et~al.}(2004)\citenamefont{{Babich},
  {Creminelli}, and {Zaldarriaga}}}]{2004JCAP...08..009B}
\bibinfo{author}{\bibfnamefont{D.}~\bibnamefont{{Babich}}},
  \bibinfo{author}{\bibfnamefont{P.}~\bibnamefont{{Creminelli}}},
  \bibnamefont{and}
  \bibinfo{author}{\bibfnamefont{M.}~\bibnamefont{{Zaldarriaga}}},
  \bibinfo{journal}{Journal of Cosmology and Astro-Particle Physics}
  \textbf{\bibinfo{volume}{8}}, \bibinfo{pages}{9} (\bibinfo{year}{2004}),
  \eprint{arXiv:astro-ph/0405356}.

\bibitem[{\citenamefont{{Komatsu} et~al.}(2002)\citenamefont{{Komatsu},
  {Wandelt}, {Spergel}, {Banday}, and {G{\'o}rski}}}]{2002ApJ...566...19K}
\bibinfo{author}{\bibfnamefont{E.}~\bibnamefont{{Komatsu}}},
  \bibinfo{author}{\bibfnamefont{B.~D.} \bibnamefont{{Wandelt}}},
  \bibinfo{author}{\bibfnamefont{D.~N.} \bibnamefont{{Spergel}}},
  \bibinfo{author}{\bibfnamefont{A.~J.} \bibnamefont{{Banday}}},
  \bibnamefont{and} \bibinfo{author}{\bibfnamefont{K.~M.}
  \bibnamefont{{G{\'o}rski}}}, \bibinfo{journal}{\apj}
  \textbf{\bibinfo{volume}{566}}, \bibinfo{pages}{19} (\bibinfo{year}{2002}),
  \eprint{arXiv:astro-ph/0107605}.

\bibitem[{\citenamefont{{Babich} and
  {Zaldarriaga}}(2004)}]{2004PhRvD..70h3005B}
\bibinfo{author}{\bibfnamefont{D.}~\bibnamefont{{Babich}}} \bibnamefont{and}
  \bibinfo{author}{\bibfnamefont{M.}~\bibnamefont{{Zaldarriaga}}},
  \bibinfo{journal}{\prd} \textbf{\bibinfo{volume}{70}},
  \bibinfo{pages}{083005} (\bibinfo{year}{2004}),
  \eprint{arXiv:astro-ph/0408455}.

\bibitem[{\citenamefont{{Komatsu} et~al.}(2005)\citenamefont{{Komatsu},
  {Spergel}, and {Wandelt}}}]{2005ApJ...634...14K}
\bibinfo{author}{\bibfnamefont{E.}~\bibnamefont{{Komatsu}}},
  \bibinfo{author}{\bibfnamefont{D.~N.} \bibnamefont{{Spergel}}},
  \bibnamefont{and} \bibinfo{author}{\bibfnamefont{B.~D.}
  \bibnamefont{{Wandelt}}}, \bibinfo{journal}{\apj}
  \textbf{\bibinfo{volume}{634}}, \bibinfo{pages}{14} (\bibinfo{year}{2005}),
  \eprint{arXiv:astro-ph/0305189}.

\bibitem[{\citenamefont{{Yadav} et~al.}(2007)\citenamefont{{Yadav}, {Komatsu},
  and {Wandelt}}}]{2007ApJ...664..680Y}
\bibinfo{author}{\bibfnamefont{A.~P.~S.} \bibnamefont{{Yadav}}},
  \bibinfo{author}{\bibfnamefont{E.}~\bibnamefont{{Komatsu}}},
  \bibnamefont{and} \bibinfo{author}{\bibfnamefont{B.~D.}
  \bibnamefont{{Wandelt}}}, \bibinfo{journal}{\apj}
  \textbf{\bibinfo{volume}{664}}, \bibinfo{pages}{680} (\bibinfo{year}{2007}),
  \eprint{arXiv:astro-ph/0701921}.

\bibitem[{\citenamefont{{Yadav} et~al.}(2008)\citenamefont{{Yadav}, {Komatsu},
  {Wandelt}, {Liguori}, {Hansen}, and {Matarrese}}}]{2008ApJ...678..578Y}
\bibinfo{author}{\bibfnamefont{A.~P.~S.} \bibnamefont{{Yadav}}},
  \bibinfo{author}{\bibfnamefont{E.}~\bibnamefont{{Komatsu}}},
  \bibinfo{author}{\bibfnamefont{B.~D.} \bibnamefont{{Wandelt}}},
  \bibinfo{author}{\bibfnamefont{M.}~\bibnamefont{{Liguori}}},
  \bibinfo{author}{\bibfnamefont{F.~K.} \bibnamefont{{Hansen}}},
  \bibnamefont{and}
  \bibinfo{author}{\bibfnamefont{S.}~\bibnamefont{{Matarrese}}},
  \bibinfo{journal}{\apj} \textbf{\bibinfo{volume}{678}}, \bibinfo{pages}{578}
  (\bibinfo{year}{2008}), \eprint{arXiv:0711.4933}.

\bibitem[{\citenamefont{{Komatsu} et~al.}(2003)\citenamefont{{Komatsu},
  {Kogut}, {Nolta}, {Bennett}, {Halpern}, {Hinshaw}, {Jarosik}, {Limon},
  {Meyer}, {Page} et~al.}}]{2003ApJS..148..119K}
\bibinfo{author}{\bibfnamefont{E.}~\bibnamefont{{Komatsu}}},
  \bibinfo{author}{\bibfnamefont{A.}~\bibnamefont{{Kogut}}},
  \bibinfo{author}{\bibfnamefont{M.~R.} \bibnamefont{{Nolta}}},
  \bibinfo{author}{\bibfnamefont{C.~L.} \bibnamefont{{Bennett}}},
  \bibinfo{author}{\bibfnamefont{M.}~\bibnamefont{{Halpern}}},
  \bibinfo{author}{\bibfnamefont{G.}~\bibnamefont{{Hinshaw}}},
  \bibinfo{author}{\bibfnamefont{N.}~\bibnamefont{{Jarosik}}},
  \bibinfo{author}{\bibfnamefont{M.}~\bibnamefont{{Limon}}},
  \bibinfo{author}{\bibfnamefont{S.~S.} \bibnamefont{{Meyer}}},
  \bibinfo{author}{\bibfnamefont{L.}~\bibnamefont{{Page}}},
  \bibnamefont{et~al.}, \bibinfo{journal}{\apjs}
  \textbf{\bibinfo{volume}{148}}, \bibinfo{pages}{119} (\bibinfo{year}{2003}),
  \eprint{astro-ph/0302223}.

\bibitem[{\citenamefont{{Curto} et~al.}(2008)\citenamefont{{Curto},
  {Macias-Perez}, {Martinez-Gonzalez}, {Barreiro}, {Santos}, {Hansen},
  {Liguori}, and {Matarrese}}}]{2008arXiv0804.0136C}
\bibinfo{author}{\bibfnamefont{A.}~\bibnamefont{{Curto}}},
  \bibinfo{author}{\bibfnamefont{J.~F.} \bibnamefont{{Macias-Perez}}},
  \bibinfo{author}{\bibfnamefont{E.}~\bibnamefont{{Martinez-Gonzalez}}},
  \bibinfo{author}{\bibfnamefont{R.~B.} \bibnamefont{{Barreiro}}},
  \bibinfo{author}{\bibfnamefont{D.}~\bibnamefont{{Santos}}},
  \bibinfo{author}{\bibfnamefont{F.~K.} \bibnamefont{{Hansen}}},
  \bibinfo{author}{\bibfnamefont{M.}~\bibnamefont{{Liguori}}},
  \bibnamefont{and}
  \bibinfo{author}{\bibfnamefont{S.}~\bibnamefont{{Matarrese}}},
  \bibinfo{journal}{ArXiv e-prints} \textbf{\bibinfo{volume}{0804.0136}}
  (\bibinfo{year}{2008}), \eprint{0804.0136}.

\bibitem[{\citenamefont{{Yadav} and {Wandelt}}(2008)}]{2008PhRvL.100r1301Y}
\bibinfo{author}{\bibfnamefont{A.~P.~S.} \bibnamefont{{Yadav}}}
  \bibnamefont{and} \bibinfo{author}{\bibfnamefont{B.~D.}
  \bibnamefont{{Wandelt}}}, \bibinfo{journal}{Physical Review Letters}
  \textbf{\bibinfo{volume}{100}}, \bibinfo{pages}{181301}
  (\bibinfo{year}{2008}).

\bibitem[{\citenamefont{{Martinez-Gonzalez}}(2008)}]{2008arXiv0805.4157M}
\bibinfo{author}{\bibfnamefont{E.}~\bibnamefont{{Martinez-Gonzalez}}},
  \bibinfo{journal}{ArXiv e-prints} \textbf{\bibinfo{volume}{0805.4157}}
  (\bibinfo{year}{2008}), \eprint{0805.4157}.

\bibitem[{\citenamefont{{Jasche} et~al.}(2009)\citenamefont{{Jasche},
  {Kitaura}, and {Ensslin}}}]{2009arXiv0901.3043J}
\bibinfo{author}{\bibfnamefont{J.}~\bibnamefont{{Jasche}}},
  \bibinfo{author}{\bibfnamefont{F.~S.} \bibnamefont{{Kitaura}}},
  \bibnamefont{and} \bibinfo{author}{\bibfnamefont{T.~A.}
  \bibnamefont{{Ensslin}}}, \bibinfo{journal}{ArXiv e-prints}
  (\bibinfo{year}{2009}), \eprint{0901.3043}.

\bibitem[{\citenamefont{{Coles} and {Jones}}(1991)}]{1991MNRAS.248....1C}
\bibinfo{author}{\bibfnamefont{P.}~\bibnamefont{{Coles}}} \bibnamefont{and}
  \bibinfo{author}{\bibfnamefont{B.}~\bibnamefont{{Jones}}},
  \bibinfo{journal}{\mnras} \textbf{\bibinfo{volume}{248}}, \bibinfo{pages}{1}
  (\bibinfo{year}{1991}).

\bibitem[{\citenamefont{{Vio} et~al.}(2001)\citenamefont{{Vio}, {Andreani}, and
  {Wamsteker}}}]{2001PASP..113.1009V}
\bibinfo{author}{\bibfnamefont{R.}~\bibnamefont{{Vio}}},
  \bibinfo{author}{\bibfnamefont{P.}~\bibnamefont{{Andreani}}},
  \bibnamefont{and}
  \bibinfo{author}{\bibfnamefont{W.}~\bibnamefont{{Wamsteker}}},
  \bibinfo{journal}{\pasp} \textbf{\bibinfo{volume}{113}},
  \bibinfo{pages}{1009} (\bibinfo{year}{2001}),
  \eprint{arXiv:astro-ph/0105107}.

\bibitem[{\citenamefont{{Neyrinck} et~al.}(2009)\citenamefont{{Neyrinck},
  {Szapudi}, and {Szalay}}}]{2009arXiv0903.4693N}
\bibinfo{author}{\bibfnamefont{M.~C.} \bibnamefont{{Neyrinck}}},
  \bibinfo{author}{\bibfnamefont{I.}~\bibnamefont{{Szapudi}}},
  \bibnamefont{and} \bibinfo{author}{\bibfnamefont{A.~S.}
  \bibnamefont{{Szalay}}}, \bibinfo{journal}{ArXiv e-prints}
  (\bibinfo{year}{2009}), \eprint{0903.4693}.

\bibitem[{\citenamefont{{Sachs} and {Wolfe}}(1967)}]{1967ApJ...147...73S}
\bibinfo{author}{\bibfnamefont{R.~K.} \bibnamefont{{Sachs}}} \bibnamefont{and}
  \bibinfo{author}{\bibfnamefont{A.~M.} \bibnamefont{{Wolfe}}},
  \bibinfo{journal}{\apj} \textbf{\bibinfo{volume}{147}}, \bibinfo{pages}{73}
  (\bibinfo{year}{1967}).

\bibitem[{\citenamefont{{Rees} and {Sciama}}(1968)}]{1968Natur.217..511R}
\bibinfo{author}{\bibfnamefont{M.~J.} \bibnamefont{{Rees}}} \bibnamefont{and}
  \bibinfo{author}{\bibfnamefont{D.~W.} \bibnamefont{{Sciama}}},
  \bibinfo{journal}{\nat} \textbf{\bibinfo{volume}{217}}, \bibinfo{pages}{511}
  (\bibinfo{year}{1968}).

\bibitem[{\citenamefont{{Fergusson} and
  {Shellard}}(2008)}]{2008arXiv0812.3413F}
\bibinfo{author}{\bibfnamefont{J.~R.} \bibnamefont{{Fergusson}}}
  \bibnamefont{and} \bibinfo{author}{\bibfnamefont{E.~P.~S.}
  \bibnamefont{{Shellard}}}, \bibinfo{journal}{ArXiv e-prints}
  (\bibinfo{year}{2008}), \eprint{0812.3413}.

\bibitem[{\citenamefont{{Komatsu} and {Spergel}}(2001)}]{2001PhRvD..63f3002K}
\bibinfo{author}{\bibfnamefont{E.}~\bibnamefont{{Komatsu}}} \bibnamefont{and}
  \bibinfo{author}{\bibfnamefont{D.~N.} \bibnamefont{{Spergel}}},
  \bibinfo{journal}{\prd} \textbf{\bibinfo{volume}{63}},
  \bibinfo{pages}{063002} (\bibinfo{year}{2001}),
  \eprint{arXiv:astro-ph/0005036}.

\bibitem[{\citenamefont{{Kogo} and {Komatsu}}(2006)}]{2006PhRvD..73h3007K}
\bibinfo{author}{\bibfnamefont{N.}~\bibnamefont{{Kogo}}} \bibnamefont{and}
  \bibinfo{author}{\bibfnamefont{E.}~\bibnamefont{{Komatsu}}},
  \bibinfo{journal}{\prd} \textbf{\bibinfo{volume}{73}},
  \bibinfo{pages}{083007} (\bibinfo{year}{2006}),
  \eprint{arXiv:astro-ph/0602099}.

\bibitem[{\citenamefont{{Babich}}(2005)}]{2005PhRvD..72d3003B}
\bibinfo{author}{\bibfnamefont{D.}~\bibnamefont{{Babich}}},
  \bibinfo{journal}{\prd} \textbf{\bibinfo{volume}{72}},
  \bibinfo{pages}{043003} (\bibinfo{year}{2005}),
  \eprint{arXiv:astro-ph/0503375}.

\bibitem[{\citenamefont{{Heavens}}(1998)}]{1998MNRAS.299..805H}
\bibinfo{author}{\bibfnamefont{A.~F.} \bibnamefont{{Heavens}}},
  \bibinfo{journal}{\mnras} \textbf{\bibinfo{volume}{299}},
  \bibinfo{pages}{805} (\bibinfo{year}{1998}), \eprint{arXiv:astro-ph/9804222}.

\bibitem[{\citenamefont{{Creminelli} et~al.}(2006)\citenamefont{{Creminelli},
  {Nicolis}, {Senatore}, {Tegmark}, and {Zaldarriaga}}}]{2006JCAP...05..004C}
\bibinfo{author}{\bibfnamefont{P.}~\bibnamefont{{Creminelli}}},
  \bibinfo{author}{\bibfnamefont{A.}~\bibnamefont{{Nicolis}}},
  \bibinfo{author}{\bibfnamefont{L.}~\bibnamefont{{Senatore}}},
  \bibinfo{author}{\bibfnamefont{M.}~\bibnamefont{{Tegmark}}},
  \bibnamefont{and}
  \bibinfo{author}{\bibfnamefont{M.}~\bibnamefont{{Zaldarriaga}}},
  \bibinfo{journal}{Journal of Cosmology and Astro-Particle Physics}
  \textbf{\bibinfo{volume}{5}}, \bibinfo{pages}{4} (\bibinfo{year}{2006}),
  \eprint{arXiv:astro-ph/0509029}.

\bibitem[{\citenamefont{{Yadav} and {Wandelt}}(2005)}]{2005PhRvD..71l3004Y}
\bibinfo{author}{\bibfnamefont{A.~P.} \bibnamefont{{Yadav}}} \bibnamefont{and}
  \bibinfo{author}{\bibfnamefont{B.~D.} \bibnamefont{{Wandelt}}},
  \bibinfo{journal}{\prd} \textbf{\bibinfo{volume}{71}},
  \bibinfo{pages}{123004} (\bibinfo{year}{2005}),
  \eprint{arXiv:astro-ph/0505386}.

\end{thebibliography}

\end{document}